\documentclass[aps,prb,showpacs,notitlepage,floatfix,twocolumn,superscriptaddress,eqsecnum]{revtex4-2}
\usepackage{amssymb}
\usepackage{amsmath}
\usepackage{amsfonts}
\usepackage{bbm}
\usepackage{bm}
\usepackage{graphicx}
\usepackage{braket}
\usepackage{mathtools}
\usepackage{dsfont}
\usepackage{physics}
\usepackage{bbold}

\usepackage[dvipsnames]{xcolor}
\usepackage[titletoc]{appendix}

\newcommand\myshade{85}
\colorlet{myurlcolor}{MidnightBlue}

\usepackage{pifont}
\usepackage{subfigure}
\usepackage{amssymb}
\usepackage{amsmath}
\usepackage{times}
\usepackage{graphicx}
\usepackage{epsfig}
\usepackage{xcolor,dsfont}
\RequirePackage{color}

\definecolor{MyDarkGreen}{rgb}{0.02,0.60,0.06}

\definecolor{kellygreen}{rgb}{0.3, 0.73, 0.09}
\definecolor{deeppink}{rgb}{1.0, 0.08, 0.58}

\usepackage[colorlinks,linkcolor=myurlcolor!\myshade!black,anchorcolor=myurlcolor!\myshade!black,citecolor=myurlcolor!\myshade!black,urlcolor=MidnightBlue]{hyperref}

\def\I{\mathbbm{I}}
\def\HH{\mathbbm{H}}
\def\H{\mathcal{H}}
\def\O{\mathcal{O}}
\def\epr{\ket{\text{EPR}}}
\def\bea{\begin{equation}\begin{aligned}}
\def\eea{\end{aligned}\end{equation}}

\graphicspath{{figures/}}

\begin{document}
\setlength{\unitlength}{1mm}

\title{Quantum Many-Body Scars from Einstein-Podolsky-Rosen States in Bilayer Systems}

\author{Julia Wildeboer}
\email{jwildeb@iastate.edu}
\affiliation{Department of Physics \& Astronomy, Iowa State University, Ames, Iowa 50011, USA}

\author{Christopher M. Langlett}
\email{clanglett85@tamu.edu}
\affiliation{Department of Physics \& Astronomy, Texas A\&M University, College Station, Texas 77843, USA}

\author{Zhi-Cheng Yang}
\email{zcyang19@pku.edu.cn}
\affiliation{Joint Quantum Institute, NIST/University of Maryland, College Park, Maryland 20742, USA}
\affiliation{Joint Center for Quantum Information and Computer Science, NIST/University of Maryland, College Park, Maryland 20742, USA}
\affiliation{School of Physics, Peking University, Beijing 100871, China}
\affiliation{Center for High Energy Physics, Peking University, Beijing 100871, China}

\author{Alexey V. Gorshkov}
\email{gorshkov@umd.edu}
\affiliation{Joint Quantum Institute, NIST/University of Maryland, College Park, Maryland 20742, USA}
\affiliation{Joint Center for Quantum Information and Computer Science, NIST/University of Maryland, College Park, Maryland 20742, USA}

\author{Thomas Iadecola}
\email{iadecola@iastate.edu}
\affiliation{Department of Physics \& Astronomy, Iowa State University, Ames, Iowa 50011, USA}
\affiliation{Ames National Laboratory, Ames, Iowa 50011, USA}

\author{Shenglong Xu}
\email{slxu@tamu.edu}
\affiliation{Department of Physics \& Astronomy, Texas A\&M University, College Station, Texas 77843, USA}

\begin{abstract}
Quantum many-body scar states are special eigenstates of nonintegrable models with distinctive entanglement features that give rise to infinitely long-lived coherent dynamics under quantum quenches from certain initial states.
We elaborate on a construction of quantum many-body scar states in which they emerge from Einstein-Podolsky-Rosen (EPR) states in systems with two layers, wherein the two layers are maximally entangled. 
We apply this construction to spin systems as well as systems of itinerant fermions and bosons and demonstrate how symmetries can be harnessed to enhance its versatility. 
We show that several well known examples of quantum many-body scars, including the tower of states in the spin-1 XY model and the $\eta$-pairing states in the Fermi-Hubbard model, can be understood within this formalism. 
We also demonstrate how an {\it infinite} tower of many-body scar states can emerge in bilayer Bose-Hubbard models with charge conservation.
\end{abstract}
\maketitle
\tableofcontents

\section{Introduction}
Quantum many-body scar~(QMBS) states are special highly excited eigenstates with atypical properties relative to other eigenstates at the same energy density~\cite{serbyn2021_review,Moudgalya22Review,Chandran22Review}.
Such eigenstates emerge in quantum many-body systems that are nonintegrable and are expected to obey the eigenstate thermalization hypothesis~(ETH)~\cite{ETH1,ETH2,GOE}, which states that all eigenstates in a small energy window should exhibit observable properties that are identical in the thermodynamic limit.
In ETH-obeying systems, quantum dynamics from typical initial states exhibits relaxation to a local thermal equilibrium dictated by the eigenstates with which the initial state has overlap~\cite{rigol1,ETH4}.
QMBS states typically constitute a vanishing fraction of all eigenstates in the thermodynamic limit and display no dynamical signature for generic initial states.
However, for special initial states that have overlap either predominantly~\cite{turnerNat} or exclusively~\cite{TomMichael,TomMichael2} with the scar states, the resulting dynamics exhibit distinctive coherent features that are absent in generic quench experiments.
QMBS states often appear in ``towers'' consisting of a number of states scaling polynomially with system size~\cite{Sanjay4,Sanjay5,turnerNat,TomMichael,TomMichael2,Shibata,mark1,mark2,Sanjay1,Chattopadhyay1,Pakrouski,Pallegar,Ren,Burnell,Tang21}.
When these states are equally spaced in energy, the resulting dynamics becomes perfectly oscillatory, leading to sharp experimental signatures of these rare eigenstates~\cite{bernien,Su22,Zhang22,Chen22}.
These coherent dynamics may enable applications of QMBS states in quantum sensing protocols~\cite{Dooley21,Desaules22,Dooley22}. 

One important research direction in this field is the development of systematic constructions of QMBS state towers with distinctive dynamical signatures. 
Techniques for achieving this include group-theoretic constructions~\cite{Pakrouski,Ren,Burnell}, matrix product state methods~\cite{scar-mps,Ren22}, and projector embeddings~\cite{embed2,embed1,ChoiTurner,OkSeulgi,wildeboer_scar_2020,wildeboer_scar_2021,Lee,Banerjee,Biswas22}.
These methods provide insight into manifestations of QMBS states in previously unexplored models, which constitutes another important direction of ongoing research that is connected to efforts to realize such states in experiments.
While QMBS states have been heavily studied in one-dimensional~(1D) systems, their two-dimensional~(2D) counterparts have also begun to attract attention~\cite{TomMichael,OkSeulgi,wildeboer_scar_2020,wildeboer_scar_2021,Ren22,Pallegar}, particularly in light of recent experimental progress~\cite{bluvstein2021}.

Generally QMBS states are identified in a many-body spectrum through their subextensive entanglement scaling, which contrasts with the volume law entanglement scaling of typical finite-energy-density eigenstates.
However, sub-volume-law entanglement entropy is not a necessary condition for an eigenstate to be a QMBS~\cite{ourrainbow,schindler2022exact,srivatsa2022mobility}.
In Ref.~\cite{ourrainbow}, it was demonstrated that there exists a class of QMBS states that are highly entangled while retaining a simple entanglement structure, and that can be embedded into many-body Hamiltonians with a suitable bipartite structure.
These states, dubbed ``rainbow scars'' for their connection to the rainbow state and its long-range entanglement structure~\cite{ourrainbow,Ram_rez_2014,Vitagliano_2010,srivatsa2022mobility}, display extensive entanglement scaling for generic bipartitions, going against the standard definition of QMBS states.  
Thus, entanglement may be a good indicator but not a definitive method to judge if a quantum state is indeed a nonthermal QMBS state. 

In this work, we apply the tools developed in Ref.~\cite{ourrainbow} to construct a wide variety of models with QMBS states. 
We focus our attention on bilayer quantum many-body systems, which can arise both in systems with two layers separated in real space as well as systems where two internal states, such as spin states, play the role of the two layers.
We show that Einstein-Podolsky-Rosen (EPR) states in which the two layers are maximally entangled give rise to QMBS states, and refer to these scar states as ``EPR scars'' 
to distinguish their geometric structure from that of the rainbow state.
We construct these EPR scars for spin systems like those considered in Ref.~\cite{ourrainbow}, as well as for itinerant fermionic and bosonic degrees of freedom, focusing in particular on 2D examples that cannot be mapped to local spin systems.
We lay out the general construction in Sec.~\ref{sec:EPR}.

Our main conclusions are twofold. First, we show that several known examples of QMBS states fall into the EPR scar framework.
In Sec.~\ref{sec:spins}, we show that scars in the spin-$1$ XY model~\cite{TomMichael} can be recast as EPR scars upon mapping the spin-$1$ system to a bilayer spin-$\frac{1}{2}$ system projected onto the local triplet sector.
In Sec.~\ref{sec:fermions}, we show that the celebrated $\eta$-pairing tower of states~\cite{YangPRL,mark1,Sanjay1,Pallegar} also falls into this construction, in addition to laying out generalizations beyond the original Fermi-Hubbard model that include spin-orbit coupling~\cite{Sanjay1,Li} and superconducting pairing~\cite{Pallegar} terms.
In both of these examples the bilayer structure is emergent, stemming from the rewriting in terms of spin-$\frac{1}{2}$'s for the spin-$1$ XY model, and from the spin index for $\eta$-pairing.
Second, in Sec.~\ref{sec:bosons}, we construct an \emph{infinite-dimensional} tower of states in bilayer Bose-Hubbard systems.
In the simplest case, this tower emerges in systems where the intra and interlayer interactions have opposite signs, a scenario that could be engineered in optical lattices for mixtures of two bosonic species.

\section{Bilayer Einstein-Podolsky-Rosen Quantum Many-Body Scar States}
In this section we describe the construction of EPR scar states in bilayer systems and consider how different choices of bipartition influence entanglement calculations.
Subsequently we describe a general construction of parent Hamiltonians in doubled Hilbert spaces that is independent of the concrete details of the single system Hamiltonian $\mathcal{H}_{1}$. 

\label{sec:EPR}
\subsection{The EPR State and its Entanglement Structure}

We begin by considering two identical quantum systems, labeled $1$ and $2$, which may either be spatially separated (e.g., the two layers of a bilayer system) or constructed using an internal degree of freedom (e.g., electron spin). 
The Hilbert space $\mathbbm{H}_{1(2)}$ of each system is spanned by an identical basis of states $\ket{n}_{1(2)}$.
The single-system Hilbert space $\mathbbm{H}_{1(2)}$ need not have a local tensor product structure, the lack of which could arise from projection into a symmetry sector or from kinetic constraints.
A general state in the doubled Hilbert space $\mathbbm{H}=\mathbbm{H}_{1}\otimes\mathbbm{H}_{2}$ is given by $\ket{\psi}=\sum_{n,\tilde{n}}\psi_{n,\tilde{n}} \ket{n}_1 \otimes \ket{\tilde n}_2$.
The many-body EPR state is a pure state defined in the doubled Hilbert space as
\begin{equation}\label{eq:EPR}
\epr = \frac{1}{\sqrt{\rm dim (\HH_1)}}\sum\limits_{n=1}^{\rm dim (\HH_1)}\ket{n}_1 \otimes \ket{n}_2\,.
\end{equation}
Because of its unique entanglement structure, the EPR state and its finite temperature variant called the thermofield-double state,
\begin{equation}
\label{eq:TDF_state}
\ket{\text{TFD}(\beta)} = \frac{1}{\sqrt{Z}}\sum_{n}e^{-\beta E_{n}/2}\ket{n}_{1}\otimes \ket{n}_{2}\, ,
\end{equation} 
where $Z = \sum_{n} e^{-\beta E_{n}}$ is the partition function at inverse temperature $\beta$, are of great research interest across multiple fields~\cite{blackhole2, hartman2013time, papadodimas2015local, schuster2021many, nezami2021quantum}. 
The EPR state that is the focus of our work is simply the infinite-temperature thermofield-double state: $\epr = \ket{\text{TFD}(\beta=0)}$. 
In the rest of paper, we drop the subscript labelling the two copies $1$ and $2$ unless ambiguity arises. 

\begin{figure}
\includegraphics[width=1.00\columnwidth]{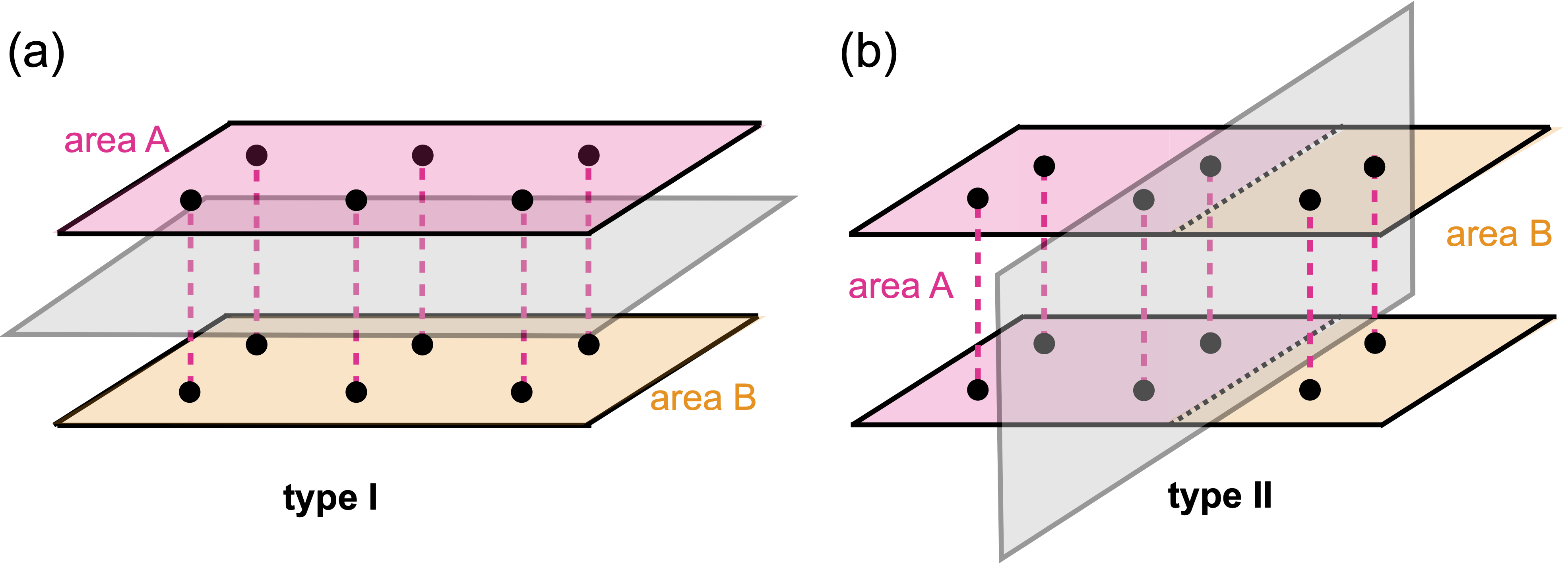}
\caption{
         {\it Entanglement bipartitions for bilayer systems}. 
         (a) A bipartition of type I, where subregions $A$ and $B$ are the top and bottom layers, respectively. 
         (b) Contrarily a bipartition of type II includes degrees of freedom from both layers in each subregion ($A$ or $B$). 
         In the 2D geometry shown here, each pair of sites 
         $i$ (from the top and the bottom layer) that interact with each other are sites with the same coordinates $(x,y)$ 
         and different $z$-coordinates. 
        }
\label{fig:area_cuts}
\end{figure}

Now we consider the entanglement structure of $\epr$. 
Throughout the paper, we quantify the entanglement between a subregion $A$ and its complement $B$ with the von-Neumann entropy $S^{\rm vN}=-\rm Tr (\rho_A \ln \rho_A)$, where $\rho_A$ is the reduced density matrix of subregion $A$ computed from an eigenstate $|E\rangle$, i.e. 
$\rho_{A} = {\rm Tr}_{B}\left(\ket{E}\bra{E}\right)$. 
We note that, by construction, the two copies of the EPR state Eq.~\eqref{eq:EPR} share the maximal entanglement entropy 
$S^{\rm vN} = \ln [\rm dim(\HH_1)]$, 
where region $A$ is defined as in  Fig.~\ref{fig:area_cuts}(a). 

In the simplest case, each Hilbert space $\mathbbm{H}_{1(2)}$ has a tensor product structure.
For concreteness, let us consider the Hilbert space of $N$ qubits with dimension $2^N$ for each copy (generalization to local dimension larger than $2$ is straightforward).
In this case, the sum in Eq.~\eqref{eq:EPR} can be carried out independently for each pair of qubits, and 
the EPR state becomes a product of local EPR pairs,
\bea\label{eq:unconstrained_EPR}
    \epr = \frac{1}{2^{N/2}} \prod_{i=1}^{N} \left(\ket{0_{i}}\otimes  \ket{0_{i}} + \ket{1_{i}}\otimes \ket{1_{i}}\right)\,.
\eea
The bipartition in Fig.~\ref{fig:area_cuts}(a) cuts through $N$ EPR pairs, leading to the maximal entanglement entropy $S^{\rm vN} = N\ln 2$ between the two copies. On the other hand, the bipartite entanglement in Fig.~\ref{fig:area_cuts}(b) is zero because of the tensor product structure of the Hilbert space. 

Now we consider the EPR state in Eq.~\eqref{eq:EPR} for more general Hilbert spaces without tensor product structure.
As mentioned before, by construction, the bipartition in Fig.~\ref{fig:area_cuts}(a) always leads to the maximal entanglement entropy $S^{\rm vN} = \ln \rm dim (\HH)$.
However, the entanglement entropy for the bipartition in Fig.~\ref{fig:area_cuts}(b) is in general nonzero for Hilbert spaces that are not factorized. 
A common approach to realizing such a Hilbert space is to add a constraint, e.g.~from symmetries or kinetic constraints, to the Hilbert space of $N$ qubits.
Suppose the constraint is implemented by fixing some set $\{p\}$ of conserved quantities in $\HH_{1(2)}$, and let $\mathcal{P}_{\{p\}}$ be the projector into the constrained Hilbert space. 
Assuming a bipartition of the type shown in Fig.~\ref{fig:area_cuts}(b), the EPR state can then be written as  
\begin{equation}
\begin{aligned}
    \epr &= \frac{1}{\sqrt{\rm dim  (\HH_1)}}\sum_n \mathcal{P}_{\{p\}}\ket{n}_1 \otimes \mathcal{P}_{\{p\}}\ket{n}_2\\
                  &= \frac{1}{\sqrt{\rm dim  (\HH_1)}}\sum_{n_A, n_B} \mathcal{P}_{\{p\}}\ket{n_A n_B}_1 \otimes \mathcal{P}_{\{p\}}\ket{n_A n_B}_2\\
\end{aligned}
\end{equation}
where $\ket{n}$ now lives in the Hilbert space of qubits and can be split into two parts $\ket{n_A}$ and $\ket{n_B}$. The only nonzero matrix elements of the density matrix $\rho_A$ of region $A$ are then given by 
\begin{equation}
    \bra{n_A}_1 \otimes \bra{n_A}_2 \rho_A \ket{n'_A }_1 \otimes \ket{n'_A}_2 =  \sum_{n_B} p_{n_A n_B}p_{n_{A'} n_B}\, ,
\end{equation}
where the matrix element $p_{n_A n_B}$ equals one up to normalization for states $\ket{n_A n_B}$ within the constrained Hilbert space and zero otherwise. 

From the above calculation, it becomes clear that the entanglement entropy of $\epr$ for the bipartition in Fig.~\ref{fig:area_cuts}(b) is the same as that of the following state defined in a single copy of the system,
\begin{equation}\label{eq:single_layer}
    \ket{\psi} = \frac{1}{\sqrt{\rm dim  (\HH_1)}}\sum_n \mathcal{P}_{\{p\}}\ket{n}\,.
\end{equation}
Without the constraint imposed by the set $\{p\}$, this state is the $x$-basis product state $\ket{+}\cdots \ket{+}$, where $\ket{+}=(\ket 0 + \ket 1)/\sqrt 2$.
Adding a constraint results in entanglement.
For instance, in the case of U(1) symmetry, the projection of this state into a generic magnetization sector (i.e., one with a finite magnetization density above a fully polarized state) has bipartite entanglement entropy of order $\ln N$. 
Another example for a system with a constrained Hilbert space is 
the well-known PXP model, which captures the physics of Rydberg atoms 
in the regime when two neighboring atoms cannot be simultaneously in the 
excited state~\cite{Lesanovsky2011many,turnerNat,bluvstein2021, scartheo8, scartheo9}. Here the single-layer state and thus the constrained EPR states display an area law entanglement entropy.

\subsection{Constructing the Parent Hamiltonian for the EPR Scar State}
Because of the maximal entanglement between the two copies, the EPR state in Eq.~\eqref{eq:EPR} has the special property that applying an operator $\mathcal{O}$ to one copy is equivalent to applying its transpose $\mathcal{O}^T$ to the other copy.
To show this, consider the single-copy operator $\mathcal{O} = \sum_{mn} \ket{m}O_{mn}\bra{n}$ and its transpose $\mathcal{O}^T = \sum_{mn} \ket{m }O_{nm}\bra{n}$, which we take to act within $\HH_1$. Using the definition of the EPR state in Eq.~\eqref{eq:EPR}, one can show that
\bea
\label{eq:EPR_property}
\mathcal{O}\otimes \I \epr & = \frac{1}{\sqrt{\rm dim (\HH_1)}} \sum\limits_{mn} O_{mn}\ket{m} \otimes \ket{n} \\  
& = \I \otimes \mathcal{O}^T \epr \,.
\eea
Based on this property, one can construct a family of interacting parent Hamiltonians for which $\epr$ is an exact eigenstate.
Since $\epr$ has a very simple entanglement structure described in the last section, it is not thermal and becomes a QMBS state, which we call the EPR scar state, when the parent Hamiltonian is nonintegrable.

We now outline the construction of the parent Hamiltonian.
Consider the Hamiltonian in the doubled Hilbert space:
\begin{equation}
\begin{aligned}
\label{eq:general_construction} 
\H = \H_{1}\otimes \I+\I\otimes \H_{2}+ \H_{12}\,.
\end{aligned}
\end{equation}
In the above, $\H_{1}$ and $\H_{2}$ are the Hamiltonians within each copy and $\H_{12}$ describes the interaction between the two copies, which can be expanded in the basis $\{\O_A\}$ of one-copy Hermitian operators as $\sum_{A,B}\lambda_{AB}\O_A\otimes \O_B$.
Using Eq.~\eqref{eq:EPR_property} to move the nontrivial actions of $\mathcal{H}$ on $|{\rm EPR}\rangle$ to a single copy, we find that the general condition for the EPR state to be an eigenstate of $\H$ with energy $E$ is that
\begin{equation}\label{eq:condition}
    \H_{1}+\H_{2}^{*} + \sum\limits_{A,B} \lambda_{AB} \O_{A} \O_{B}^{*} = E \mathbbm{I}\,.
\end{equation}
Notice that $\sum_{A,B} \lambda_{AB} \mathcal{O}_A \mathcal{O}_B$ now acts on a single copy, and hence this is an operator equation in a single copy of the Hilbert space. 
The construction Eq.~\eqref{eq:general_construction} is valid in arbitrary spatial dimensions.
In our previous work~\cite{ourrainbow}, we considered a less general construction also requiring $\H_{1} + \H_{2}^{*} = 0$. 
The general construction can be further enriched by considering local unitary transformations and symmetries.
In the following sections, we demonstrate the general construction using specific examples with different Hilbert space structures including spins, fermions, and bosons and uncover EPR scars in some well-studied many-body Hamiltonians. 

\section{Spin Systems}
\label{sec:spins}
In the first part of this section devoted to  
spin systems we formulate the EPR scar construction for spin-$S$ degrees of freedom.  
We then proceed to demonstrate that the tower of QMBS states in the spin-$1$ XY model~\cite{TomMichael} can be recast as EPR scars in a bilayer spin-$\frac{1}{2}$ system, which in 1D reduces to a ladder geometry.
In the process, we illustrate how the general construction outlined in Sec.~\ref{sec:EPR} can be enriched by unitary transformations and U(1) symmetry.
Finally, we consider a bilayer Heisenberg model on the triangular lattice and illustrate the even richer interplay of SU(2) symmetry with our construction.

\subsection{Spin EPR State}
\label{subsec:spinepr}
We start by considering a spin-$S$ model on a lattice of $N$ sites with $S^{z}_{i} \in \{-S, \dots, +S\}$, $\forall$ $i = 1, \ldots, N$. 
For such a system, the EPR state in the doubled Hilbert space takes the form 
\begin{widetext}
\bea
\label{spin_epr}
\epr = \frac{1}{\left(2S+1\right)^{N/2}} \bigotimes_{i=1}^{N}
\sum_{m=-S}^{+S}\ket{m_{i}}\otimes\ket{m_{i}}
=
\frac{1}{\left(2S+1\right)^{N/2}} \bigotimes_{i=1}^{N}\sum\limits_{m=-S}^{S}\Gamma(S,m)
\left(S_i^+\otimes S_i^+\right)^{m+S}\ket{-S_i}\otimes\ket{-S_i}
\eea
\end{widetext}
with 
$\Gamma(S,m)=\frac{(S-m)!}{(2S)!(m+S)!}$ and $S^+_j = S^x_j + iS^y_j$. 
In this spin EPR state, spins from Hilbert spaces $\HH_1$ and $\HH_2$ that share the same site index $i$ are perfectly correlated.

In cases where the parent Hamiltonian of the EPR state \eqref{spin_epr} respects a global symmetry, the projection of $\epr$ into each symmetry sector becomes an independent eigenstate of the Hamiltonian.
For example, it is easy to see that, when multiplied out, the state $\epr$~\eqref{spin_epr} consists of several terms that each can be associated with an eigenvalue of the total magnetization operator
\bea
\label{eq:Sztot}
S^{z}_{\rm tot} = \sum_{i}\left(S_{i}^{z}\otimes \I + \I \otimes S_{i}^{z}\right).
\eea
Consequently, in models with a U(1) symmetry generated by $S^z_{\rm tot}$, the projections of the EPR state~\eqref{spin_epr} into each magnetization sector become scar states themselves.
More generally, as we will see in Secs.~\ref{subsec:XY} and ~\ref{subsec:Heisenberg} and in Appendix~\ref{appB}, the total 
number of scar states into which Eq.~\eqref{spin_epr} separates 
depends on the symmetries of the underlying bilayer construction. 
For example, for a bilayer spin-$\frac{1}{2}$ system with only U(1) symmetry, we will see that the number of EPR scars is $N+1$.

\subsection{Spin-1 XY Model: U(1) Symmetry}
\label{subsec:XY}
An early example of an exact QMBS tower and the associated periodic dynamics was found in Ref.~\cite{TomMichael}, which studied the spin-1 XY model on a $D$-dimensional hypercubic lattice.
The model hosts a tower of scar states with anomalously low entanglement entropy, with each scar state obtained from the previous one through a raising operator due to an emergent $SU(2)$ algebra.
Furthermore, time evolution from an initial state has finite 
overlap with the tower and exhibits 
perfect periodic revivals, while generic initial states rapidly thermalize as dictated by ETH. 
We will show how these scars are in fact EPR scars.

\begin{figure}[t!]
\includegraphics[width=1.00\columnwidth]{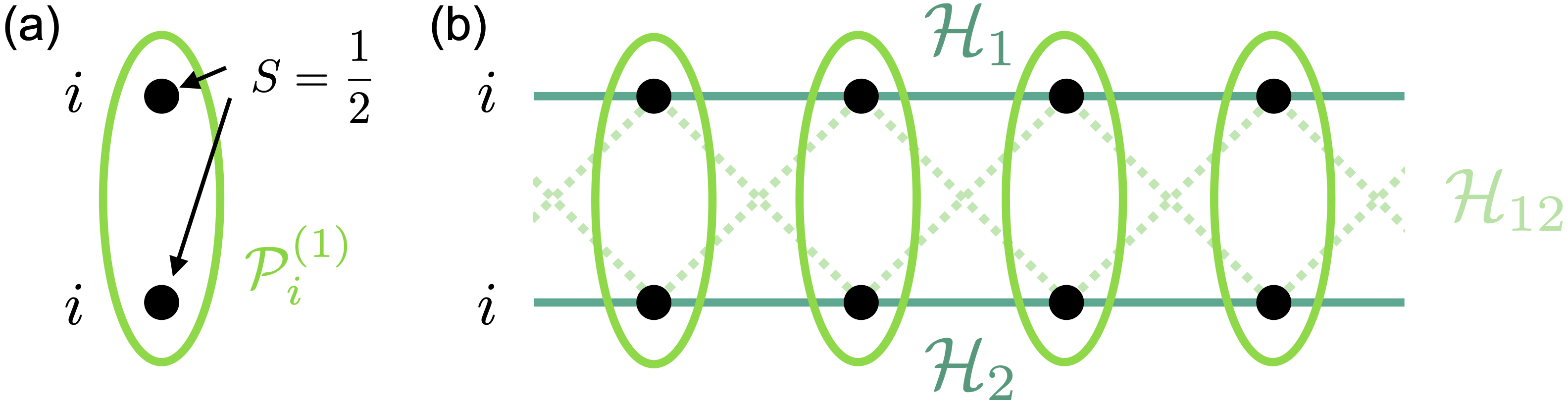}
\caption{
         {\it The spin-1 XY chain.} 
         (a) A spin-$1$ degree of freedom expressed in terms of two spin-$\frac{1}{2}$ particles. 
         The operator $\mathcal{P}^{(1)}_{i}$ projects a pair of spins $i$ onto the triplet subspace 
         to recover 
         the spin-$1$ degree of freedom. 
         (b) A one-dimensional chain of $S=1$ spins expressed as a ladder of spin-$\frac{1}{2}$ particles. 
         Each leg is composed of spin-$\frac{1}{2}$ degrees of  freedom, where sites on the upper~(lower) leg are 
         denoted by $i$. 
         The green ovals denote the local projections $\mathcal{P}^{(1)}_{i}$ onto the spin-$1$ space, while the solid horizontal lines depict the intra-leg coupling, i.e. $\H_{1}$ and $\H_{2}$. 
         Dashed lines indicate the coupling between the two legs, i.e. $\H_{12}$.
         }
\label{fig:spinXY}
\end{figure}

For concreteness, let us first consider the spin-1 XY model on a 1D chain with $N$ sites given by 
\bea
\label{eq:XYham}
\H^{S=1}_{\text{XY}} = \sum_{i=1}^{N-1} \left(S^{x}_{i}S^{x}_{i+1}+ S^{y}_{i}S^{y}_{i+1} 
\right)\, .
\eea
We take $N$ to be even for convenience.
In the following, we map the spin-$1$ chain onto a ladder composed of spin-$\frac{1}{2}$ degrees of freedom~(see Fig.~\ref{fig:spinXY}) and demonstrate that the new Hamiltonian obeys the operator equation in Eq.~\eqref{eq:condition} after a local unitary transformation.
As a result, the EPR state emerges as a scarred eigenstate of the original Hamiltonian. 

First, we replace each spin-$1$ operator by a sum of two spin-$\frac{1}{2}$ Pauli operators:
\begin{equation}
    S_i^{\alpha} \mapsto \frac{1}{2}(\sigma_i^{\alpha}\otimes \I  + \I\otimes \sigma_{i}^{\alpha})\, ,
\end{equation}
where $\sigma^{\alpha}_i\otimes \I $ and $\I\otimes \sigma^{\alpha}_{i}$ ($\alpha=x,y,z$) are the Pauli operators on site $i$ of the top and bottom leg. 

The spin-$1$ Hamiltonian then becomes a ladder Hamiltonian 
\begin{align}
\begin{split}
    \H_{\text{XY}}^{S=1/2} = \sum_{i=1}^{N-1}& (\sigma^{x}_{i}\sigma^{x}_{i+1}\otimes \I + \I \otimes \sigma^{x}_{i}\sigma^{x}_{i+1}) \\
    + & (\sigma^{x}_{i}\otimes \sigma^{x}_{i+1} + \sigma^{x}_{i+1}\otimes \sigma^{x}_{i} ) + 
    (x\rightarrow y)\,.\label{eq:onetohalf}
\end{split}
\end{align}
By construction, the ladder Hamiltonian commutes with the total spin on each rung, which takes value $0$ (singlet) or $1$ (triplet).
The subsector with all triplets coincides with the original spin-$1$ Hamiltonian. 
This subsector is defined by the global projector $\mathcal{P}^{(1)} = \prod_{i=1}^{N}\mathcal{P}^{(1)}_{i}$ 
with 
\bea
\label{projector}
\mathcal{P}^{(1)}_{i} = \frac{1}{4} 
\sum_{\alpha=x,y,z}\sigma^\alpha_{i}\otimes \sigma^\alpha_{i}
+ \frac{3}{4}\,.
\eea
Equipped with Eqs.~\eqref{eq:XYham}-\eqref{projector}, it is easy to see that 
\bea
\label{eq:Spinchange}
\H_{\text{XY}}^{S=1} = \mathcal{P}^{(1)} \left( \H_{\text{XY}}^{S=1/2}\right)  \mathcal{P}^{(1)}\,.
\eea
At first glance, the Hamiltonian in Eq.~\eqref{eq:onetohalf}  does not obey the general construction in Eq.~\eqref{eq:condition}. However, there exists a simple local unitary transformation bringing it to the required form. Consider the unitary transformation
\bea
\mathcal{C} = \I \otimes \prod_{i=1}^{N/2}\sigma^{z}_{2 i-1}\,.
\eea
Under this transformation, $\I \otimes \sigma_{i}^x$ and $\I \otimes \sigma_{i}^y$ acquire a minus sign for odd 
$i$.
Now we split the transformed ladder Hamiltonian into terms within each leg and terms connecting the two legs, i.e. 
\begin{eqnarray}
\mathcal{C}\,\H_{\rm XY}^{S=1/2}\,\mathcal{C} = \H_1\otimes \I + \I\otimes \H_2 + \H_{12}\, , 
\end{eqnarray}
where
\begin{align}
\begin{split}
   &\H_1 =  \frac{1}{4}\sum\limits_i \left(\sigma_i^x \sigma_{i+1}^x + \sigma_i^y \sigma_{i+1}^y \right), \\
   &\H_2 = - \frac{1}{4}\sum\limits_{i} \left(\sigma_{i}^x \sigma_{i+1}^x + \sigma_{i}^y \sigma_{i+1}^y \right), \\
   &\H_{12} =  \frac{1}{4}\sum\limits_{i\in \mathrm{odd}} \left(\sigma_i^x \otimes \sigma_{i+1}^x - \sigma_{i+1}^x \otimes \sigma_{i}^x \right)\\
   &\,\,\,\,\,\,\,\,\,\,-\frac{1}{4}\sum\limits_{i\in \mathrm{even}} \left(\sigma_i^x \otimes \sigma_{i+1}^x - \sigma_{i+1}^x \otimes \sigma_{i}^x \right) + (x\rightarrow y)\,.
\end{split}
\end{align}
This transformed Hamiltonian obeys Eq.~\eqref{eq:condition} with $E=0$. 
Therefore, the unconstrained EPR state in Eq.~\eqref{eq:unconstrained_EPR} is a zero-energy eigenstate of the transformed ladder Hamiltonian, i.e., $\mathcal C\H_{\text{XY}}^{S=1/2} \mathcal{C} \epr = 0$.
Multiplying both sides by $\mathcal C$ and using $\mathcal C^2=\I$, we find that the transformed EPR state
\bea
\label{c_epr}
\mathcal{C} \epr = & \frac{1}{2^{N/2}} \bigotimes_{i=1}^{N} (\ket{0_i}\otimes\ket{0_{i}} + (-1)^ i \ket{1_i}\otimes\ket{1_{i}})
\eea
is an eigenstate of the original ladder Hamiltonian,
\bea
\H_{\text{XY}}^{S=1/2} \mathcal{C} \epr = 0\,.
\eea
Note that $\mathcal{C} \epr$ lives in the subsector of $\H_{\text{XY}}^{S=1/2}$ that 
coincides with the spin-$1$ Hamiltonian. 
It turns out the above construction can be generalized to the spin-$1$ XY model on any bipartite lattice, including the $D$-dimensional hypercubic lattice studied in Ref.~\cite{TomMichael}. 
The only required modification is that the product in $\mathcal{C}$ should be taken only over one of the sublattices.
We henceforth restrict to the 1D case for simplicity. 

To connect the EPR scar state $\mathcal C\epr$ to the tower of states found in Ref.~\cite{TomMichael}, we rewrite Eq.~\eqref{c_epr} in the spin-1 language:
\bea
\mathcal{C} \epr = \frac{1}{2^{N/2}} \bigotimes_{i=1}^{N} (\ket{+1_i} + (-1)^ i \ket{-1_{i}} )\,.
\eea
Next, note that $\H_{\text{XY}}^{S=1}$ $(\H_{\text{XY}}^{S=1/2})$ conserves the total magnetization $\sum_i S^z_i$. 
The projection of $\mathcal{C} \epr$ into each magnetization sector therefore remains an eigenstate.
Notice that $\mathcal{C}\epr$ overlaps with only the even magnetization sectors, leading to $N+1$ degenerate scarred eigenstates including the two fully polarized states $\ket{+1\cdots +1}$ and $\ket{-1\cdots -1}$. 
Explicitly, we can write $\mathcal{C} \epr$ as
\bea
\label{c_epr2}
\mathcal{C} \epr = (-1)^{N/2}\frac{1}{2^{N/2}}\sum_{n=0}^{N}\sqrt{\binom{N}{n}}\ket{\mathcal{S}_n},
\eea
where 
\bea
\label{eq:scarxy1}
\ket{\mathcal{S}_n}=\frac{1}{n!\sqrt{\binom{N}{n}}}(J^{+})^{n}\ket{-1\cdots -1}
\eea
is the (normalized) projection of $\mathcal C\epr$ into the sector with magnetization $2n-N$.
The raising operator $J^{+}$ is given by 
\bea
J^{+} = \frac{1}{2}\sum_{i=1}^{N}(-1)^{i}(S_{i}^{+})^{2}\,.
\eea
The states $\ket{\mathcal S_n}$ are precisely the scar states constructed by other means in Ref.~\cite{TomMichael}.
They are degenerate eigenstates of $\H^{S=1}_{\rm XY}$, but adding a magnetic field $h_z\sum_i S^z_i$ lifts the degeneracy and results in a tower of states with equal energy spacing $2 h_z$.
The state $\mathcal C \epr$ belongs to the family of initial states shown in Ref.~\cite{TomMichael} to exhibit perfect periodic revivals with period $\pi/h_z$ under a quantum quench.

Finally, we remark that the $(N+1)$-fold degeneracy is a generic feature of the EPR scars in systems with U(1) symmetry.
In Appendix~\ref{appB}, we consider two square-lattice Heisenberg layers coupled by an Ising interaction as another example of a spin system with a U(1) EPR scar tower.

\subsection{Bilayer Triangular-Lattice Heisenberg Model: SU(2) Symmetry} 
\label{subsec:Heisenberg}
\begin{figure*}[t]
\includegraphics[width=1.00\textwidth]{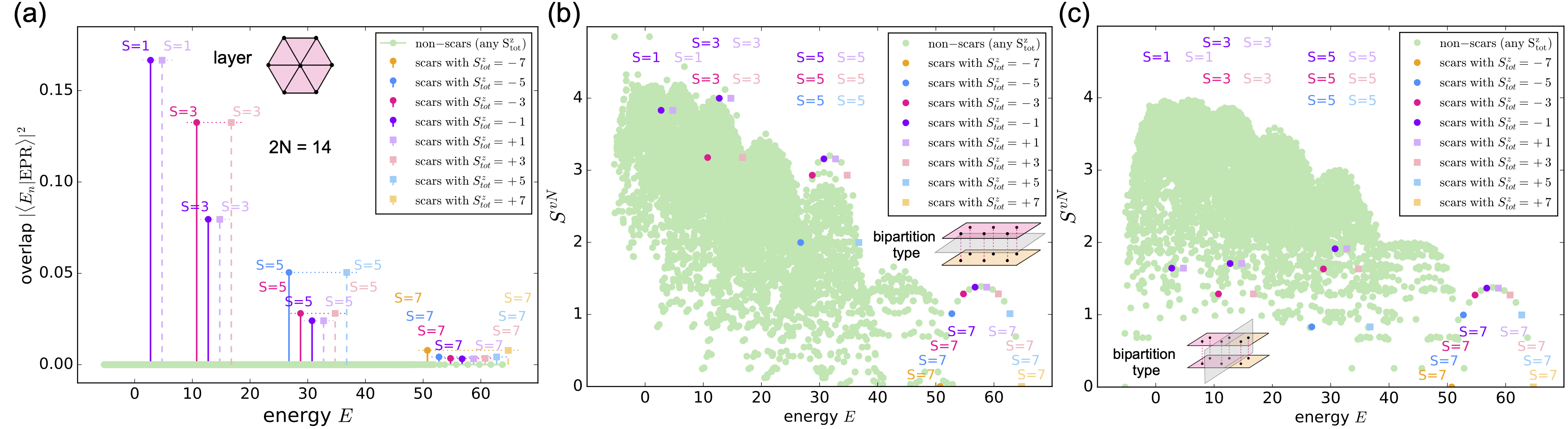}
\caption{
         {\it Heisenberg spin-$1/2$ bilayer system on a triangular lattice}.  Results were obtained from exact diagonalization of Eq.~\eqref{eq:Heisenberg-num} with random $J_{ij} \in [0.9,1.1]$ and $\lambda=1$ for a $2N=14$-site bilayer with hexagon-shaped layers [see inset of panel (a)].
         (a) Overlap $|\langle E_{n}| {\rm EPR}\rangle|^{2}$ between the EPR scar and each eigenstate.
         Each eigenstate is characterized by the quantum numbers $(S_{\rm tot},S_{\rm tot}^{z})$; for the scar states, $S_{\rm tot} \in \{1,3,5,7\}$ and $S^{z}_{\rm tot} \in \{-S, \dots, +S\}$ in increments of 2 with $S=7$ 
         (see Table~\ref{table2}). Colors are used to identify 
         multiplets of scar states with fixed $S^{z}_{\rm tot}$ and all allowed corresponding $S_{\rm tot}$. 
         We do not resolve the symmetries $(S_{\rm tot},S^{z}_{\rm tot})$ for the non-scar states. 
         (b) The von-Neumann entanglement entropy $S^{\rm vN}$ for a bipartition (see inset) parallel to the two layers reveals that the scar states have maximal entanglement entropy for their symmetry sector. 
         (c) The entanglement entropy $S^{\rm vN}$ for a bipartition perpendicular to the two layers reveals that the scar states have atypically low entanglement. 
         A dome-shaped structure formed by eigenstates associated with a fixed $S_{\rm tot}$ is also clearly visible. 
}
\label{fig:tria_Heisenberg}
\end{figure*}
Having studied a U(1) symmetric spin model in the previous section, we now move on to an SU(2) symmetric spin model and show how the SU(2) symmetry leads to an EPR scar tower with a rich structure.
We consider the following bilayer triangular-lattice Heisenberg model consisting of exchange-coupled ferromagnetic and antiferromagnetic layers,
\begin{align}
\label{eq:ham1}
\begin{split}
\H &= \H_{1}\otimes \I + \I \otimes\H_{2} + 
\H_{12}\,.
\end{split}
\end{align}
The individual Hamiltonians take the form 
\bea
\label{eq:ham1_parts}
&\H_{1} = \sum_{\langle ij\rangle}J_{ij}\,\vec{S}_{i}\cdot\vec{S}_{j}\,,
\,\,\,\,\,\H_{2} = -\H_{1}, \\  
&\H_{12} = \lambda \sum_{i}\vec{S}_{i} \otimes \vec{S}_{i}\,.
\eea
The operators $S^{z}_{i}$ and $\vec{S}_{i} = (S^{x}_{i},S^{y}_{i}, S^{z}_{i})$ are spin-$\frac{1}{2}$ operators acting on sites $i$ in either the top or bottom layer. 
 
This model has an SU(2) symmetry with commuting generators
\bea
\label{eq:conserved_spins_TRIA}
&S_{\rm tot}^{z} = S^{z,1}_{\rm tot} + S^{z,2}_{\rm tot} = \sum_{i}S_{i}^{z}\otimes \I + \I \otimes S_{i}^{z}, \\
&S_{\rm tot}^{2} = \left( \sum_{i}\vec{S}_{i}\otimes \I + \I \otimes \vec{S}_{i}\right)^{2},
\eea
which correspond respectively to the magnetization and the total spin. 
With SU(2) symmetry, the unconstrained EPR state of $2N$ qubits, Eq.~\eqref{eq:unconstrained_EPR}, can be projected into each sector with fixed $S^2_{\rm tot}=S(S+1)$ and $S^z_{\rm tot}$ to yield a scar state in that sector.
Generally speaking for a system of $2N$ spin-$1/2$ particles 
we can have $S\in\{0, 1,\ldots, N\}$ and $S^z_{\rm tot}\in\{-S,-S+1,\dots,S-1, S\}$.
However, as is evident from Eq.~(\ref{eq:unconstrained_EPR}), the EPR scars can only reside in sectors with $S_{\rm tot}^{z,1}=S_{\rm tot}^{z,2}=\{-N/2, -N/2+1, \ldots, N/2\}$.
Hence, the allowed values of $S^{z}_{\rm tot}$ for the EPR scars to exist are $S^z_{\rm tot} = \{-N, -N+2, \ldots, N\}$ in increments of 2.
Likewise, the allowed values of the total $S$ also decrease from $N$ in steps of 2.
 
To count the total number of EPR scars, it is necessary to distinguish the cases where the number $N$ of spins in each layer is even or odd.
For $N$ odd the allowed values of $S$ are $\{1,3, \ldots, N\}$, whereas for even $N$ they are $\{0,2, \ldots, N\}$.
For each $S$, the EPR state overlaps with $S+1$ of the $2S+1$ possible magnetization sectors. 
We therefore conclude that the total number of scars in an SU(2)-invariant bilayer spin system is $N_{\rm scars}=\sum_{\text{allowed }S} (S+1)$, i.e.,
\begin{align}\label{number_scars_spin}
N_{\rm scars} = 
\frac{1}{4}
\times
\begin{cases}
(N+1)(N+3) & \text{for } N \text{ odd}\\
(N+2)^2 & \text{for } N \text{ even} 
\end{cases}\,.
\end{align}

As a concrete example, we now consider a bilayer system living on a triangular lattice with $2N=14$ spins. 
The shape of each $N=7$ layer is shown in the inset of Fig.~\ref{fig:tria_Heisenberg}(a). 
Since $N$ is odd, the EPR state has overlap with sectors of spin quantum number, $S_{\rm tot} \in \{1,3,5,7\}$ and total magnetization
\begin{eqnarray}
S_{\rm tot}^{z} = \pm p \,\,\, {\rm with}\,\,\, 
p = 1,3,5,7\,.
\end{eqnarray}
Each of these magnetization sectors contains $4$, $3$, $2$, or $1$ scars with corresponding values for $S_{\rm tot}$ (see Table~\ref{table2}). 
We visualize this set of EPR scars by performing an exact diagonalization study. 
To fully resolve all scar states, we diagonalize the Hamiltonian
\begin{align}
\label{eq:Heisenberg-num}
    \H+ S^z_{\rm tot} + S^{2}_{\rm tot}\, ,
\end{align}
which ensures that all scar states are nondegenerate.
In our numerics, we draw the exchange couplings $J_{ij}$ uniformly from the interval $[0.9,1.1]$ to break the rotational symmetry of the lattice, and set the interlayer coupling $\lambda=1$.
\begin{table}[b]
    \begin{tabular}{ | c | c | c | c | c | c | c | c | c | }
    \hline
    \hline
    \multicolumn{9}{|c|}{\bf{Bilayer Triangular Lattice Heisenberg Model}} \\
    \hline
       $S_{\rm tot}^{z}$ & $-7$ & $-5$ & $-3$  & $-1$ & $+1$ & $+3$ & $+5$ & $+7$  \\ \hline \hline
      {\rm total spin $S_{\rm tot}$} & 7  & 5,7  &  3,5,7 & 1,3,5,7 & 1,3,5,7 & 3,5,7 & 5,7 & 7  \\ \hline
      $N_{\rm scars}$  & 1        & 2      &  3   & 4 & 4 & 3 & 2 & 1 \\ \hline
    \end{tabular}
\caption{
{\it EPR scar states in the Heisenberg bilayer model on the triangular lattice.} Due to $S_{\rm tot}^{z}$ and $S_{\rm tot}^{2}=S(S+1)$ being conserved quantities we distinguish the single members of the EPR scar tower by their respective quantum numbers $(S, S_{\rm tot}^{z})$. For a bilayer consisting of $2N=14$ sites, the allowed values of $S_{tot}^{z}$ are $S_{\rm tot}^{z} = \pm 1,\,\pm 3,\,\pm 5,\, \pm 7$ with each $S_{\rm tot}^{z}$-sector containing $4$, $3$, $2$, or $1$ scars with corresponding values for $S$. 
In total, we have $20$ scar states.
}
\label{table2}
\end{table}

In Fig.~\ref{fig:tria_Heisenberg}(a) we compute the overlap  $|\langle E_{n}|{\rm EPR} \rangle|^{2}$ between the EPR state \eqref{eq:unconstrained_EPR} and each energy eigenstate of Eq.~\eqref{eq:Heisenberg-num}.
We utilize different colors (see legend) to label scars with quantum numbers $(S_{\rm tot}, S_{\rm tot}^{z})$ according to their $S^z_{\rm tot}$ sector.
One clearly sees that the entropically most likely allowed magnetization sectors $S^z_{\rm tot}=\pm1$ (light and dark purple, respectively) have the highest total weight when summing over allowed values of $S_{\rm tot}$, as should be expected from the 
fact that $\epr$ is an equal amplitude superposition of allowed 
spin configurations.

In Fig.~\ref{fig:tria_Heisenberg}(b) and (c) we plot the entanglement entropy $S^{\rm vN}$ of all eigenstates for
bipartitions of types I and II, respectively (see insets and Fig.~\ref{fig:area_cuts}).
The scar states are highlighted using the color scheme of Fig.~\ref{fig:tria_Heisenberg}(a).
In Fig.~\ref{fig:tria_Heisenberg}(b), the entanglement with respect to the type-I cut scales extensively. 
Indeed, each scar state is a projection of $\epr$ into a symmetry sector, so the two layers retain the maximal entanglement allowed by the dimension of that sector, consistent with the discussion in Sec.~\ref{sec:EPR}. 
In comparison, Fig.~\ref{fig:tria_Heisenberg}(c) utilizes a type-II cut for which the scar states appear as states with anomalously low entanglement.
Note that we have not resolved any symmetries for the non-scar states in Fig.~\ref{fig:tria_Heisenberg}(b) and (c) which is why the distribution of entanglement entropy appears broad. 
Nonetheless, the dome-like structure for each set of states with fixed $S_{\rm tot}$ is clearly visible. 

We stress that the 2D bilayer Heisenberg model in  Eqs.~\eqref{eq:ham1} and~\eqref{eq:ham1_parts} with and even without random intralayer exchange couplings is nonintegrable. 
To confirm this, we study its level statistics. We utilize the same coupling parameters $J_{ij}$ and $\lambda$ as before and fix the magnetization to $S^{z}_{\rm tot} = +5$. 
We find the mean level-spacing ratio  $\langle r \rangle \approx 0.50379$, which is in reasonably close agreement with the Gaussian orthogonal ensemble~(GOE)~\cite{GOE}, which according to random matrix theory has $\langle r \rangle_{\rm GOE}\approx 0.536$. 
More details on the level statistics are presented in Appendix~\ref{appB}, where we also discuss a spin-$\frac{1}{2}$ bilayer system with U(1) symmetry on the square lattice.

\section{Fermionic Systems}
\label{sec:fermions}

In this section, we apply the general construction outlined in Sec.~\ref{sec:EPR} to spinful fermionic systems, where we take advantage of the internal spin degree of freedom to obtain the desired 
doubled Hilbert space structure.

\subsection{Fermionic EPR State}
A generic spinful fermionic Hamiltonian $\H$ is written in terms of the fermionic creation~(annihilation) operators $c^\dagger_{i,\sigma}$~($c^{}_{i,\sigma}$) on site $i$ with spin $\sigma=\{\uparrow, \downarrow
\}$. 
To directly apply the formalism in Sec.~\ref{sec:EPR} and decompose the full Hilbert space as $\HH_\uparrow \otimes \HH_\downarrow$, one has to be careful about fermionic anticommutation; fermionic operators with opposite spins anticommute with each other, therefore they do not act totally independently on the two copies of the Hilbert space.
To mitigate this problem, we introduce two flavors of spinless fermionic operators $\psi_{i,1}$ and $\psi_{i,2}$.
They are related to the original spinful fermions via
\bea
&\psi_{i,1} = c_{i,\uparrow},  \\
&\psi_{i,2} = (-1)^{N_{\uparrow}} c_{i,\downarrow}, 
\label{eq:fermion_mapping}
\eea
where $N_{\uparrow} = \sum_j c^\dagger_{j,\uparrow}c_{j,\uparrow}$, and $(-1)^{N_{\uparrow}}\equiv \mathcal{F}_{\uparrow}$ is the parity of the spin-up fermions. One can readily check using $\{c_{i,\uparrow},\mathcal{F}_{\uparrow}\}=0$ that they satisfy the commutation relations:
\bea
    \{\psi_{i,1}, \psi_{j,1}^\dagger \} =\delta_{ij}, \quad 
    \{\psi_{i,2}, \psi_{j,2}^\dagger \} =\delta_{ij}, \quad
    [\psi_{i,1}, \psi_{j,2}]= 0\,.
\eea
Therefore, $\psi_{i,1}$ and $\psi_{i,2}$ have \textit{fermionic} self-statistics and \textit{bosonic} mutual statistics.
Hence, one can treat the Hilbert space spanned by the two flavors of fermions as independent and construct EPR scars in a way analagous to spin systems.
The inverse mapping of Eq.~(\ref{eq:fermion_mapping}) is given by
\bea
&c_{i,\uparrow} = \psi_{i,1}, \\
&c_{i,\downarrow} = (-1)^{N_1} \psi_{i,2}, 
\eea
where $N_1 = \sum_j \psi_{j,1}^\dagger \psi_{j,1}$. In what follows, we shall represent these two flavors of fermions using notation consistent with previous sections, i.e.
\bea
 &\psi_{i,1} \rightarrow \psi_i \otimes \mathbb{I}, \\
 &\psi_{i,2} \rightarrow \mathbb{I} \otimes \psi_i\,.
\eea

The unconstrained EPR state Eq.~\eqref{eq:unconstrained_EPR} in $\HH_\uparrow \otimes \HH_\downarrow$ can be written as
\bea
\label{eq:Fermi_epr}
\epr = & \frac{1}{2^{N/2}}\prod\limits_{i=1}^N (\I\otimes\I + \psi^\dagger_i \otimes \psi^\dagger_i )\ket{0} 
\\=& \frac{1}{2^{N/2}}e^{ \sum \limits_i \psi_i^\dagger \otimes \psi_i^\dagger }\ket{0}
\\
=&  \frac{1}{2^{N/2}}e^{ \sum_i-\mathcal{F}_\uparrow c^\dagger_{i,\uparrow} c^\dagger_{i,\downarrow} }\ket{0},
\eea
where $\psi_i^\dagger \otimes \psi_i^\dagger \equiv \psi_{i,1}^\dagger \psi_{i,2}^\dagger = -\mathcal{F}_\uparrow c_{i,\uparrow}^\dagger c_{i,\downarrow}^\dagger$, and 
$\ket{0} \equiv \bigotimes^N_{i=1}\ket{0_i}\otimes\ket{0_i}$ 
is the 
vacuum for $2N$ sites on the two layers. 
In the presence of U(1) charge conservation, the projection of $\epr$ into each charge sector also becomes an eigenstate, leading to a U(1) tower of EPR scars of the following form upon expanding the exponential function in Eq.~\eqref{eq:Fermi_epr}:
\bea
\label{eq:FermiEPRTower}
\epr_n = \frac{1}{n!\sqrt{\binom{N}{n}}}\left(\sum_i c_{i,\uparrow}^\dagger 
c_{i,\downarrow}^\dagger\right)^n \ket{0}\,.
\eea
Note that in each charge sector the factor $-\mathcal F_{\uparrow}$ becomes an overall phase of $\pm 1$ and can therefore be omitted.
The particle number of the state $\epr_n$ is $2n$ for $0\leq n\leq N$. As a result, there are $N+1$ states in the U(1) tower. 

In the following, we show that applying appropriate transformations to $\epr$ yields two towers of exact eigenstates in the Fermi-Hubbard model.
One unitary transformation, which is well defined only on bipartite lattices, yields the celebrated $\eta$-pairing states~\cite{YangPRL}.
The other transformation, which can be defined on any lattice, yields another tower of eigenstates that was also considered in Refs.~\cite{Pakrouski,Pallegar}.
We further give the necessary conditions for the EPR state to remain an eigenstate of the most general quadratic interaction, including spin-orbit coupling~\cite{Li} and superconducting pairing~\cite{Pallegar} terms.

\subsection{Fermi-Hubbard Model}
\label{sec:FHmodel}

The Fermi-Hubbard model on an arbitrary lattice with $N$ sites is given by
\bea
\label{fermiH0}
\H &= -t \sum_{\sigma}\sum_{\langle ij \rangle}\left(c^{\dagger}_{\sigma, i}c_{\sigma, j} + \text{H.c.} \right) \\
&+ U\sum_{i}\left(n_{\uparrow, i}-\frac{1}{2}\right)\,
\left(n_{\downarrow, i}-\frac{1}{2}\right)\,,
\eea
where we label sites with indices $i,j$ and where $\langle ij\rangle$ denotes a pair of nearest-neighbor sites.
The operators $n_{\sigma, i}=c^\dagger_{\sigma,i} c_{\sigma,i}$ are the standard number operators. 
The Hubbard onsite interaction in Eq.~\eqref{fermiH0} is  formulated in such a way that it invariant under particle-hole symmetry as is the kinetic part. 
To check whether $\H$ obeys the general criterion in Eq.~\eqref{eq:condition}, it is easier to rewrite it in terms of the spinless fermion operators $\psi_i$. We also define the spinless number operator $n_i = \psi^\dagger_i \psi_i$.
With these operators, 
\bea
\H = \H_\uparrow \otimes \I + \I \otimes \H_\downarrow + \H_{\uparrow\downarrow} + \frac{NU}{4}\, ,
\eea
where
\bea\label{eq:hubbard_general}
&\H_\uparrow = \H_\downarrow =-t \sum\limits_{\langle ij \rangle} (\psi^\dagger_i \psi_j + \text{H.c.}) -\frac{U}{2} \sum_i n_i, \\
&\H_{\uparrow\downarrow} = U\sum_i n_i \otimes n_i\, .
\eea
When deriving these terms, we used the identity $\mathcal{F}^2=\I$ so $\mathcal{F}$ does not appear.
Although these terms do not obey the criterion~\eqref{eq:condition}, we demonstrate below that there exist simple unitary transformations bringing the Hamiltonian to the required form.
We consider two transformations.
The first one, denoted $\mathcal{C}$, is similar to the chiral transformation used in our treatment of the spin-1 XY model in Sec.~\ref{subsec:XY} and works for bipartite lattices.
The second one, denoted $\pi$, is a particle-hole transformation and works for any lattice.

\subsubsection{Chiral Transformation}
\label{trafo}
For a bipartite lattice, we can associate each site $i$ with one of the two sublattices $A$ or $B$.
We define the chiral transformation $\mathcal{C}$ to act on one of the spin species (say $\sigma=\downarrow$) and on one of the sublattices (say $B$), so that
$\mathcal{C}: c_{\downarrow, i} \rightarrow - c_{\downarrow, i}$ for $i \in B$, 
or equivalently
\bea
\mathcal{C}: \I \otimes \psi_i \rightarrow - \I \otimes \psi_i \quad \text{for }i \in B\,.
\eea
The nearest neighbor hopping term in $\H_\downarrow$ links the two sublattices and therefore acquires a minus sign under the transformation.
The Hamiltonian in Eq.~\eqref{eq:hubbard_general} becomes
\bea\label{eq:hubbard_c}
&\mathcal C\H_\uparrow\mathcal C  = -t \sum\limits_{\langle ij \rangle} (\psi^\dagger_i \psi_j + \text{H.c.}) -\frac{U}{2} \sum_i n_i, \\
&\mathcal C\H_\downarrow\mathcal C  = +t \sum\limits_{\langle ij \rangle} (\psi^\dagger_i \psi_j + \text{H.c.}) -\frac{U}{2} \sum_i n_i, \\
&\mathcal C\H_{\uparrow\downarrow}\mathcal C = U\sum_i n_i \otimes n_i\,.
\eea
Using the property~\eqref{eq:EPR_property} of the EPR state and $n_i^2=n_i$, one can verify that Eq.~\eqref{eq:hubbard_c} obeys the general criterion Eq.~\eqref{eq:condition} with $E=0$.
Hence, taking account the constant piece $NU/4$, we have demonstrated that the $\mathcal{C}$-transformed EPR state is an eigenstate of the Hubbard model on arbitrary bipartite lattices, i.e.,
\bea
\H  \mathcal{C}\epr = \frac{NU}{4} \mathcal{C}\epr\,.
\eea
The $\mathcal{C}$-transformed EPR state takes the following form in terms of the original $c$ operators:
\bea
\label{eq:CEPR}
\epr_{\mathcal{C}} = \mathcal{C} \epr = \frac{1}{2^{N/2}}\prod\limits_{i=1}^N (\mathbbm 1 -\xi_i \mathcal{F}_\uparrow c^\dagger_{i,\uparrow}
c^\dagger_{i,\downarrow} )\ket{0},
\eea
where $\xi_i = +1$ ($-1$) for $i\in A$ ($B$).

The Hubbard model separately conserves the number of spin-up fermions $N_\uparrow$ and the number of spin-down fermions $N_\downarrow$.
In the $\mathcal{C}$-transformed EPR state, a spin up creation operator and a spin down creation operator always appear together. As a result, $N_\uparrow=N_\downarrow = n$, ranging from 0 to $N$.
Projecting $\mathcal{C} \epr$ into each allowed total charge sector yields the following modification of Eq.~\eqref{eq:FermiEPRTower}:
\bea
\label{eq:CEPRn}
\mathcal C\epr_n = \frac{1}{n!\sqrt{\binom{N}{n}}}\left(\sum_i \xi_i\, c_{i,\uparrow}^\dagger 
c_{i,\downarrow}^\dagger\right)^n \ket{0}\, .
\eea
This tower of $N+1$ states is nothing but the $\eta$-pairing states.
The construction of this tower in the Hubbard model is very similar to that in the spin 1 XY model discussed in Sec.~\ref{subsec:XY}, consistent with the connection found in Ref.~\cite{mark1}.

\subsubsection{Particle-Hole Transformation}

Now we consider another transformation, the particle-hole transformation $\pi$, which we apply to one of the spin species (say, $\sigma=\downarrow$) .
This transformation exchanges creation and annihilation operators, $\pi : c^\dagger_{\downarrow, i} \leftrightarrow c_{\downarrow,i}$, or equivalently
\bea
\mathcal{\pi}: \I \otimes \psi^\dagger_i \leftrightarrow \I \otimes \psi_i\, .
\eea
This transformation flips the sign of $t$ for the spin-down fermions as well as that of the interaction strength $U$. Under the transformation, the Hamiltonian becomes
\bea\label{eq:hubbard_pi}
&\pi\H_\uparrow\pi  =-t \sum\limits_{\langle ij \rangle} (\psi^\dagger_i \psi_j + \text{H.c.}) +\frac{U}{2} \sum_i n_i, \\
&\pi\H_\downarrow\pi  =+t \sum\limits_{\langle ij \rangle} (\psi^\dagger_i \psi_j + \text{H.c.}) + \frac{U}{2} \sum_i n_i, \\
&\pi\H_{\uparrow\downarrow}\pi = -U\sum_i n_i \otimes 
n_i\,,
\eea
which obeys the general criterion~\eqref{eq:condition} with $E=0$. Therefore, the $\mathcal{\pi}$-transformed EPR state is an eigenstate of the Hubbard model on arbitrary lattices,
\bea
\H  \mathcal{\pi}\epr = - \frac{NU}{4} \mathcal{\pi}
\epr\,.
\eea
The minus sign comes from the fact that the constant piece in the Hamiltonian changes sign under $\pi$. 
The $\mathcal{\pi}$-transformed EPR state takes the following form in terms of the original $c$ operators:
\bea
\label{eq:piEPR}
\epr_\pi = \mathcal{\pi} \epr = \frac{1}{2^{N/2}}\prod\limits_{i=1}^N (\mathbb{1} -\mathcal{F}_\uparrow c^\dagger_{i,\uparrow} c_{i,\downarrow} )\ket{\Omega}\,,
\eea
where the particle-hole-transformed vacuum
\bea
\ket{\Omega} = \prod_i c^\dagger_{i,\downarrow}\ket{0}
\eea
is the spin polarized state in which each site is occupied by a spin-down fermion. 
Since $c^\dagger_\uparrow$ appears together with another $c_\downarrow$ operator, the $\mathcal{\pi}$ transformed state has a fixed total particle number $N_\uparrow + N_\downarrow=N$. However, the total magnetization $N_\uparrow - N_\downarrow$ varies from $-N$ to $N$ in steps of 2, leading to a new tower of $N+1$ states~\cite{Pakrouski,Pallegar},
\bea
\mathcal \pi\epr_n = \frac{1}{n!\sqrt{\binom{N}{n}}}\left(\sum_i c^\dagger_{i,\uparrow} c_{i,\downarrow}\right)^n \ket{\Omega}
\eea
with total magnetization $-N+2n$. That is distinct from the $\mathcal{C}$-transformed EPR states which instead has varying charge but zero magnetization.

We remark that in the standard Hubbard model, both the $\mathcal C$ and $\mathcal \pi$ towers of states are not regarded as many-body scars, since their existence is enforced by symmetry.
For instance, the Hubbard model has a spin SU(2) symmetry, and the $\mathcal{\pi}$ tower of $N+1$ states are the unique states with maximal total spin $N/2$ and their respective magnetizations.
Furthermore, on bipartite lattices, the Hubbard model acquires another SU(2) symmetry of the charge degrees of freedom~\cite{YangPRL}, and the $\mathcal C$ tower of $N+1$ states are enforced in an analogous fashion.
However, a variety of extended Hubbard models are known in which these enforcing symmetries are broken such that these two towers become genuine scar towers~\cite{mark1,Sanjay1,Pallegar}.

\subsubsection{Two Towers of States on the Square Lattice}
\begin{figure*}[t!]
\includegraphics[width=1.00\textwidth]{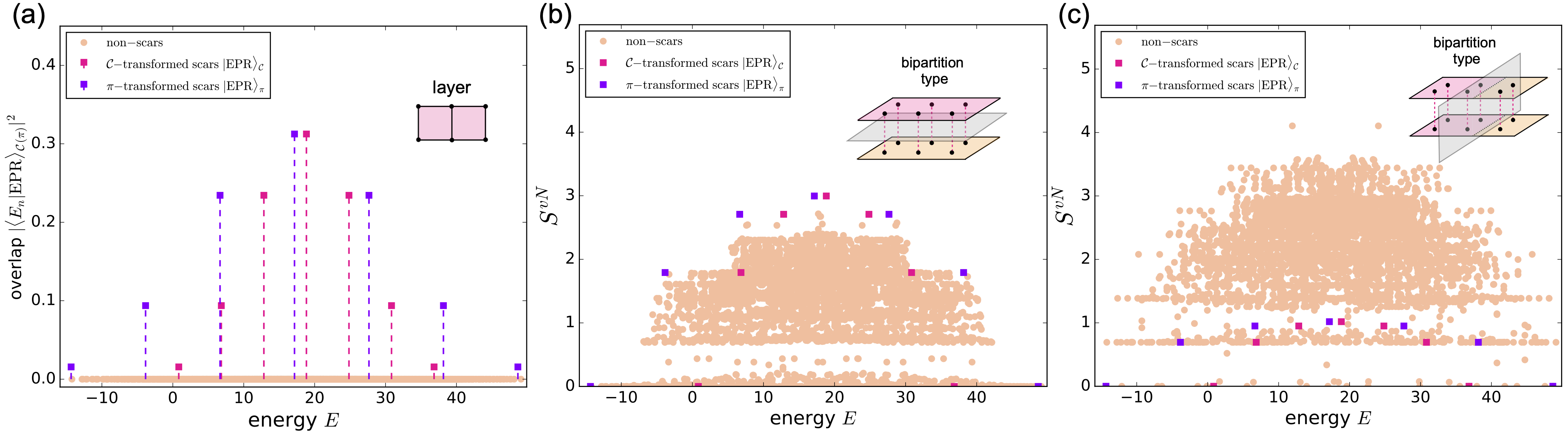}
\caption{{\it Fermi-Hubbard bilayer system on a square lattice.}
        Results are obtained from exact diagonalization of Eq.~\eqref{eq:Hubbard-num} for a square lattice of dimension $(L_x,L_y)=(3,2)$, with hopping amplitude $t=1$ and interaction strength $U = 0.56$, chemical potential $\mu = 3.0$, and magnetic field $h=5.25$.
        (a) Overlap between each energy eigenstate of Eq.~\eqref{fermiH0} and the $\mathcal C$- (magenta) and $\pi$-transformed (purple) EPR state, i.e. 
        $\sum_{n}|\langle E_{n}|\epr_{\mathcal{C}(\pi)} \rangle|^{2} = 1$. 
        (b) Von-Neumann entanglement entropy as a function of energy for a type-I bipartition~(see inset) for which the EPR towers exhibit high entanglement.
        (c) Same as in (b) except for a type-II bipartition (see inset) for which the EPR towers exhibit low entanglement.
}
\label{fig:fermion1}
\end{figure*}
We have shown that the standard Hubbard model on a bipartite lattice hosts two towers of states from the $\mathcal{C}$- and $\mathcal{\pi}$-transformed EPR states, respectively.
To visualize these towers, we consider the Hubbard model on the square lattice in the presence of a chemical potential $\mu$ and a magnetic field $h$ in the $z$-direction:
\bea
\label{eq:Hubbard-num}
\H &= -t \sum_{\sigma}\sum_{\langle ij \rangle}\left(c^{\dagger}_{\sigma, i}c_{\sigma, j} + \text{H.c.} \right) \\
&+ U\sum_{i}\left(n_{\uparrow, i}-\frac{1}{2}\right)\,
\left(n_{\downarrow, i}-\frac{1}{2}\right) \\ 
&+ \mu \sum_i ( n_{i,\uparrow} + n_{i,\downarrow}) +h \sum_i ( n_{i,\uparrow} - n_{i,\downarrow})\,.
\eea
The chemical potential $\mu$ and magnetic field $h$ lift the degeneracies of the $\mathcal{C}$ and $\pi$ EPR towers, respectively. From the analysis of the last two subsections, the energies of the two towers of states are
\bea
E^\mathcal{C}_n &= \frac{NU}{4} -2\mu n, \\
E^\mathcal{\pi}_n &= \left(-\frac{U}{4} -\mu  + h\right) N - 2 h n
\eea
for $0\leq n \leq N$. 

We explicitly verify the two towers by diagonalizing Eq.~\eqref{eq:Hubbard-num} on a $3\times 2$ square lattice, choosing $t=1$, $U=0.56$, $\mu=3.0$ and $h=5.25$.
In Fig.~\ref{fig:fermion1}(a), we plot the overlap between each eigenstate and the $\mathcal C$- (magenta) and $\pi$-transformed (purple) EPR states, highlighting both towers of states with the expected equal energy spacings. Fig.~\ref{fig:fermion1}(b) and (c) show the entanglement entropy for bipartitions of types I and II, respectively.
Note that the entanglement entropy for the type-I bipartition is interpreted in this setting as an entanglement in spin space, rather than real space.
As expected, the entanglement entropy for the type-I bipartition takes the maximum value allowed within each symmetry sector, while that for the type-II bipartition is markedly smaller for the scar states than for typical eigenstates.

\subsection{Generalized Fermi-Hubbard Models}

In this section we generalize the Fermi-Hubbard model \eqref{fermiH0} to include spin-orbit coupling and superconducting pairing terms and derive the conditions under which the EPR state and its $\mathcal C$- and $\pi$-transformed variants are eigenstates.
We work on an arbitrary lattice and consider a generic Hamiltonian of the following form
\begin{equation}
\label{eq:FermHam}
\H = \H_{\rm k} + \H_{\rm int}\,,
\end{equation}
where $\H_{\rm k}$ is quadratic in fermion operators and $\H_{\rm int} = \sum_i U_i (n_{i,\uparrow}-1/2) (n_{i,\downarrow}-1/2)$.
$\epr$ is an eigenstate of $\H_{\rm int}$ with eigenvalue $\sum_i U_i/4$.
The quadratic part $\H_{\rm k}$ contains hopping, spin-orbit coupling, and superconducting pairing terms:
\bea
\label{eq:HopHam}
    \H_{\rm k} = & \sum_{\sigma, \sigma^\prime}\sum_{i,j}\left(t^{\sigma \sigma^\prime}_{ij}c^{\dagger}_{i, \sigma}c_{j \sigma^\prime}+ \text{H.c.}\right)\\
    +&\sum_{\sigma, \sigma^\prime}\sum_{i, j}\left(\Delta^{\sigma \sigma^\prime}_{ij}c^{\dagger}_{i, \sigma}c^{\dagger}_{j, \sigma^\prime}+ \text{H.c.}\right)\,.
\eea
Here $t^{\sigma \sigma^\prime}_{ij}~(\Delta^{\sigma \sigma^\prime}_{ij})$ are spin- and position-dependent hopping~(pairing) strengths for $i\neq j$, and the Hamiltonian is manifestly Hermitian in this form.
We emphasize that the terms $t^{\sigma\sigma^\prime}_{ii}$ correspond not to hopping but rather to a chemical potential when $\sigma=\sigma^\prime$ or a magnetic field for $\sigma \neq \sigma^\prime$.
The parameterization of $t$ and $\Delta$ contains a redundancy that is removed by setting
\bea
t^{\sigma \sigma^\prime}_{ij} = \left(t^{\sigma' \sigma}_{ji}\right)^{*},  \quad \Delta^{\sigma\sigma'}_{ij} = -\Delta^{\sigma'\sigma}_{ji}\,.
\eea
Demanding that Eq.~\eqref{eq:HopHam} satisfy Eq.~\eqref{eq:condition} when written in terms of the $\psi$ operators, we find the following constraints on $t$ and $\Delta$ such that $\mathcal H\epr$ is confined to a two dimensional subspace spanned by $\epr$ and its parity partner $\mathcal{F}_{\uparrow}\epr$~(see Appendix~\ref{appC} for details):
\bea
\label{eq:FermConst}
&t^{\uparrow \uparrow}_{ij} + t^{\downarrow \downarrow}_{ji} = 0,\quad\quad \Delta^{\uparrow \uparrow}_{ij} + \Delta^{\downarrow \downarrow}_{ji} = 0,
\\
&t^{\uparrow \downarrow}_{ij} - t^{\uparrow \downarrow}_{ji} = 0,\quad\quad \Delta^{\uparrow \downarrow}_{ij} + \Delta^{\downarrow \uparrow}_{ij}  = 0\,.
\eea
Note that the above criteria are not satisfied for the standard Fermi-Hubbard model \eqref{fermiH0}---for example, the above conditions imply that the two spin species have hopping amplitudes of opposite signs.
To ensure that $\epr$ is an exact eigenstate there is an additional constraint, namely, $\sum_{i}\Delta^{\downarrow \uparrow}_{ii}=0$.

In the most general case, the model \eqref{eq:FermHam} does not conserve U(1) charge owing to the pairing terms $\Delta^{\sigma \sigma^\prime}_{ij}$, and only the global fermion parity is conserved.
In this case, the only eigenstates guaranteed by Eq.~\eqref{eq:FermConst} are $\epr$ and its parity partner.
If instead $\Delta^{\sigma \sigma^\prime}_{ij}=0$ and charge is conserved, we recover the U(1) EPR towers obtained from projecting these states onto each U(1) sector.

In order to connect the general Hamiltonian Eq.~\eqref{eq:FermHam} to the standard Fermi-Hubbard model, we consider the effect of the particle-hole transformation $\pi$ and the chiral transformation $\mathcal C$ on the above conditions.
We first discuss the transformation $\pi: c_{i, \downarrow}\leftrightarrow c^{\dagger}_{i, \downarrow}$, whose action on the EPR state is shown in Eq.~\eqref{eq:piEPR}.
Under this transformation, the constraints Eq.~\eqref{eq:FermConst} become 
\bea
\label{eq:PHFermConst}
&t^{\uparrow \uparrow}_{ij} - t^{\downarrow \downarrow}_{ij} = 0,\quad\quad \Delta^{\uparrow \uparrow}_{ij} + \Delta^{\downarrow \downarrow}_{ji} = 0,
\\
&t^{\uparrow \downarrow}_{ij} + t^{\downarrow \uparrow}_{ij} = 0,\quad\quad \Delta^{\uparrow \downarrow}_{ij} - \Delta^{\uparrow \downarrow}_{ji} = 0\,.
\eea
Notice that, under this transformation, the spin-orbit coupling and pairing between opposite spins exchange with one another.
As with the constraints \eqref{eq:FermConst}, the constraints \eqref{eq:PHFermConst} only ensure that $\H\pi\epr$ resides in a two-dimensional subspace spanned by $\pi\epr$ and $\pi\mathcal F^{\uparrow}\epr$. To ensure that $\pi\epr$ is an eigenstate we must add the condition $\sum_{i}t^{\downarrow \uparrow}_{ii}=0$.

We now make a connection with the $\mathcal C$-EPR tower and the $\eta$-pairing states, and specialize to the case of a Bravais lattice in which site $i$ is located at position $\bm r_i$.
We consider the transformation
\bea
\mathcal C_{\bm Q}
:
\begin{cases}
c_{i, \downarrow}\\
c^\dagger_{i, \downarrow}
\end{cases}
\to
\quad
\begin{cases}
e^{i\bm Q\cdot \bm r_i}\, c_{i, \downarrow}\\
e^{-i\bm Q\cdot\bm r_i}\, c^\dagger_{i, \downarrow}
\end{cases}\,,
\eea
where $\bm Q$ is a generic momentum vector.
The transformed EPR state becomes,
\bea
\label{eq:EtaScar}
\mathcal C_{\bm Q}\epr = \frac{1}{2^{N/2}}\prod\limits_{i=1}^N (\mathbbm 1 -e^{-i\bm Q\cdot\bm r_i} \mathcal{F}_\uparrow c^\dagger_{i,\uparrow}
\otimes 
c^\dagger_{i,\downarrow} )\ket{0}\,.
\eea
This transformed state generalizes the $\mathcal C$-transformed EPR state in Eq.~\eqref{eq:CEPR}---for example, on the 2D square lattice, choosing $\bm Q = (\pi,\pi)$ (in units where the lattice spacing is one) results in Eq.~\eqref{eq:CEPR}.
Under this transformation, the constraints in Eq.~\eqref{eq:FermConst} become 
\bea
\label{eq:EtaFermConst}
&t^{\uparrow \uparrow}_{ij} = -t^{\downarrow \downarrow}_{ji}e^{i\bm Q \cdot (\bm r_i-\bm r_j)},\quad &\Delta^{\uparrow \uparrow}_{ij} = -\Delta^{\downarrow \downarrow}_{ji}e^{i\bm Q\cdot (\bm r_i+\bm r_j)},\,\,\,\\
\\
&t^{\uparrow \downarrow}_{ij} = t^{\uparrow \downarrow}_{ji}e^{i\bm Q\cdot(\bm r_i-\bm r_j)},\quad &\Delta^{\uparrow \downarrow}_{ij} = -\Delta^{\downarrow \uparrow}_{ij}e^{i\bm Q\cdot(\bm r_i+\bm r_j)}\,.
\eea
The EPR state $\mathcal C_{\bm Q}\epr$ is an exact eigenstate when $\sum_{i}\Delta^{\downarrow \uparrow}_{ii}e^{i\bm{Q}\cdot \bm r_{i}}=0$.
Note that, when $\bm Q=0$, the above constraints reduce to those in Eq.~\eqref{eq:FermConst}.

As a specific example of applying Eq.~\eqref{eq:EtaFermConst}, first demand that the couplings are translation invariant, i.e, $t^{\sigma \sigma^\prime}_{ij}\equiv t^{\sigma \sigma^\prime}_{i-j}$.
Since the phase factors on the $\Delta$ constraints oscillate with $(\bm r_i + \bm r_j)$, breaking the assumption of translation invariance, we set the pairing terms to zero.
We now rewrite the remaining two conditions in momentum space as follows:
\bea
\label{eq:MomConst}
&t^{\uparrow \uparrow}_{\bm{k}} = -t^{\downarrow \downarrow}_{-\bm{k}+\bm{Q}}, \quad t^{\uparrow \downarrow}_{\bm{k}} = t^{\uparrow \downarrow}_{-\bm{k}+\bm{Q}}\,.
\eea
These constraints are the same as the ones found in Ref.~\cite{Li} guaranteeing that the $\eta$-pairing states are eigenstates of the Fermi-Hubbard model with spin-orbit coupling.
More generic Fermi-Hubbard models are realized through the use of the constraints in conjunction with unitary transformations, such as the Hirsch model~\cite{HIRSCH1989326,Marsiglio,duan2007general,mark1,bhattacharyya1999hubbard}.  

\section{Bosonic Systems}
\label{sec:bosons}
\begin{figure*}[t]
    \includegraphics[width=1.00\textwidth]{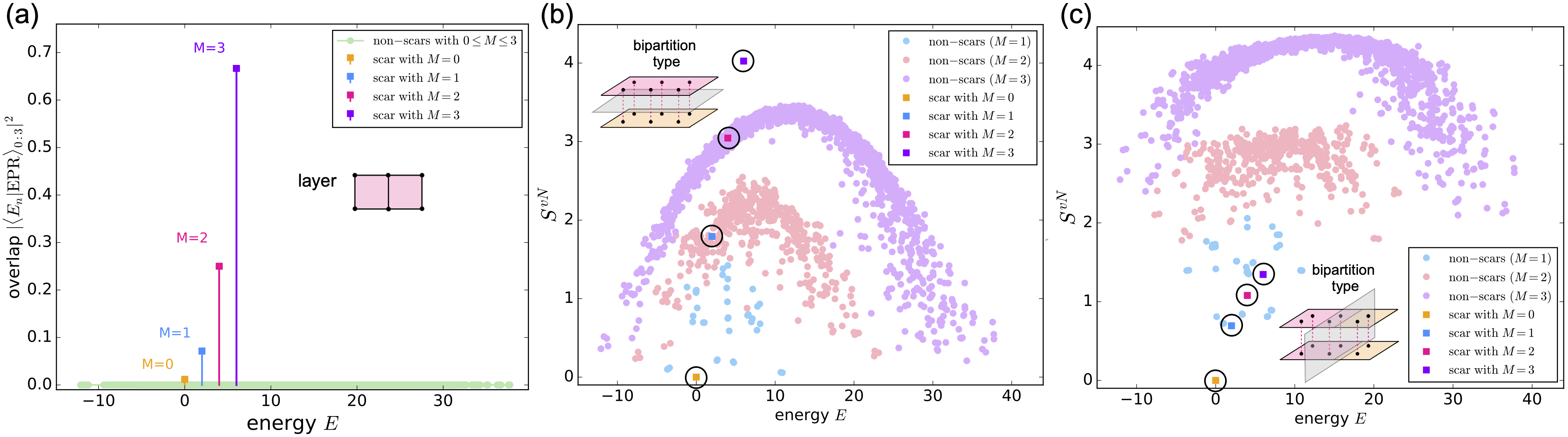}
    \caption{
    {\it Bose-Hubbard bilayer system on a square lattice}. Results shown are from exact diagonalization of $\mathcal C\H\mathcal C$ [see Eqs.~\eqref{eq:bh_ham0}--\eqref{eq:Cboson}] for two square-lattice layers of size $(L_x,L_y)=(3,2)$, for a total of $2N=12$ sites.
    $U_{i}$ and $t_{ij}$ are uniformly drawn from the interval $[0.9, 1.1]$ while the interlayer interaction $\lambda=1$. 
    The chemical potential $\mu=1.0$ is responsible for the position of the four scars in the energy spectrum (see Eq.~\eqref{ham_mu}).
    (a) The overlap between the EPR scar $\mathcal C\epr_{0:3}$ [see Eq.~\eqref{boseEPR}] and each energy eigenstate. Each $M$-sector contains exactly one scar state.
    (b) Entanglement entropy utilizing the the type-I
    bipartition shown in the inset for $M=0,\ldots,3$. The respective single scar state in each $M$-sector is identified with a closed square highlighted by a circle. 
    (c) Same as in (b) but with a type-II bipartition.
    }
    \label{fig:bosehubbard3}
\end{figure*}
We now focus on bilayer systems consisting of bosonic degrees of freedom. Advances and new possibilities of engineering lattice models with ultra-cold gases in  optical lattices have delivered a plethora of new 
discoveries experimentally and theoretically~\cite{trefzger2009pair,sun2015tunneling,sowi2012dipolar,wang2007quantum,young22}. Phonons in trapped-ion crystals (see, e.g., \cite{porras04,deng08, serafini09, debnath18, katz22c}) and optomechanical arrays (see, e.g., \cite{bhattacharya08,chang11b,raeisi20}), as well as photons in multi-mode cavities (see, e.g., \cite{mehta22b}) and cavity arrays (see, e.g., \cite{hartmann09,tomadin10,peropadre16}), are also often well-described by bosonic lattice Hamiltonians. 
The aim of the present section is to formulate the EPR scar construction in bilayer systems of itinerant bosonic degrees of freedom. 
In the case of U(1) symmetry, we uncover an {\it infinite} tower of EPR scar states labeled by the total U(1) charge, which is unique to the bosonic case. 
We then discuss different options for the interlayer interaction and comment on experimental settings for possible realization and detection. 

\subsection{Bosonic EPR State}
Consider a bosonic model on a lattice of $N$ sites with annihilation (creation) operators $b_i$~($b_i^\dagger$) defined on each site $i$ that satisfy the canonical commutation relation $[b_{i}, b_{j}^{\dagger}]=\delta_{ij}$.
Defining the normalized local number states for individual sites in $\HH_{1(2)}$,
\begin{align}
    \ket{M_i} = \frac{(b^\dagger_i)^M}{\sqrt{M!}}\ket{0_i}\,,
\end{align}
we can write the EPR state in the doubled Hilbert space as
\bea\label{epr_bosons}
    \epr &= \bigotimes^N_{i=1}\left(\sum^\infty_{M=0} \ket{M_i}\otimes \ket{M_i}\right)\\
    &= \exp (\sum_{i=1}^{N} b_{i}^{\dagger} \otimes b_{i}^{\dagger}) \ket{0}\,,
\eea
where $\ket{0} \equiv \bigotimes^N_{i=1}\ket{0_i}\otimes\ket{0_i}$ is the vacuum state. 
Note that this state is not normalizable, a feature unique to bosonic systems due to their infinite-dimensional local Hilbert space.
However, in the presence of U(1) symmetry, the projection of this state into each charge sector is normalizable. 
Expanding the exponential in the second line of Eq.~\eqref{epr_bosons}, we find an {\it infinite} tower of states spanned by
\bea\label{eq:boson_epr}
\epr_{M} =\frac{1}{M! \sqrt{\binom{N+M-1}{M}}} \left( \sum_{i=1}^{N} b_{i}^{\dagger} \otimes b_{i}^{\dagger} 
\right )^{M} \ket{0}\,,
\eea
which are normalized states with fixed boson number $2M$.

\subsection{Bose-Hubbard Models}
\label{sec:BHmodel}

Now we are ready to construct a bosonic many-body 
bilayer model that realizes this infinite tower of states.
Consider the following bilayer Bose-Hubbard model 
\bea
\label{eq:bh_ham0}
\H = \H_1\otimes \I + \I\otimes\H_2 +\H_{12}
\eea
with
\bea
\label{eq:bh_ham1}
\H_{1} 
&= -\sum_{\langle ij \rangle} \left(
t_{ij} b^{\dagger}_{i}b_{j} + \text{H.c.}\right)
+ \sum_{i} U_{i} n_{i}\left(n_{i}-1\right)\\
\H_{2} 
&= +\sum_{\langle ij \rangle} \left(
t^*_{ij} b^{\dagger}_{i}b_{j} + \text{H.c.}\right)
+ \sum_{i} U_{i} n_{i}\left(n_{i}-1\right)
\eea
where $n_{i} = b_{i}^{\dagger} b_{i}$ is the number operator, $t_{ij}$ are nearest-neighbor hoppings which in general can be complex, and $U_{i}$ is the onsite interaction strength.
The Hamiltonian Eq.~\eqref{eq:bh_ham1} conserves the boson number $\mathcal{M} = \sum_{i}n_{i}$ in each layer.
One choice of interlayer coupling such that $\epr$ is an eigenstate is
\bea
\label{eq:bh_ham1a}
\H_{12} = \lambda \sum_{i} \left(n_{i}\otimes \I - \I \otimes n_{i}\right)^{2}\,,
\eea
which explicitly penalizes differences in particle number between corresponding sites in different layers.
It is then straightforward to show, e.g.~using Eq.~\eqref{eq:EPR_property}, that $\H_{12}\epr=0$, such that Eq.~\eqref{eq:condition} is satisfied with $E=0$.
 
The hopping terms in the two layers are related by complex conjugation with a minus sign, and thus become the same if $t_{ij}$ is purely imaginary. When $t_{ij}$ is real, the 
relative minus sign can be partially removed on bipartite lattices with sublattices $A$ and $B$ by performing a chiral transformation on bosons in one of the layers,
\bea
\label{eq:Cboson}
\mathcal{C}: \I \otimes b_i \rightarrow - \I \otimes b_i  \quad \text{for }i \in B\,.
\eea
This transformation commutes with $\H_1$ and $\H_{12}$ but changes the sign of the hopping term in $\H_2$:
\bea
\label{ham_BH2}
&\mathcal C\H_{2}\mathcal C = 
-\sum_{\langle ij\rangle} 
t_{i j}\left( b^{\dagger}_{i}b_{j} + \text{H.c.}\right)
-  \sum_{i} U_{i} n_{i}\left(n_{i}-1\right)\,.
\eea 
The $\mathcal C$-transformed EPR state
\bea
 \mathcal C\epr &= \exp (\sum_{i=1}^{N} \xi_i b_{i}^{\dagger} \otimes b_{i}^{\dagger}) \ket{0}\,,
\eea
where $\xi_i = +1$ ($-1$) for $i\in A$ ($B$), and the associated U(1) tower of states,
\bea
\label{eq:CEPR_tower_bose}
\mathcal C\epr_{M} =\frac{1}{M! \sqrt{\binom{N+M-1}{M}}} \left( \sum_{i=1}^{N} \xi_i b_{i}^{\dagger} \otimes b_{i}^{\dagger} 
\right )^{M} \ket{0}\,,
\eea
are then eigenstates of $\mathcal C \H \mathcal C$.

To visualize the U(1) scar tower of $\mathcal C\H\mathcal C$, we perform an 
exact diagonalization study. For practical reasons, we limit the number of bosons to 
a maximum of $M_{\rm max}$ per layer, i.e., we consider the $\mathcal{C}$-transformed version of the truncated EPR state
\begin{align}
\label{boseEPR}
\epr_{0:M_{\rm max}} = 
\Gamma(N,M_{\rm Max})
\sum_{M=0}^{M_{\rm max}} \sqrt{\binom{N+M-1}{M}} \epr_{M}
\end{align}
with a normalization factor of 
$\Gamma(N,M_{\rm Max})=\frac{1}{\sqrt{\binom{N+M_{\rm max}}{M_{\rm max}}}}$. 
For the numerical study, we set $M_{\rm max}=3$; we then expect four EPR scar states with $0 \leq M \leq 3$. We consider a bilayer square lattice where each layer is of dimensions $(L_x,L_y)=(3,2)$.
The site-dependent Hubbard interaction $U_{i}$ and hopping matrix elements $t_{ij}$ are uniformly drawn from the interval $[0.9, 1.1]$ while the interlayer interaction strength $\lambda = 1$. 
Since the states $\epr_{M}$ described by Eq.~\eqref{eq:boson_epr} are eigenstates of Eq.~\eqref{eq:bh_ham0} with the same energy, we numerically scrutinize a slightly modified Hamiltonian by adding a chemical potential to Eq.~\eqref{eq:bh_ham0} so that the newly designed Hamiltonian 
\bea 
\label{ham_mu}
\mathcal{H} + \mu\sum_{i=1}^{N} n_{i}\otimes \I + \I\otimes n_{i},
\eea
with $\mu = 1.0$, hosts a tower of EPR scars with equal energy spacing. 

In Fig.~\ref{fig:bosehubbard3}(a), we plot the overlap between Eq.~\eqref{boseEPR} and each eigenstate of the bilayer system as a function of energy. 
The four scars with boson numbers $2M$ ($M=0,1,2,3$, with data from different sectors plotted in different colors) are clearly visible and are located at energies $\mu M$.  
Note that the overlap of the scar states with $\mathcal C\epr_{0:3}$ increases monotonically with particle number,
since the total number of configurations grows as more bosons are added.
We also compute the entanglement entropy for bipartitions of types I and II in Fig.~\ref{fig:bosehubbard3}(b) and (c), respectively.
As in Secs.~\ref{sec:spins} and \ref{sec:fermions}, the bipartition-dependence of the entanglement entropy characteristic of EPR scars is again clearly visible.

We close this section by summarizing two alternative parent Hamiltonians for EPR scar states in Bose-Hubbard systems.
We present these Hamiltonians assuming a bipartite lattice so that hoppings of the same sign can be used in both layers; the relevant scar state is then the $\mathcal C$-transformed EPR state.

The first parent Hamiltonian we consider is defined by 
\begin{align}
\label{ham_BH3}
\H_{1} &= -\sum\limits_{\langle ij\rangle} t_{ij} (b_{i}^{\dagger} b_{j} + \text{H.c.})  + U\sum\limits_{i} n_{i} (n_{i}-1) \nonumber\\
&= \H_{2} \\
\H_{12} &= -2U \sum_{i} n_{i} \otimes n_{i}\,.\nonumber
\end{align}
This is simply a system of two identical Bose-Hubbard layers coupled by an interlayer density-density interaction.
The hopping terms in $\H_1$ and $\H_2$ cancel when acting on the state $\mathcal C\epr$, however the interaction terms do not as they are invariant under $\mathcal C$.
However, the coefficient of the interaction term in $\H_{12}$ is chosen such that it cancels the intralayer interaction terms upon applying Eq.~\eqref{eq:EPR_property}:
\begin{align}
    (\H_1+\H_2^*-2U\sum_i n_i^2)\mathcal C\epr = -2U\sum_i n_i\mathcal C\epr\,.
\end{align}
The leftover term is just a chemical potential that is fixed within each U(1) charge sector.
Therefore the states in the infinite U(1) $\mathcal C$-EPR tower~Eq.~\eqref{eq:CEPR_tower_bose} are eigenstates of the Hamiltonian defined by Eq.~\eqref{ham_BH3}:
\begin{eqnarray}
\H \mathcal{C} \epr_{M} = -2 U M\mathcal{C}\epr_{M}\,.
\end{eqnarray}

The second parent Hamiltonian we consider is defined by 
\bea
\label{ham_BH4}
\H_{1} &= -\sum\limits_{\langle ij\rangle} t_{ij} (b_{i}^{\dagger} b_{j} + \text{H.c.})  + U\sum\limits_{i} n_{i} (n_{i}-1)\\
&\qquad +\frac{\lambda}{2}\sum_i \left[(b_i^\dagger)^2+b_i^2\right]\\
&= \H_{2}, \\
\H_{12}&= -\lambda \sum_{i} (b_i^\dagger\otimes b_i + b_i\otimes b_i^\dagger)\,. 
\eea
In this case, the layers are coupled by an interlayer tunneling, which is present in experimental realizations of Bose-Hubbard bilayers~\cite{Preiss15}.
The U(1)-nonconserving term $\frac{\lambda}{2}\sum_i \left[(b_i^\dagger)^2+b_i^2\right]$ in both layers is necessary to retain $\mathcal C\epr$ as an eigenstate in the presence of interlayer tunneling. Such pair creation-annihilation Hamiltonians can be engineered, for example, in photonic \cite{leghtas15} and atomic \cite{zhang21f} systems using a coherent state of another species of bosons that can be coherently converted into a pair of $b_i$ bosons.  
Since the operators $b_i$ and $b_i^\dagger$ are real in the Fock basis, mapping $\H_{12}$ onto a single layer using Eq.~\eqref{eq:EPR_property} yields $-\lambda\sum_i \left[(b_i^\dagger)^2+b_i^2\right]$, which then cancels against the corresponding terms in $\H_{1(2)}$ when $\H$ is applied to $\mathcal C\epr$.
Note that the model \eqref{ham_BH4} conserves the parity of the total number of bosons---thus, there are two EPR scar states corresponding to the projection of $\mathcal C\epr$ into each parity sector.
 
\section{Conclusion}\label{sec:conclusions}
In this work, we elaborated on a construction~\cite{ourrainbow} in which maximally entangled EPR states between two copies of a quantum many-body system give rise to scar states with a distinctive entanglement structure. 
We applied this construction to systems of spins, fermions, and bosons and demonstrated that it can be used to obtain several well-known examples of scar states, including the towers of states in the spin-1 XY~\cite{TomMichael} and Fermi-Hubbard~\cite{YangPRL,mark1,Sanjay1,Pakrouski,Pallegar} models. 
We also demonstrated a qualitatively distinct 
{\it infinite} tower of states in number-conserving bosonic models.

Our work motivates several directions of future research.
One direction is to apply the construction to constrained degrees of freedom such as anyons and dimers, the latter of which can be realized experimentally in systems of Rydberg atoms~\cite{Verresen21,Semeghini21}.
Our results on bosonic systems also raise the possibility of finding scar states in Bose-Hubbard bilayers~\cite{Preiss15} or multispecies mixtures.
An important concern in experimental realizations is the ability to prepare a state having large overlap with the EPR state for a subsequent quantum quench, which reveals coherent dynamical signatures of these scar states~\cite{ourrainbow}.
A variety of state preparation protocols are possible depending on the nature of the experimental apparatus.
For example, in hybrid analog-digital setups such as the one detailed in Ref.~\cite{bluvstein2022epr}, the EPR state can be prepared using digital gates before performing an analog quantum quench.
An alternative possibility is to prepare the EPR states adiabatically using an appropriately engineered Hamiltonian.
Another intriguing option is to use non-unitary methods, e.g. as described in Ref.~\cite{Agarwal22}.

\begin{acknowledgments}
This work was supported in part by the National Science Foundation under Grant No.~DMR-2143635 (T.I.).
Z.-C.Y. acknowledges a startup fund at Peking University.
The work of T.I.~was performed in part at the Aspen Center for Physics, which is supported by National Science Foundation grant PHY-1607611. Z.-C.Y.~and A.V.G.~acknowledge funding by AFOSR, AFOSR MURI, NSF PFCQC program, DoE ASCR Quantum Testbed Pathfinder program (award No.~DE-SC0019040), DoE QSA, NSF QLCI (award No.~OMA-2120757), DoE ASCR Accelerated Research in Quantum Computing program (award No.~DE-SC0020312), ARO MURI, and DARPA SAVaNT ADVENT.  
\end{acknowledgments}

\appendix
\begin{figure*}[t!]
\includegraphics[width=1.00\textwidth]{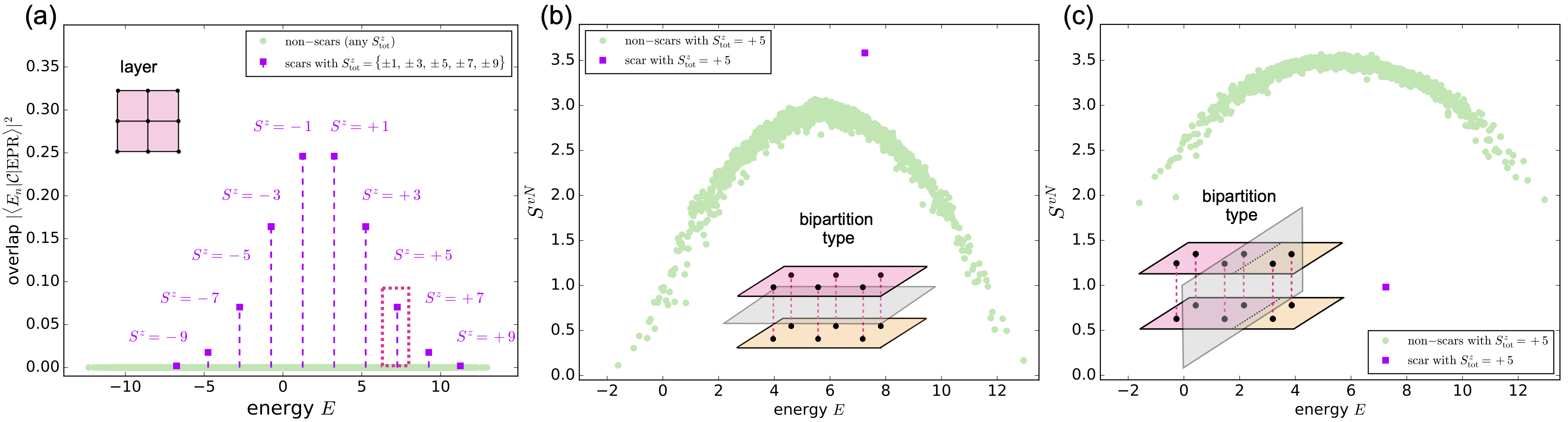}
\caption{
         {\it Heisenberg spin-$1/2$ bilayer system on a square lattice}.
         (a) Overlap $|\langle E_{n}| \mathcal{C}|{\rm EPR}\rangle|^{2}$ between the EPR scar and each eigenstate $\ket{E_{n}}$ of Eq.~\eqref{eq:HeisSq} as a function of energy.
         The full EPR scar~\eqref{c_epr_app} is projected 
         into each non-zero magnetization sector in the 
         range $[-N, N]$ in steps of  $2$ with $N=9$ leading 
         to $N+1=10$ scars. Each of the two layers is of size  $(L_x,L_y) = (3,3)$ (see inset). 
         (b) The Von-Neunmann entanglement entropy $S^{\rm vN}$ utilizing the bipartition given in the inset within 
         the magnetization sector $S_{\rm tot}^{z} = +5$ where the single EPR scar is denoted with a square.
         (c) The same as in (b) with a different bipartition to demonstrate that the EPR entanglement scaling depends on the chosen system bipartition. Here we choose a cut (see inset) that does not cut the interlayer interaction $\H_{12}$. 
}
\label{fig:square_overlapN6_abc}
\end{figure*}
\section{Bilayer Square-Lattice Heisenberg Model: U(1) symmetry}
\label{appA}
To emphasize that the number of EPR scars does not depend on the geometry but rather on the symmetries of the Hamiltonian, we study in this Appendix the Heisenberg model on a square lattice utilizing our bilayer prescription.
For this example, we take the Hamiltonian of the form
\bea
\label{eq:HeisSq}
\H  = \H_{1}\otimes \I + \I\otimes\H_{2} + \H_{12}\,, 
\eea
where 
\bea
\label{eq:scar_ham3a}
&\H_{1} =
\sum_{\langle ij\rangle}J_{ij}\,\left(S_{i}^{+} S_{j}^{-} 
+ S_{i}^{-}S_{j}^{+}\right) 
+ \sum_{\langle ij\rangle}\Delta_{ij}S_{i}^{z}S_{j}^{z},
\\
&\H_{2} = 
\sum_{\langle ij\rangle}J_{ij}\,
\left(S_{i}^{+} S_{j}^{-} + S_{i}^{-}S_{j}^{+}\right) 
- \sum_{\langle ij\rangle}\Delta_{ij}S_{i}^{z}S_{j}^{z},
\\ 
&\H_{12} = \lambda\sum_{i}S_{i}^{z} \otimes S_{i}^{z}\,.
\eea
Notice that, unlike in Eq.~\eqref{eq:ham1_parts}, we assume interlayer Ising---rather than Heisenberg---coupling. 
The minus sign from the construction relating the single-copy Hamiltonians is partially gauged away under the transformation
\bea
\label{eq:Cspin}
\mathcal{C}: \I \otimes S^{\pm}_{i} \rightarrow - \I \otimes S^{\pm}_{i} \quad \text{for }i \in B\,,
\eea
which acts on {\it one} sublattice ($B$) of {\it one} layer.  
Consequently, the intralayer couplings $J_{ij}$ in Eq.~\eqref{eq:scar_ham3a} appear with the same sign, while the parameters $\Delta_{ij}$ in the two layers are still required to appear with opposite signs. 
Similar to the case described in Sec.~\ref{subsec:XY} the $\mathcal{C}$-transformed EPR state is 
\bea
\label{c_epr_app}
\mathcal{C} \epr = & \frac{1}{2^{N/2}} \bigotimes_{i=1}^{N} (\ket{0_i}\otimes\ket{0_{i}} + (-1)^ i \ket{1_i}\otimes\ket{1_{i}})\,.
\eea

The Hamiltonian Eq.~\eqref{eq:HeisSq} has two independent U(1) symmetries associated with $S_{\rm tot}^{z,1} = \sum_{i}S_{i}^{z}\otimes \I$ and $S_{\rm tot}^{z,2} =  \sum_{i} \I \otimes S_{i}^{z}$. 
However, we stress that despite the two independent symmetries present in Eq.~\eqref{eq:scar_ham3a}, only the U(1) symmetry associated with the total $S^{z}_{\rm tot} = S_{\rm tot}^{z,1} + S_{\rm tot}^{z,2}$ is relevant for the tower of scar states.  

To visualize this tower of EPR scar states we perform an exact diagonalization study. 
To fully resolve all scar states, 
we add to Eq.~\eqref{eq:HeisSq} a term proportional to $S_{\rm tot}^{z} = S_{\rm tot}^{z,1} + S_{\rm tot}^{z,2}$ 
to guarantee that each of the $N+1$ EPR 
scar states arising from projecting Eq.~\eqref{c_epr_app} 
into different $S_{\rm tot}^{z}$ sectors 
has a distinctive energy offset. Specifically, we diagonalize the Hamiltonian
\bea
\label{eq:hh}
\H  
+ h \sum_{i} \left( S_{i}^{z} \otimes \I + \I \otimes S_{i}^{z} \right)\,, 
\eea
with $\H$ defined in Eq.~\eqref{eq:HeisSq}. 
We study a system of $2N=18$ spins on a square lattice bilayer, where each layer is of size $(L_x,L_y)=(3,3)$ (see Fig.~\ref{fig:square_overlapN6_abc}(a) inset). 
The intralayer exchange couplings $J_{ij}$ and $\Delta_{ij}$ 
are uniformly drawn from the interval $[1.0, 2.0]$ to break 
the  $\pi/2$ rotational symmetry of the square 
lattice. The interlayer coupling and the magnetic field are set to one, i.e. $\lambda = 1$ and $h = 1$.   

In Fig.~\ref{fig:square_overlapN6_abc}(a) we compute the fidelity $|\langle E_{n}|\mathcal{C}|\text{EPR}\rangle|^{2}$ between each energy eigenstate $\ket{E_n}$ of the Hamiltonian~\eqref{eq:hh} and the EPR state~\eqref{c_epr_app}. 
Due to the correlated nature of the EPR state, 
the total magnetization $S^{z}_{\rm tot}$ of the states in the scar tower is restricted to values 
\begin{eqnarray}
S_{\rm tot}^{z} = \left\{\pm 1, \pm 3, \pm 5, \pm 7, \pm 9\right\}
\end{eqnarray}
associated with $N+1=10$ scar states projected into each non-zero total magnetization sector as seen in Fig.~\ref{fig:square_overlapN6_abc}(a). 
In analogy with the bilayer system discussed in Sec.~\ref{subsec:Heisenberg} we clearly see that the entropically most likely allowed magnetization sectors 
$S^z_{\rm tot} = \pm 1$ have the highest total weight, as 
should be expected from the fact that $\mathcal{C}\epr$ is an equal  amplitude superposition of allowed spin configurations. 
\begin{figure}[b]
\includegraphics[width=1.00\columnwidth]{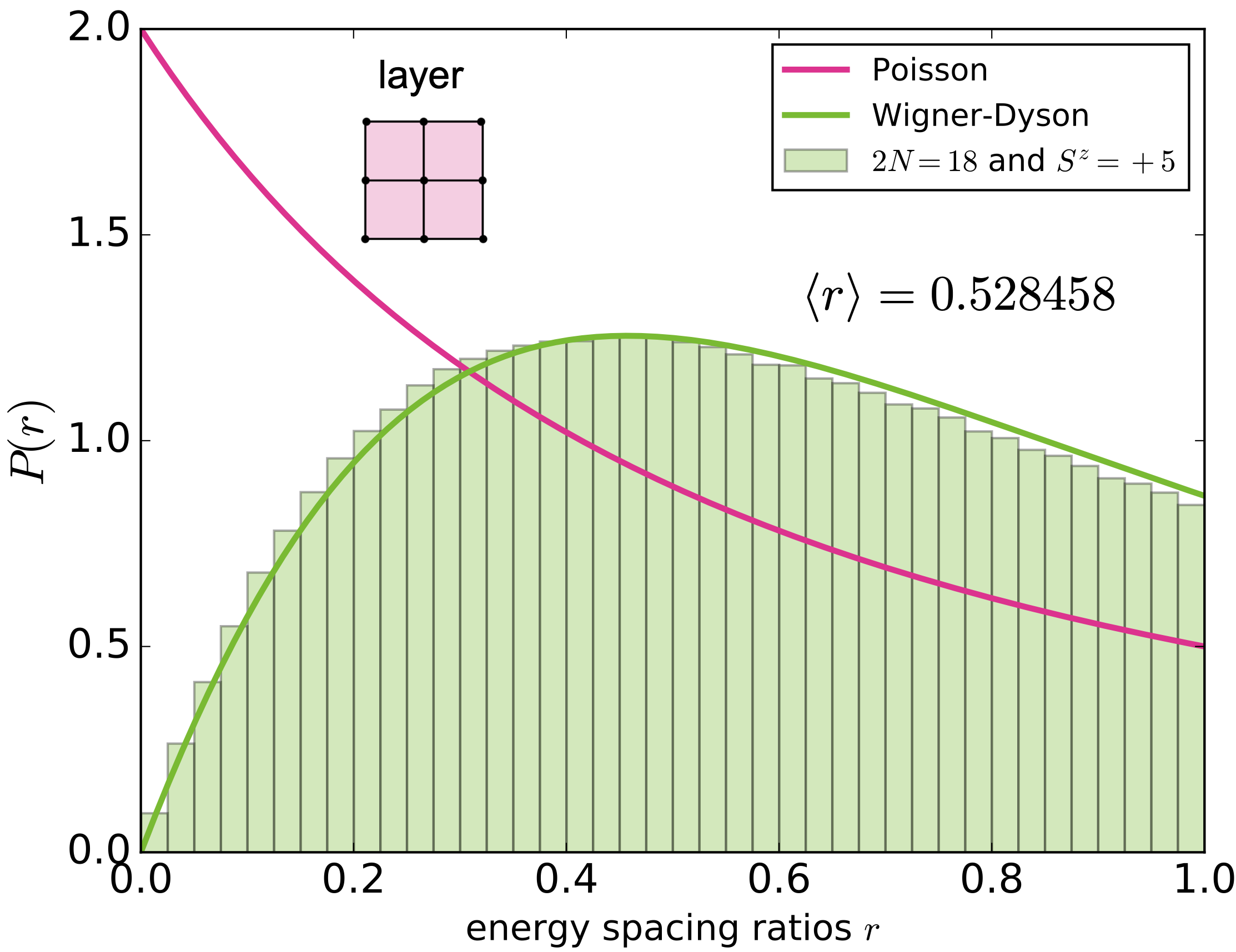}
\caption{
         {\it Heisenberg spin-$1/2$ bilayer system on a square lattice}.
         Probability distribution $P(r)$ for the energy level spacing parameter $r$ from exact  diagonalization. 
         The curves represent the analytical predication from Wigner-Dyson statistics~(green) and Poisson statistics~(magenta).
         We also find the average level spacing parameter $\langle r \rangle \approx 0.528$, which is near the GOE predication $\langle r \rangle_{\rm GOE}  \approx 0.53$ indicating a thermalized system.
         The parameters used here are the same as in the data shown in Fig.~\ref{fig:square_overlapN6_abc}, i.e. a square lattice bilayer of size $2N=18$ with layer dimensions $(L_x,L_y) = (3,3)$, restricted to the magnetization sector
         $S^{z}_{\rm tot} = +5$. The results are averaged over $2000$ independent disorder 
         ($J_{ij},\Delta_{ij} \in [1.0,2.0]$) realizations.   
}
\label{fig:square_overlapN6_d}
\end{figure}

We study the entanglement properties of the eigenstates in Fig.~\ref{fig:square_overlapN6_abc}(b) and (c) for two different bipartitions, type I and II, respectively (see insets and Fig.~\ref{fig:area_cuts}) in the $S_{\rm tot}^{z}=+5$ 
sector. 
The single scar state associated with $S^{z}_{\rm tot} = +5$ 
is highlighted by a closed purple square. 
In Fig.~\ref{fig:square_overlapN6_abc}(b), the entanglement entropy of the scar state
with respect to the type-I cut takes the maximum value allowed by the dimension of its $S^z_{\rm tot}$ sector, consistent with the discussion in Sec.~\ref{sec:EPR}. 
In comparison, Fig.~\ref{fig:square_overlapN6_abc}(c) utilizes a type-II cut for which the scar state appears as a state with anomalously low entanglement. 
The dome-like structure for each set of states with fixed $S_{\rm tot}$ is clearly visible. 

We point out that the 2D bilayer Heisenberg model in  Eqs.~\eqref{eq:HeisSq} and~\eqref{eq:scar_ham3a} is nonintegrable with and without random intralayer exchange couplings.  
To provide evidence for this, we study the level statistics.
Specifically, we utilize the same coupling parameters $J_{ij}, \Delta_{ij}$ and $\lambda$ as before and fix the magnetization to $S^{z}_{\rm tot} = +5$. 
In Fig.~\ref{fig:square_overlapN6_d}, we plot the probability distribution $P(r)$ of the ratio $r$ between the spacings of adjacent eigenvalues of Eq.~\eqref{eq:HeisSq}, defined as
\begin{align}
 r_{n} = \frac{\text{min}(E_{n+1}-E_{n},E_{n+2}-E_{n+1})}{\text{max}(E_{n+1}-E_{n},E_{n+2}-E_{n+1})}\,. 
\end{align}
Here the $\{E_{n}\}$ form an ordered list of energy eigenvalues.
We compute the mean level-spacing ratio to be 
$\langle r \rangle = 0.528$, which is in excellent agreement with the Gaussian orthogonal ensemble~(GOE)~\cite{GOE}, which according to random matrix theory has 
$\langle r \rangle_{\rm GOE}\approx 0.536$.
We stress that this is markedly different from the Poisson distribution value 
$\langle r \rangle_{\rm Poisson}\approx 0.386$~\cite{Poisson} found in integrable~(localized) models. 

\section{Additional Information on the Spin-$\frac{1}{2}$ Bilayer Models and the Bose-Hubbard Model}
\label{appB}
\begin{figure}[b]
\includegraphics[width=1.00\columnwidth]{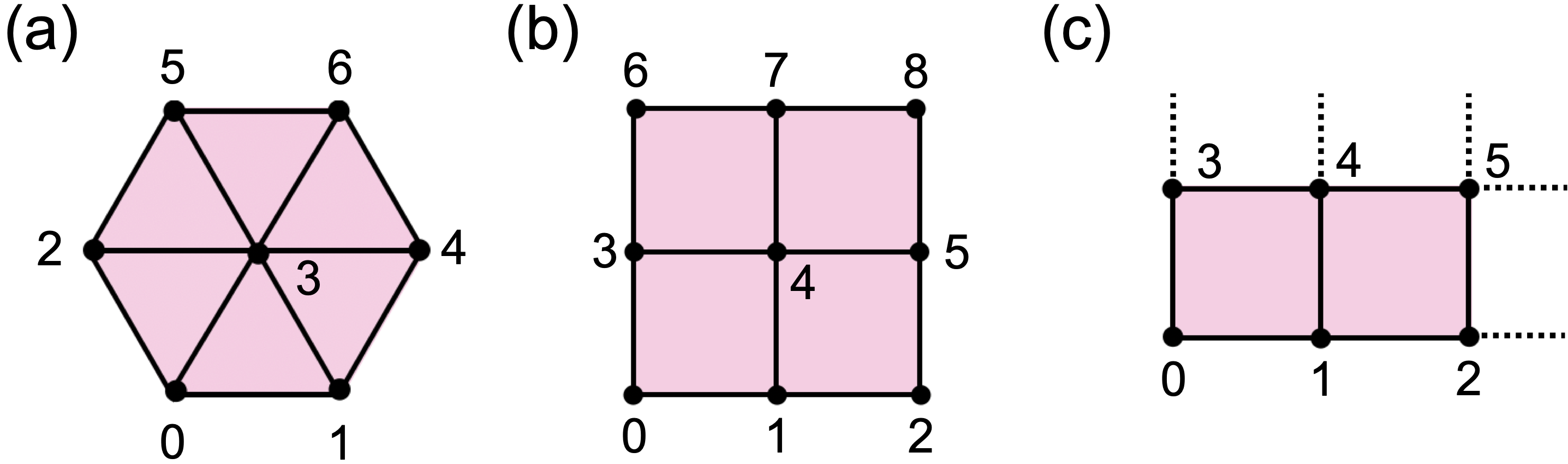}
\caption{
         {\it Square and triangular lattices for the bilayer systems}. (a) shows the $N=7$ single layer triangular lattice that in its doubled form is utilized for the SU(2) symmetric spin-$\frac{1}{2}$ Heisenberg bilayer system discussed in Sec.~\ref{subsec:Heisenberg} while (b) shows 
         the $(L_{x},L_{y})=(3,3)$  
         square lattice that is used for the U(1) symmetric spin-$\frac{1}{2}$ bilayer system scrutinized in Appendix~\ref{appA}. (c) shows the $(L_x,L_y)=(3,2)$ square lattice that the Bose-Hubbard bilayer model presented in Sec.~\ref{sec:BHmodel} 
         lives on. 
         Links representing the periodic boundaries are given by dotted lines.  
         Sites number in connection with couplings listed in Tables~\ref{table3},~\ref{table4}, and ~\ref{table5} give information on the parameters used in the numerics. 
}
\label{fig:parameter}
\end{figure}
In this Appendix we provide the full set of parameters for the spin-$\frac{1}{2}$ models scrutinized in Sec.~\ref{subsec:Heisenberg} and in Appendix~\ref{appA} and the Bose-Hubbard model in Sec.~\ref{sec:BHmodel}. 

The numerics presented for the SU(2) symmetric spin-$\frac{1}{2}$ bilayer model in Sec.~\ref{subsec:Heisenberg} are obtained from a system which consists of two $N=7$ triangular lattice layers (see inset Fig.~\ref{fig:tria_Heisenberg}(a) or Fig.~\ref{fig:parameter}(a). 
The intralayer couplings $J_{ij}$ acting on a pair of sites $(i,j)$ are given in Table~\ref{table3}. Labeling of the sites is shown in Fig.~\ref{fig:parameter}(a). The Heisenberg-type interlayer interaction is given by $\lambda=1$.  

In Appendix~\ref{appA} we scrutinize another spin-$\frac{1}{2}$ model, namely a Heisenberg model living on two square lattice layers of size $(L_x,L_y)=(3,3)$. The two layers interact via an Ising interaction resulting in a bilayer model with U(1) symmetry.  
The intralayer interaction couplings $J_{ij}$ and $\Delta_{ij}$ [pairs of sites $(i,j)$ are as shown in Fig.~\ref{fig:parameter}(b)] for the U(1) symmetric model are uniformly drawn from the interval $[1.0,2.0]$, and the specific values are given in Table~\ref{table4}. The interlayer coupling parameter $\lambda$ is set to one. 

In Sec.~\ref{sec:BHmodel} we present a Bose-Hubbard bilayer model that hosts an infinite U(1) tower of scar states. 
We offer numerical data on a bilayer system where each layer 
is a square lattice of size $(L_x,L_y)=(3,2)$ utilizing periodic boundary conditions~\footnote{The Bose-Hubbard bilayer model is the only model that utilizes periodic boundary conditions in the respective numerics. For all other bilayer models we use open  boundary conditions.}.  
The hopping parameters $t_{ij}$ and the onsite Hubbard interaction $U_{i}$ are uniformly drawn from the interval $[0.9,1.1]$, and the specific values are given in Table~\ref{table5}. Pairs of sites $(i,j)$ are as shown in Fig.~\ref{fig:parameter}(c). 
\begin{table*}[b]
    \begin{tabular}{ | c | c || c | c |}
    \hline
    \hline
    \multicolumn{4}{|c|}{\bf{Bilayer Triangular Lattice Spin-1/2  Model}} \\
    \hline
       \,\,sites $(i,j)$\,\, & \,\,coupling $J_{ij}$\,\, & \,\,sites  $(i,j)$\,\, & \,\,coupling $J_{ij}$\,\,  \\ \hline \hline
             $(1,0)$  & 1.02907  & $(4,3)$  & 1.03556  \\ \hline
             $(2,0)$  & 1.06241  & $(5,2)$  & 1.04165  \\ \hline
             $(3,0)$  & 0.99934  & $(5,3)$  & 1.07462  \\ \hline
             $(3,1)$  & 1.04354  & $(6,3)$  & 0.94805  \\ \hline
             $(3,2)$  & 1.05231  & $(6,4)$  & 0.958189 \\ \hline
             $(4,1)$  & 1.07878  & $(6,5)$  & 0.921175 \\ \hline
    \end{tabular}
\caption{
{\it Random intralayer couplings $J_{ij}$ for the SU(2) 
symmetric spin-$\frac{1}{2}$ Heisenberg bilayer model on the triangular lattice.} Shown are all coupling parameters $J_{ij}$ uniformly drawn from the interval $[0.9,1.1]$ for each 
of the twelve links, i.e. pairs of sites $(i,j)$ as depicted in Fig.~\ref{fig:parameter}(a). 
}
\label{table3}
\end{table*}
\begin{table*}[b]
    \begin{tabular}{ | c | c | c || c | c | c |}
    \hline
    \hline
    \multicolumn{6}{|c|}{\bf{Bilayer Square Lattice Spin-1/2  Model}} \\
    \hline
       \,sites $(i,j)$\, & \,coupling $J_{ij}$\, & \,coupling $\Delta_{ij}$\, & \,sites $(i,j)$\, & \,coupling $J_{ij}$\, & \,coupling $\Delta_{ij}$\,  \\ \hline \hline
             $(1,0)$  & 1.09727 & 1.76540 & $(5,4)$  & 1.23907 & 1.64768 \\ \hline
             $(2,1)$  & 1.54445 & 1.52591 & $(6,3)$  & 1.90745 & 1.22202 \\ \hline
             $(3,0)$  & 1.45301 & 1.31678 & $(7,4)$  & 1.24303 & 1.58075 \\ \hline
             $(4,1)$  & 1.12546 & 1.68808 & $(7,6)$  & 1.20747 & 1.08961 \\ \hline
             $(4,3)$  & 1.44898 & 1.13261 & $(8,5)$  & 1.12325 & 1.45412 \\ \hline
             $(5,2)$  & 1.06910 & 1.89085 & $(8,7)$  & 1.88908 & 1.55184 \\ \hline
    \end{tabular}
\caption{
{\it Random intralayer couplings $J_{ij}$ and $\Delta_{ij}$ for the U(1) symmetric spin-$\frac{1}{2}$ Heisenberg bilayer model on the square lattice.} Shown are all coupling parameters $J_{ij}$ and $\Delta_{ij}$ uniformly drawn from the interval $[1.0,2.0]$ for each of the twelve links, i.e. pairs of 
sites $(i,j)$ as depicted in Fig.~\ref{fig:parameter}(b).  
}
\label{table4}
\end{table*}
\begin{table*}[b]
    \begin{tabular}{ | c | c || c | c ||| c | c |}
    \hline
    \hline
    \multicolumn{6}{|c|}{\bf{Bilayer Bose-Hubbard Model}} \\
    \hline
       \,\,sites $(i,j)$\,\, & \,\,hopping $t_{ij}$\,\, &
       \,\,sites $(i,j)$\,\, & \,\,hopping $t_{ij}$\,\, & 
       \,\,site  $i$\,\, & \,\,coupling $U_{i}$\,\,  \\ \hline \hline
             $(0,1)$  & 1.08498  &  $(0,3)$   & 0.98946      & 0   & 1.07920   \\ \hline
             $(1,2)$  & 1.04128  &  $(3,0)$   & 0.90974      & 1   & 1.03249   \\ \hline
             $(2,0)$  & 1.00495  &  $(1,4)$   & 1.05341      & 2   & 1.08896   \\ \hline
             $(3,4)$  & 0.97037  &  $(4,1)$   & 1.04650      & 3   & 1.01060   \\ \hline
             $(4,5)$  & 1.07180  &  $(2,5)$   & 0.99541      & 4   & 1.00565   \\ \hline
             $(5,3)$  & 0.90285  &  $(5,2)$   & 1.06707      & 5   & 0.90678   \\ \hline
    \end{tabular}
\caption{
{\it Random hopping strengths $t_{ij}$ and onsite Hubbard repulsion $U_{i}$ 
for the U(1) symmetric Bose-Hubbard bilayer model on the 
square lattice.} Shown are all hopping parameters $t_{ij}$ and onsite Hubbard interaction strength $U_{i}$ uniformly drawn
from the interval $[0.9,1.1]$ for each of the twelve links, 
i.e. pairs of sites $(i,j)$ as depicted in Fig.~\ref{fig:parameter}(c). 
Pairs of sites $(i,j)$ with $i>j$ refer to links representing  
the periodic boundaries (see dashed links in  Fig.~\ref{fig:parameter}(c)).  
}
\label{table5}
\end{table*}

\section{Generalized Fermion Constraints}
\label{appC}

\subsection{Deriving the EPR constraints.}
\label{appC2}
In this appendix, we give an explicit example of how the constraints in Eqs.~(\ref{eq:FermConst})--(\ref{eq:EtaFermConst}) are derived and how the action of $\H$ on the EPR state is constrained to a two-dimensional EPR scar subspace unless an additional condition is imposed.
As our example, we will derive the third constraint in Eq.~\eqref{eq:PHFermConst} on the spin-orbit coupling constants $t^{\uparrow\downarrow}_{ij}$.
Consider the Hamiltonian under the $\pi$ transformation:
\bea
\label{eq:ConstHam}
\pi\mathcal{H}^{\uparrow \downarrow}\pi = \sum_{i,j}t_{ij}^{\uparrow \downarrow}c_{i,\uparrow}^{\dagger}c_{j,\downarrow}^{\dagger}+ (t_{ij}^{\uparrow\downarrow})^{*}c_{j,\downarrow}c_{i,\uparrow}\,.
\eea
In the following derivation, we will use the following forms of Eq.~\eqref{eq:EPR_property}: 
\bea
\label{eq:fermEPR_ident}
&c_{i, \downarrow}\epr = \mathcal{F}_{\uparrow}c_{i, \uparrow}^{\dagger}\epr\\
&c_{i, \downarrow}^{\dagger}\epr = \mathcal{F}_{\uparrow}c_{i, \uparrow}\epr\,,
\eea
which follows from applying Eq.~\eqref{eq:EPR_property} to the $\psi$ fermions and rewriting the result in terms of $c$ fermions
We now give a step-by-step procedure for determining the constraints on $t^{\uparrow \downarrow}_{ij}$:
\begin{widetext}
\bea
\pi\mathcal{H}^{\uparrow \downarrow}\pi\epr &= \sum_{i,j}\bigg[t_{ij}^{\uparrow \downarrow}c_{i,\uparrow}^{\dagger}c_{j,\downarrow}^{\dagger}- (t_{ij}^{\uparrow\downarrow})^{*}c_{i,\uparrow}c_{j,\downarrow}\bigg]\epr\\
&= -\sum_{i,j}\bigg[t_{ij}^{\uparrow \downarrow}\mathcal{F}_{\uparrow}c_{i,\uparrow}^{\dagger}c_{j,\uparrow}- (t_{ij}^{\uparrow\downarrow})^{*}\mathcal{F}_{\uparrow}c_{i,\uparrow}c_{j,\uparrow}^{\dagger}\bigg]\epr\\
&= -\sum_{i,j}\bigg[t_{ij}^{\uparrow \downarrow}\mathcal{F}_{\uparrow}c_{i,\uparrow}^{\dagger}c_{j,\uparrow}+ (t_{ij}^{\uparrow\downarrow})^{*}\mathcal{F}_{\uparrow}c_{j,\uparrow}^{\dagger}c_{i,\uparrow}\bigg]\epr 
+ \sum_{i}(t_{ii}^{\uparrow\downarrow})^{*}\mathcal{F}_{\uparrow}\epr\\
&= -\sum_{i,j}\bigg[t_{ij}^{\uparrow \downarrow}+(t_{ji}^{\uparrow \downarrow})^{*} \bigg]\mathcal{F}_{\uparrow}c_{i,\uparrow}^{\dagger}c_{j,\uparrow}\epr 
+ \sum_{i}(t_{ii}^{\uparrow\downarrow})^{*}\mathcal{F}_{\uparrow}\epr\,.
\eea
\end{widetext}
In the first line above we anticommuted the second set of fermionic operators. In the second line we applied the identity Eq.~\eqref{eq:fermEPR_ident} and moved the parity operator to the left using the relations $[\mathcal{F}_{\uparrow}, c_{i,\downarrow}~(c_{i,\downarrow}^{\dagger})]=0$ and $\{ \mathcal{F}_{\uparrow}, c_{i,\uparrow}~(c_{i,\uparrow}^{\dagger})\}=0$.
In the third line, we normal ordered the second term, giving rise to an extra term $\sim \mathcal{F}_{\uparrow}\epr$.
To arrive at the constraint, we set the first term in the final line to zero and find that $\left(t_{ij}^{\uparrow \downarrow}+t_{ij}^{\downarrow \uparrow} \right)=0$, which matches the main text.
When this constraint is satisfied, we find that $\pi\H^{\uparrow\downarrow}\pi$ maps $\epr$ to its parity partner $\mathcal F^\uparrow\epr$ (up to a constant).
To obtain the EPR state as a true eigenstate, we must additionally demand that $\sum_{i}t_{ii}^{\uparrow \downarrow}=0$.
All other constraints found in the main text are derived in a similar manner to the above.

\bibliography{YYY_References}

\begin{thebibliography}{87}%
\makeatletter
\providecommand \@ifxundefined [1]{%
 \@ifx{#1\undefined}
}%
\providecommand \@ifnum [1]{%
 \ifnum #1\expandafter \@firstoftwo
 \else \expandafter \@secondoftwo
 \fi
}%
\providecommand \@ifx [1]{%
 \ifx #1\expandafter \@firstoftwo
 \else \expandafter \@secondoftwo
 \fi
}%
\providecommand \natexlab [1]{#1}%
\providecommand \enquote  [1]{``#1''}%
\providecommand \bibnamefont  [1]{#1}%
\providecommand \bibfnamefont [1]{#1}%
\providecommand \citenamefont [1]{#1}%
\providecommand \href@noop [0]{\@secondoftwo}%
\providecommand \href [0]{\begingroup \@sanitize@url \@href}%
\providecommand \@href[1]{\@@startlink{#1}\@@href}%
\providecommand \@@href[1]{\endgroup#1\@@endlink}%
\providecommand \@sanitize@url [0]{\catcode `\\12\catcode `\$12\catcode
  `\&12\catcode `\#12\catcode `\^12\catcode `\_12\catcode `\%12\relax}%
\providecommand \@@startlink[1]{}%
\providecommand \@@endlink[0]{}%
\providecommand \url  [0]{\begingroup\@sanitize@url \@url }%
\providecommand \@url [1]{\endgroup\@href {#1}{\urlprefix }}%
\providecommand \urlprefix  [0]{URL }%
\providecommand \Eprint [0]{\href }%
\providecommand \doibase [0]{https://doi.org/}%
\providecommand \selectlanguage [0]{\@gobble}%
\providecommand \bibinfo  [0]{\@secondoftwo}%
\providecommand \bibfield  [0]{\@secondoftwo}%
\providecommand \translation [1]{[#1]}%
\providecommand \BibitemOpen [0]{}%
\providecommand \bibitemStop [0]{}%
\providecommand \bibitemNoStop [0]{.\EOS\space}%
\providecommand \EOS [0]{\spacefactor3000\relax}%
\providecommand \BibitemShut  [1]{\csname bibitem#1\endcsname}%
\let\auto@bib@innerbib\@empty
\bibitem [{\citenamefont {Serbyn}\ \emph {et~al.}(2021)\citenamefont {Serbyn},
  \citenamefont {Abanin},\ and\ \citenamefont {Papi{\'c}}}]{serbyn2021_review}%
  \BibitemOpen
  \bibfield  {author} {\bibinfo {author} {\bibfnamefont {M.}~\bibnamefont
  {Serbyn}}, \bibinfo {author} {\bibfnamefont {D.~A.}\ \bibnamefont {Abanin}},\
  and\ \bibinfo {author} {\bibfnamefont {Z.}~\bibnamefont {Papi{\'c}}},\
  }\bibfield  {title} {\bibinfo {title} {Quantum many-body scars and weak
  breaking of ergodicity},\ }\href {https://doi.org/10.1038/s41567-021-01230-2}
  {\bibfield  {journal} {\bibinfo  {journal} {Nature Physics}\ }\textbf
  {\bibinfo {volume} {17}},\ \bibinfo {pages} {675} (\bibinfo {year}
  {2021})}\BibitemShut {NoStop}%
\bibitem [{\citenamefont {Moudgalya}\ \emph {et~al.}(2022)\citenamefont
  {Moudgalya}, \citenamefont {Bernevig},\ and\ \citenamefont
  {Regnault}}]{Moudgalya22Review}%
  \BibitemOpen
  \bibfield  {author} {\bibinfo {author} {\bibfnamefont {S.}~\bibnamefont
  {Moudgalya}}, \bibinfo {author} {\bibfnamefont {B.~A.}\ \bibnamefont
  {Bernevig}},\ and\ \bibinfo {author} {\bibfnamefont {N.}~\bibnamefont
  {Regnault}},\ }\bibfield  {title} {\bibinfo {title} {Quantum many-body scars
  and hilbert space fragmentation: a review of exact results},\ }\href
  {https://doi.org/10.1088/1361-6633/ac73a0} {\bibfield  {journal} {\bibinfo
  {journal} {Reports on Progress in Physics}\ }\textbf {\bibinfo {volume}
  {85}},\ \bibinfo {pages} {086501} (\bibinfo {year} {2022})}\BibitemShut
  {NoStop}%
\bibitem [{\citenamefont {Chandran}\ \emph {et~al.}(2022)\citenamefont
  {Chandran}, \citenamefont {Iadecola}, \citenamefont {Khemani},\ and\
  \citenamefont {Moessner}}]{Chandran22Review}%
  \BibitemOpen
  \bibfield  {author} {\bibinfo {author} {\bibfnamefont {A.}~\bibnamefont
  {Chandran}}, \bibinfo {author} {\bibfnamefont {T.}~\bibnamefont {Iadecola}},
  \bibinfo {author} {\bibfnamefont {V.}~\bibnamefont {Khemani}},\ and\ \bibinfo
  {author} {\bibfnamefont {R.}~\bibnamefont {Moessner}},\ }\href@noop {}
  {\bibinfo {title} {Quantum many-body scars: A quasiparticle perspective}}
  (\bibinfo {year} {2022}),\ \Eprint {https://arxiv.org/abs/arXiv:2206.11528}
  {arXiv:2206.11528} \BibitemShut {NoStop}%
\bibitem [{\citenamefont {Deutsch}(1991)}]{ETH1}%
  \BibitemOpen
  \bibfield  {author} {\bibinfo {author} {\bibfnamefont {J.~M.}\ \bibnamefont
  {Deutsch}},\ }\bibfield  {title} {\bibinfo {title} {{Quantum statistical
  mechanics in a closed system}},\ }\href
  {https://doi.org/10.1103/PhysRevA.43.2046} {\bibfield  {journal} {\bibinfo
  {journal} {Phys. Rev. A}\ }\textbf {\bibinfo {volume} {43}},\ \bibinfo
  {pages} {2046} (\bibinfo {year} {1991})}\BibitemShut {NoStop}%
\bibitem [{\citenamefont {Srednicki}(1994)}]{ETH2}%
  \BibitemOpen
  \bibfield  {author} {\bibinfo {author} {\bibfnamefont {M.}~\bibnamefont
  {Srednicki}},\ }\bibfield  {title} {\bibinfo {title} {{Chaos and quantum
  thermalization}},\ }\href {https://doi.org/10.1103/PhysRevE.50.888}
  {\bibfield  {journal} {\bibinfo  {journal} {Phys. Rev. E}\ }\textbf {\bibinfo
  {volume} {50}},\ \bibinfo {pages} {888} (\bibinfo {year} {1994})}\BibitemShut
  {NoStop}%
\bibitem [{\citenamefont {D'Alessio}\ \emph {et~al.}(2016)\citenamefont
  {D'Alessio}, \citenamefont {Kafri}, \citenamefont {Polkovnikov},\ and\
  \citenamefont {Rigol}}]{GOE}%
  \BibitemOpen
  \bibfield  {author} {\bibinfo {author} {\bibfnamefont {L.}~\bibnamefont
  {D'Alessio}}, \bibinfo {author} {\bibfnamefont {Y.}~\bibnamefont {Kafri}},
  \bibinfo {author} {\bibfnamefont {A.}~\bibnamefont {Polkovnikov}},\ and\
  \bibinfo {author} {\bibfnamefont {M.}~\bibnamefont {Rigol}},\ }\bibfield
  {title} {\bibinfo {title} {From quantum chaos and eigenstate thermalization
  to statistical mechanics and thermodynamics},\ }\href
  {https://doi.org/10.1080/00018732.2016.1198134} {\bibfield  {journal}
  {\bibinfo  {journal} {Advances in Physics}\ }\textbf {\bibinfo {volume}
  {65}},\ \bibinfo {pages} {239} (\bibinfo {year} {2016})},\ \Eprint
  {https://arxiv.org/abs/https://doi.org/10.1080/00018732.2016.1198134}
  {https://doi.org/10.1080/00018732.2016.1198134} \BibitemShut {NoStop}%
\bibitem [{\citenamefont {Rigol}\ \emph {et~al.}(2008)\citenamefont {Rigol},
  \citenamefont {Dunjko},\ and\ \citenamefont {Olshanii}}]{rigol1}%
  \BibitemOpen
  \bibfield  {author} {\bibinfo {author} {\bibfnamefont {M.}~\bibnamefont
  {Rigol}}, \bibinfo {author} {\bibfnamefont {V.}~\bibnamefont {Dunjko}},\ and\
  \bibinfo {author} {\bibfnamefont {M.}~\bibnamefont {Olshanii}},\ }\bibfield
  {title} {\bibinfo {title} {Thermalization and its mechanism for generic
  isolated quantum systems},\ }\href {https://doi.org/10.1038/nature06838}
  {\bibfield  {journal} {\bibinfo  {journal} {Nature}\ }\textbf {\bibinfo
  {volume} {452}},\ \bibinfo {pages} {854} (\bibinfo {year}
  {2008})}\BibitemShut {NoStop}%
\bibitem [{\citenamefont {Polkovnikov}\ \emph {et~al.}(2011)\citenamefont
  {Polkovnikov}, \citenamefont {Sengupta}, \citenamefont {Silva},\ and\
  \citenamefont {Vengalattore}}]{ETH4}%
  \BibitemOpen
  \bibfield  {author} {\bibinfo {author} {\bibfnamefont {A.}~\bibnamefont
  {Polkovnikov}}, \bibinfo {author} {\bibfnamefont {K.}~\bibnamefont
  {Sengupta}}, \bibinfo {author} {\bibfnamefont {A.}~\bibnamefont {Silva}},\
  and\ \bibinfo {author} {\bibfnamefont {M.}~\bibnamefont {Vengalattore}},\
  }\bibfield  {title} {\bibinfo {title} {Colloquium: Nonequilibrium dynamics of
  closed interacting quantum systems},\ }\href
  {https://doi.org/10.1103/RevModPhys.83.863} {\bibfield  {journal} {\bibinfo
  {journal} {Rev. Mod. Phys.}\ }\textbf {\bibinfo {volume} {83}},\ \bibinfo
  {pages} {863} (\bibinfo {year} {2011})}\BibitemShut {NoStop}%
\bibitem [{\citenamefont {Turner}\ \emph {et~al.}(2018)\citenamefont {Turner},
  \citenamefont {Michailidis}, \citenamefont {Abanin}, \citenamefont {Serbyn},\
  and\ \citenamefont {Papi{\'c}}}]{turnerNat}%
  \BibitemOpen
  \bibfield  {author} {\bibinfo {author} {\bibfnamefont {C.~J.}\ \bibnamefont
  {Turner}}, \bibinfo {author} {\bibfnamefont {A.~A.}\ \bibnamefont
  {Michailidis}}, \bibinfo {author} {\bibfnamefont {D.~A.}\ \bibnamefont
  {Abanin}}, \bibinfo {author} {\bibfnamefont {M.}~\bibnamefont {Serbyn}},\
  and\ \bibinfo {author} {\bibfnamefont {Z.}~\bibnamefont {Papi{\'c}}},\
  }\bibfield  {title} {\bibinfo {title} {Weak ergodicity breaking from quantum
  many-body scars},\ }\href {https://doi.org/10.1038/s41567-018-0137-5}
  {\bibfield  {journal} {\bibinfo  {journal} {Nature Physics}\ }\textbf
  {\bibinfo {volume} {14}},\ \bibinfo {pages} {745} (\bibinfo {year}
  {2018})}\BibitemShut {NoStop}%
\bibitem [{\citenamefont {Schecter}\ and\ \citenamefont
  {Iadecola}(2019)}]{TomMichael}%
  \BibitemOpen
  \bibfield  {author} {\bibinfo {author} {\bibfnamefont {M.}~\bibnamefont
  {Schecter}}\ and\ \bibinfo {author} {\bibfnamefont {T.}~\bibnamefont
  {Iadecola}},\ }\bibfield  {title} {\bibinfo {title} {Weak ergodicity breaking
  and quantum many-body scars in spin-1 $xy$ magnets},\ }\href
  {https://doi.org/10.1103/PhysRevLett.123.147201} {\bibfield  {journal}
  {\bibinfo  {journal} {Phys. Rev. Lett.}\ }\textbf {\bibinfo {volume} {123}},\
  \bibinfo {pages} {147201} (\bibinfo {year} {2019})}\BibitemShut {NoStop}%
\bibitem [{\citenamefont {Iadecola}\ and\ \citenamefont
  {Schecter}(2020)}]{TomMichael2}%
  \BibitemOpen
  \bibfield  {author} {\bibinfo {author} {\bibfnamefont {T.}~\bibnamefont
  {Iadecola}}\ and\ \bibinfo {author} {\bibfnamefont {M.}~\bibnamefont
  {Schecter}},\ }\bibfield  {title} {\bibinfo {title} {Quantum many-body scar
  states with emergent kinetic constraints and finite-entanglement revivals},\
  }\href {https://doi.org/10.1103/PhysRevB.101.024306} {\bibfield  {journal}
  {\bibinfo  {journal} {Phys. Rev. B}\ }\textbf {\bibinfo {volume} {101}},\
  \bibinfo {pages} {024306} (\bibinfo {year} {2020})}\BibitemShut {NoStop}%
\bibitem [{\citenamefont {Moudgalya}\ \emph
  {et~al.}(2018{\natexlab{a}})\citenamefont {Moudgalya}, \citenamefont
  {Rachel}, \citenamefont {Bernevig},\ and\ \citenamefont
  {Regnault}}]{Sanjay4}%
  \BibitemOpen
  \bibfield  {author} {\bibinfo {author} {\bibfnamefont {S.}~\bibnamefont
  {Moudgalya}}, \bibinfo {author} {\bibfnamefont {S.}~\bibnamefont {Rachel}},
  \bibinfo {author} {\bibfnamefont {B.~A.}\ \bibnamefont {Bernevig}},\ and\
  \bibinfo {author} {\bibfnamefont {N.}~\bibnamefont {Regnault}},\ }\bibfield
  {title} {\bibinfo {title} {Exact excited states of nonintegrable models},\
  }\href {https://doi.org/10.1103/PhysRevB.98.235155} {\bibfield  {journal}
  {\bibinfo  {journal} {Phys. Rev. B}\ }\textbf {\bibinfo {volume} {98}},\
  \bibinfo {pages} {235155} (\bibinfo {year} {2018}{\natexlab{a}})}\BibitemShut
  {NoStop}%
\bibitem [{\citenamefont {Moudgalya}\ \emph
  {et~al.}(2018{\natexlab{b}})\citenamefont {Moudgalya}, \citenamefont
  {Regnault},\ and\ \citenamefont {Bernevig}}]{Sanjay5}%
  \BibitemOpen
  \bibfield  {author} {\bibinfo {author} {\bibfnamefont {S.}~\bibnamefont
  {Moudgalya}}, \bibinfo {author} {\bibfnamefont {N.}~\bibnamefont
  {Regnault}},\ and\ \bibinfo {author} {\bibfnamefont {B.~A.}\ \bibnamefont
  {Bernevig}},\ }\bibfield  {title} {\bibinfo {title} {Entanglement of exact
  excited states of affleck-kennedy-lieb-tasaki models: Exact results,
  many-body scars, and violation of the strong eigenstate thermalization
  hypothesis},\ }\href {https://doi.org/10.1103/PhysRevB.98.235156} {\bibfield
  {journal} {\bibinfo  {journal} {Phys. Rev. B}\ }\textbf {\bibinfo {volume}
  {98}},\ \bibinfo {pages} {235156} (\bibinfo {year}
  {2018}{\natexlab{b}})}\BibitemShut {NoStop}%
\bibitem [{\citenamefont {Shibata}\ \emph {et~al.}(2020)\citenamefont
  {Shibata}, \citenamefont {Yoshioka},\ and\ \citenamefont
  {Katsura}}]{Shibata}%
  \BibitemOpen
  \bibfield  {author} {\bibinfo {author} {\bibfnamefont {N.}~\bibnamefont
  {Shibata}}, \bibinfo {author} {\bibfnamefont {N.}~\bibnamefont {Yoshioka}},\
  and\ \bibinfo {author} {\bibfnamefont {H.}~\bibnamefont {Katsura}},\
  }\bibfield  {title} {\bibinfo {title} {Onsager's scars in disordered spin
  chains},\ }\href {https://doi.org/10.1103/PhysRevLett.124.180604} {\bibfield
  {journal} {\bibinfo  {journal} {Phys. Rev. Lett.}\ }\textbf {\bibinfo
  {volume} {124}},\ \bibinfo {pages} {180604} (\bibinfo {year}
  {2020})}\BibitemShut {NoStop}%
\bibitem [{\citenamefont {Mark}\ and\ \citenamefont {Motrunich}(2020)}]{mark1}%
  \BibitemOpen
  \bibfield  {author} {\bibinfo {author} {\bibfnamefont {D.~K.}\ \bibnamefont
  {Mark}}\ and\ \bibinfo {author} {\bibfnamefont {O.~I.}\ \bibnamefont
  {Motrunich}},\ }\bibfield  {title} {\bibinfo {title}
  {$\ensuremath{\eta}$-pairing states as true scars in an extended hubbard
  model},\ }\href {https://doi.org/10.1103/PhysRevB.102.075132} {\bibfield
  {journal} {\bibinfo  {journal} {Phys. Rev. B}\ }\textbf {\bibinfo {volume}
  {102}},\ \bibinfo {pages} {075132} (\bibinfo {year} {2020})}\BibitemShut
  {NoStop}%
\bibitem [{\citenamefont {Mark}\ \emph {et~al.}(2020)\citenamefont {Mark},
  \citenamefont {Lin},\ and\ \citenamefont {Motrunich}}]{mark2}%
  \BibitemOpen
  \bibfield  {author} {\bibinfo {author} {\bibfnamefont {D.~K.}\ \bibnamefont
  {Mark}}, \bibinfo {author} {\bibfnamefont {C.-J.}\ \bibnamefont {Lin}},\ and\
  \bibinfo {author} {\bibfnamefont {O.~I.}\ \bibnamefont {Motrunich}},\
  }\bibfield  {title} {\bibinfo {title} {Unified structure for exact towers of
  scar states in the affleck-kennedy-lieb-tasaki and other models},\ }\href
  {https://doi.org/10.1103/PhysRevB.101.195131} {\bibfield  {journal} {\bibinfo
   {journal} {Phys. Rev. B}\ }\textbf {\bibinfo {volume} {101}},\ \bibinfo
  {pages} {195131} (\bibinfo {year} {2020})}\BibitemShut {NoStop}%
\bibitem [{\citenamefont {Moudgalya}\ \emph
  {et~al.}(2020{\natexlab{a}})\citenamefont {Moudgalya}, \citenamefont
  {Regnault},\ and\ \citenamefont {Bernevig}}]{Sanjay1}%
  \BibitemOpen
  \bibfield  {author} {\bibinfo {author} {\bibfnamefont {S.}~\bibnamefont
  {Moudgalya}}, \bibinfo {author} {\bibfnamefont {N.}~\bibnamefont
  {Regnault}},\ and\ \bibinfo {author} {\bibfnamefont {B.~A.}\ \bibnamefont
  {Bernevig}},\ }\bibfield  {title} {\bibinfo {title}
  {$\ensuremath{\eta}$-pairing in hubbard models: From spectrum generating
  algebras to quantum many-body scars},\ }\href
  {https://doi.org/10.1103/PhysRevB.102.085140} {\bibfield  {journal} {\bibinfo
   {journal} {Phys. Rev. B}\ }\textbf {\bibinfo {volume} {102}},\ \bibinfo
  {pages} {085140} (\bibinfo {year} {2020}{\natexlab{a}})}\BibitemShut
  {NoStop}%
\bibitem [{\citenamefont {Chattopadhyay}\ \emph {et~al.}(2020)\citenamefont
  {Chattopadhyay}, \citenamefont {Pichler}, \citenamefont {Lukin},\ and\
  \citenamefont {Ho}}]{Chattopadhyay1}%
  \BibitemOpen
  \bibfield  {author} {\bibinfo {author} {\bibfnamefont {S.}~\bibnamefont
  {Chattopadhyay}}, \bibinfo {author} {\bibfnamefont {H.}~\bibnamefont
  {Pichler}}, \bibinfo {author} {\bibfnamefont {M.~D.}\ \bibnamefont {Lukin}},\
  and\ \bibinfo {author} {\bibfnamefont {W.~W.}\ \bibnamefont {Ho}},\
  }\bibfield  {title} {\bibinfo {title} {Quantum many-body scars from virtual
  entangled pairs},\ }\href {https://doi.org/10.1103/PhysRevB.101.174308}
  {\bibfield  {journal} {\bibinfo  {journal} {Phys. Rev. B}\ }\textbf {\bibinfo
  {volume} {101}},\ \bibinfo {pages} {174308} (\bibinfo {year}
  {2020})}\BibitemShut {NoStop}%
\bibitem [{\citenamefont {Pakrouski}\ \emph {et~al.}(2020)\citenamefont
  {Pakrouski}, \citenamefont {Pallegar}, \citenamefont {Popov},\ and\
  \citenamefont {Klebanov}}]{Pakrouski}%
  \BibitemOpen
  \bibfield  {author} {\bibinfo {author} {\bibfnamefont {K.}~\bibnamefont
  {Pakrouski}}, \bibinfo {author} {\bibfnamefont {P.~N.}\ \bibnamefont
  {Pallegar}}, \bibinfo {author} {\bibfnamefont {F.~K.}\ \bibnamefont
  {Popov}},\ and\ \bibinfo {author} {\bibfnamefont {I.~R.}\ \bibnamefont
  {Klebanov}},\ }\bibfield  {title} {\bibinfo {title} {Many-body scars as a
  group invariant sector of hilbert space},\ }\href
  {https://doi.org/10.1103/PhysRevLett.125.230602} {\bibfield  {journal}
  {\bibinfo  {journal} {Phys. Rev. Lett.}\ }\textbf {\bibinfo {volume} {125}},\
  \bibinfo {pages} {230602} (\bibinfo {year} {2020})}\BibitemShut {NoStop}%
\bibitem [{\citenamefont {Pakrouski}\ \emph {et~al.}(2021)\citenamefont
  {Pakrouski}, \citenamefont {Pallegar}, \citenamefont {Popov},\ and\
  \citenamefont {Klebanov}}]{Pallegar}%
  \BibitemOpen
  \bibfield  {author} {\bibinfo {author} {\bibfnamefont {K.}~\bibnamefont
  {Pakrouski}}, \bibinfo {author} {\bibfnamefont {P.~N.}\ \bibnamefont
  {Pallegar}}, \bibinfo {author} {\bibfnamefont {F.~K.}\ \bibnamefont
  {Popov}},\ and\ \bibinfo {author} {\bibfnamefont {I.~R.}\ \bibnamefont
  {Klebanov}},\ }\bibfield  {title} {\bibinfo {title} {Group theoretic approach
  to many-body scar states in fermionic lattice models},\ }\href
  {https://doi.org/10.1103/PhysRevResearch.3.043156} {\bibfield  {journal}
  {\bibinfo  {journal} {Phys. Rev. Research}\ }\textbf {\bibinfo {volume}
  {3}},\ \bibinfo {pages} {043156} (\bibinfo {year} {2021})}\BibitemShut
  {NoStop}%
\bibitem [{\citenamefont {Ren}\ \emph {et~al.}(2021)\citenamefont {Ren},
  \citenamefont {Liang},\ and\ \citenamefont {Fang}}]{Ren}%
  \BibitemOpen
  \bibfield  {author} {\bibinfo {author} {\bibfnamefont {J.}~\bibnamefont
  {Ren}}, \bibinfo {author} {\bibfnamefont {C.}~\bibnamefont {Liang}},\ and\
  \bibinfo {author} {\bibfnamefont {C.}~\bibnamefont {Fang}},\ }\bibfield
  {title} {\bibinfo {title} {Quasisymmetry groups and many-body scar
  dynamics},\ }\href {https://doi.org/10.1103/PhysRevLett.126.120604}
  {\bibfield  {journal} {\bibinfo  {journal} {Phys. Rev. Lett.}\ }\textbf
  {\bibinfo {volume} {126}},\ \bibinfo {pages} {120604} (\bibinfo {year}
  {2021})}\BibitemShut {NoStop}%
\bibitem [{\citenamefont {O'Dea}\ \emph {et~al.}(2020)\citenamefont {O'Dea},
  \citenamefont {Burnell}, \citenamefont {Chandran},\ and\ \citenamefont
  {Khemani}}]{Burnell}%
  \BibitemOpen
  \bibfield  {author} {\bibinfo {author} {\bibfnamefont {N.}~\bibnamefont
  {O'Dea}}, \bibinfo {author} {\bibfnamefont {F.}~\bibnamefont {Burnell}},
  \bibinfo {author} {\bibfnamefont {A.}~\bibnamefont {Chandran}},\ and\
  \bibinfo {author} {\bibfnamefont {V.}~\bibnamefont {Khemani}},\ }\bibfield
  {title} {\bibinfo {title} {From tunnels to towers: Quantum scars from lie
  algebras and $q$-deformed lie algebras},\ }\href
  {https://doi.org/10.1103/PhysRevResearch.2.043305} {\bibfield  {journal}
  {\bibinfo  {journal} {Phys. Rev. Research}\ }\textbf {\bibinfo {volume}
  {2}},\ \bibinfo {pages} {043305} (\bibinfo {year} {2020})}\BibitemShut
  {NoStop}%
\bibitem [{\citenamefont {Tang}\ \emph {et~al.}(2021)\citenamefont {Tang},
  \citenamefont {O'Dea},\ and\ \citenamefont {Chandran}}]{Tang21}%
  \BibitemOpen
  \bibfield  {author} {\bibinfo {author} {\bibfnamefont {L.-H.}\ \bibnamefont
  {Tang}}, \bibinfo {author} {\bibfnamefont {N.}~\bibnamefont {O'Dea}},\ and\
  \bibinfo {author} {\bibfnamefont {A.}~\bibnamefont {Chandran}},\ }\href@noop
  {} {\bibinfo {title} {Multi-magnon quantum many-body scars from tensor
  operators}} (\bibinfo {year} {2021}),\ \Eprint
  {https://arxiv.org/abs/arXiv:2110.11448} {arXiv:2110.11448} \BibitemShut
  {NoStop}%
\bibitem [{\citenamefont {Bernien}\ \emph {et~al.}(2017)\citenamefont
  {Bernien}, \citenamefont {Schwartz}, \citenamefont {Keesling}, \citenamefont
  {Levine}, \citenamefont {Omran}, \citenamefont {Pichler}, \citenamefont
  {Choi}, \citenamefont {Zibrov}, \citenamefont {Endres}, \citenamefont
  {Greiner} \emph {et~al.}}]{bernien}%
  \BibitemOpen
  \bibfield  {author} {\bibinfo {author} {\bibfnamefont {H.}~\bibnamefont
  {Bernien}}, \bibinfo {author} {\bibfnamefont {S.}~\bibnamefont {Schwartz}},
  \bibinfo {author} {\bibfnamefont {A.}~\bibnamefont {Keesling}}, \bibinfo
  {author} {\bibfnamefont {H.}~\bibnamefont {Levine}}, \bibinfo {author}
  {\bibfnamefont {A.}~\bibnamefont {Omran}}, \bibinfo {author} {\bibfnamefont
  {H.}~\bibnamefont {Pichler}}, \bibinfo {author} {\bibfnamefont
  {S.}~\bibnamefont {Choi}}, \bibinfo {author} {\bibfnamefont {A.~S.}\
  \bibnamefont {Zibrov}}, \bibinfo {author} {\bibfnamefont {M.}~\bibnamefont
  {Endres}}, \bibinfo {author} {\bibfnamefont {M.}~\bibnamefont {Greiner}},
  \emph {et~al.},\ }\bibfield  {title} {\bibinfo {title} {Probing many-body
  dynamics on a 51-atom quantum simulator},\ }\href
  {https://doi.org/10.1038/nature24622} {\bibfield  {journal} {\bibinfo
  {journal} {Nature}\ }\textbf {\bibinfo {volume} {551}},\ \bibinfo {pages}
  {579} (\bibinfo {year} {2017})}\BibitemShut {NoStop}%
\bibitem [{\citenamefont {Su}\ \emph {et~al.}(2022)\citenamefont {Su},
  \citenamefont {Sun}, \citenamefont {Hudomal}, \citenamefont {Desaules},
  \citenamefont {Zhou}, \citenamefont {Yang}, \citenamefont {Halimeh},
  \citenamefont {Yuan}, \citenamefont {Papi\'c},\ and\ \citenamefont
  {Pan}}]{Su22}%
  \BibitemOpen
  \bibfield  {author} {\bibinfo {author} {\bibfnamefont {G.-X.}\ \bibnamefont
  {Su}}, \bibinfo {author} {\bibfnamefont {H.}~\bibnamefont {Sun}}, \bibinfo
  {author} {\bibfnamefont {A.}~\bibnamefont {Hudomal}}, \bibinfo {author}
  {\bibfnamefont {J.-Y.}\ \bibnamefont {Desaules}}, \bibinfo {author}
  {\bibfnamefont {Z.-Y.}\ \bibnamefont {Zhou}}, \bibinfo {author}
  {\bibfnamefont {B.}~\bibnamefont {Yang}}, \bibinfo {author} {\bibfnamefont
  {J.~C.}\ \bibnamefont {Halimeh}}, \bibinfo {author} {\bibfnamefont {Z.-S.}\
  \bibnamefont {Yuan}}, \bibinfo {author} {\bibfnamefont {Z.}~\bibnamefont
  {Papi\'c}},\ and\ \bibinfo {author} {\bibfnamefont {J.-W.}\ \bibnamefont
  {Pan}},\ }\href@noop {} {\bibinfo {title} {Observation of unconventional
  many-body scarring in a quantum simulator}} (\bibinfo {year} {2022}),\
  \Eprint {https://arxiv.org/abs/arXiv:2201.00821} {arXiv:2201.00821}
  \BibitemShut {NoStop}%
\bibitem [{\citenamefont {Zhang}\ \emph {et~al.}(2022)\citenamefont {Zhang},
  \citenamefont {Dong}, \citenamefont {Gao}, \citenamefont {Zhao},
  \citenamefont {Hao}, \citenamefont {Guo}, \citenamefont {Chen}, \citenamefont
  {Deng}, \citenamefont {Liu}, \citenamefont {Ren}, \citenamefont {Yao},
  \citenamefont {Zhang}, \citenamefont {Xu}, \citenamefont {Wang},
  \citenamefont {Jin}, \citenamefont {Zhu}, \citenamefont {Li}, \citenamefont
  {Song}, \citenamefont {Wang}, \citenamefont {Liu}, \citenamefont {Papi\'c},
  \citenamefont {Ying}, \citenamefont {Wang},\ and\ \citenamefont
  {Lai}}]{Zhang22}%
  \BibitemOpen
  \bibfield  {author} {\bibinfo {author} {\bibfnamefont {P.}~\bibnamefont
  {Zhang}}, \bibinfo {author} {\bibfnamefont {H.}~\bibnamefont {Dong}},
  \bibinfo {author} {\bibfnamefont {Y.}~\bibnamefont {Gao}}, \bibinfo {author}
  {\bibfnamefont {L.}~\bibnamefont {Zhao}}, \bibinfo {author} {\bibfnamefont
  {J.}~\bibnamefont {Hao}}, \bibinfo {author} {\bibfnamefont {Q.}~\bibnamefont
  {Guo}}, \bibinfo {author} {\bibfnamefont {J.}~\bibnamefont {Chen}}, \bibinfo
  {author} {\bibfnamefont {J.}~\bibnamefont {Deng}}, \bibinfo {author}
  {\bibfnamefont {B.}~\bibnamefont {Liu}}, \bibinfo {author} {\bibfnamefont
  {W.}~\bibnamefont {Ren}}, \bibinfo {author} {\bibfnamefont {Y.}~\bibnamefont
  {Yao}}, \bibinfo {author} {\bibfnamefont {X.}~\bibnamefont {Zhang}}, \bibinfo
  {author} {\bibfnamefont {S.}~\bibnamefont {Xu}}, \bibinfo {author}
  {\bibfnamefont {K.}~\bibnamefont {Wang}}, \bibinfo {author} {\bibfnamefont
  {F.}~\bibnamefont {Jin}}, \bibinfo {author} {\bibfnamefont {X.}~\bibnamefont
  {Zhu}}, \bibinfo {author} {\bibfnamefont {H.}~\bibnamefont {Li}}, \bibinfo
  {author} {\bibfnamefont {C.}~\bibnamefont {Song}}, \bibinfo {author}
  {\bibfnamefont {Z.}~\bibnamefont {Wang}}, \bibinfo {author} {\bibfnamefont
  {F.}~\bibnamefont {Liu}}, \bibinfo {author} {\bibfnamefont {Z.}~\bibnamefont
  {Papi\'c}}, \bibinfo {author} {\bibfnamefont {L.}~\bibnamefont {Ying}},
  \bibinfo {author} {\bibfnamefont {H.}~\bibnamefont {Wang}},\ and\ \bibinfo
  {author} {\bibfnamefont {Y.-C.}\ \bibnamefont {Lai}},\ }\href@noop {}
  {\bibinfo {title} {Many-body hilbert space scarring on a superconducting
  processor}} (\bibinfo {year} {2022}),\ \Eprint
  {https://arxiv.org/abs/arXiv:2201.03438} {arXiv:2201.03438} \BibitemShut
  {NoStop}%
\bibitem [{\citenamefont {Chen}\ \emph {et~al.}(2022)\citenamefont {Chen},
  \citenamefont {Burdick}, \citenamefont {Yao}, \citenamefont {Orth},\ and\
  \citenamefont {Iadecola}}]{Chen22}%
  \BibitemOpen
  \bibfield  {author} {\bibinfo {author} {\bibfnamefont {I.-C.}\ \bibnamefont
  {Chen}}, \bibinfo {author} {\bibfnamefont {B.}~\bibnamefont {Burdick}},
  \bibinfo {author} {\bibfnamefont {Y.}~\bibnamefont {Yao}}, \bibinfo {author}
  {\bibfnamefont {P.~P.}\ \bibnamefont {Orth}},\ and\ \bibinfo {author}
  {\bibfnamefont {T.}~\bibnamefont {Iadecola}},\ }\href@noop {} {\bibinfo
  {title} {Error-mitigated simulation of quantum many-body scars on quantum
  computers with pulse-level control}} (\bibinfo {year} {2022}),\ \Eprint
  {https://arxiv.org/abs/arXiv:2203.08291} {arXiv:2203.08291} \BibitemShut
  {NoStop}%
\bibitem [{\citenamefont {Dooley}(2021)}]{Dooley21}%
  \BibitemOpen
  \bibfield  {author} {\bibinfo {author} {\bibfnamefont {S.}~\bibnamefont
  {Dooley}},\ }\bibfield  {title} {\bibinfo {title} {Robust quantum sensing in
  strongly interacting systems with many-body scars},\ }\href
  {https://doi.org/10.1103/PRXQuantum.2.020330} {\bibfield  {journal} {\bibinfo
   {journal} {PRX Quantum}\ }\textbf {\bibinfo {volume} {2}},\ \bibinfo {pages}
  {020330} (\bibinfo {year} {2021})}\BibitemShut {NoStop}%
\bibitem [{\citenamefont {Desaules}\ \emph {et~al.}(2022)\citenamefont
  {Desaules}, \citenamefont {Pietracaprina}, \citenamefont
  {Papi\ifmmode~\acute{c}\else \'{c}\fi{}}, \citenamefont {Goold},\ and\
  \citenamefont {Pappalardi}}]{Desaules22}%
  \BibitemOpen
  \bibfield  {author} {\bibinfo {author} {\bibfnamefont {J.-Y.}\ \bibnamefont
  {Desaules}}, \bibinfo {author} {\bibfnamefont {F.}~\bibnamefont
  {Pietracaprina}}, \bibinfo {author} {\bibfnamefont {Z.}~\bibnamefont
  {Papi\ifmmode~\acute{c}\else \'{c}\fi{}}}, \bibinfo {author} {\bibfnamefont
  {J.}~\bibnamefont {Goold}},\ and\ \bibinfo {author} {\bibfnamefont
  {S.}~\bibnamefont {Pappalardi}},\ }\bibfield  {title} {\bibinfo {title}
  {Extensive multipartite entanglement from su(2) quantum many-body scars},\
  }\href {https://doi.org/10.1103/PhysRevLett.129.020601} {\bibfield  {journal}
  {\bibinfo  {journal} {Phys. Rev. Lett.}\ }\textbf {\bibinfo {volume} {129}},\
  \bibinfo {pages} {020601} (\bibinfo {year} {2022})}\BibitemShut {NoStop}%
\bibitem [{\citenamefont {Dooley}\ \emph {et~al.}(2022)\citenamefont {Dooley},
  \citenamefont {Pappalardi},\ and\ \citenamefont {Goold}}]{Dooley22}%
  \BibitemOpen
  \bibfield  {author} {\bibinfo {author} {\bibfnamefont {S.}~\bibnamefont
  {Dooley}}, \bibinfo {author} {\bibfnamefont {S.}~\bibnamefont {Pappalardi}},\
  and\ \bibinfo {author} {\bibfnamefont {J.}~\bibnamefont {Goold}},\
  }\href@noop {} {\bibinfo {title} {Entanglement enhanced metrology with
  quantum many-body scars}} (\bibinfo {year} {2022}),\ \Eprint
  {https://arxiv.org/abs/arXiv:2207.13521} {arXiv:2207.13521} \BibitemShut
  {NoStop}%
\bibitem [{\citenamefont {Moudgalya}\ \emph
  {et~al.}(2020{\natexlab{b}})\citenamefont {Moudgalya}, \citenamefont
  {O'Brien}, \citenamefont {Bernevig}, \citenamefont {Fendley},\ and\
  \citenamefont {Regnault}}]{scar-mps}%
  \BibitemOpen
  \bibfield  {author} {\bibinfo {author} {\bibfnamefont {S.}~\bibnamefont
  {Moudgalya}}, \bibinfo {author} {\bibfnamefont {E.}~\bibnamefont {O'Brien}},
  \bibinfo {author} {\bibfnamefont {B.~A.}\ \bibnamefont {Bernevig}}, \bibinfo
  {author} {\bibfnamefont {P.}~\bibnamefont {Fendley}},\ and\ \bibinfo {author}
  {\bibfnamefont {N.}~\bibnamefont {Regnault}},\ }\bibfield  {title} {\bibinfo
  {title} {Large classes of quantum scarred hamiltonians from matrix product
  states},\ }\href {https://doi.org/10.1103/PhysRevB.102.085120} {\bibfield
  {journal} {\bibinfo  {journal} {Phys. Rev. B}\ }\textbf {\bibinfo {volume}
  {102}},\ \bibinfo {pages} {085120} (\bibinfo {year}
  {2020}{\natexlab{b}})}\BibitemShut {NoStop}%
\bibitem [{\citenamefont {Ren}\ \emph {et~al.}(2022)\citenamefont {Ren},
  \citenamefont {Liang},\ and\ \citenamefont {Fang}}]{Ren22}%
  \BibitemOpen
  \bibfield  {author} {\bibinfo {author} {\bibfnamefont {J.}~\bibnamefont
  {Ren}}, \bibinfo {author} {\bibfnamefont {C.}~\bibnamefont {Liang}},\ and\
  \bibinfo {author} {\bibfnamefont {C.}~\bibnamefont {Fang}},\ }\bibfield
  {title} {\bibinfo {title} {Deformed symmetry structures and quantum many-body
  scar subspaces},\ }\href {https://doi.org/10.1103/PhysRevResearch.4.013155}
  {\bibfield  {journal} {\bibinfo  {journal} {Phys. Rev. Research}\ }\textbf
  {\bibinfo {volume} {4}},\ \bibinfo {pages} {013155} (\bibinfo {year}
  {2022})}\BibitemShut {NoStop}%
\bibitem [{\citenamefont {Shiraishi}\ and\ \citenamefont
  {Mori}(2017)}]{embed2}%
  \BibitemOpen
  \bibfield  {author} {\bibinfo {author} {\bibfnamefont {N.}~\bibnamefont
  {Shiraishi}}\ and\ \bibinfo {author} {\bibfnamefont {T.}~\bibnamefont
  {Mori}},\ }\bibfield  {title} {\bibinfo {title} {Systematic construction of
  counterexamples to the eigenstate thermalization hypothesis},\ }\href
  {https://doi.org/10.1103/PhysRevLett.119.030601} {\bibfield  {journal}
  {\bibinfo  {journal} {Phys. Rev. Lett.}\ }\textbf {\bibinfo {volume} {119}},\
  \bibinfo {pages} {030601} (\bibinfo {year} {2017})}\BibitemShut {NoStop}%
\bibitem [{\citenamefont {Shiraishi}(2019)}]{embed1}%
  \BibitemOpen
  \bibfield  {author} {\bibinfo {author} {\bibfnamefont {N.}~\bibnamefont
  {Shiraishi}},\ }\bibfield  {title} {\bibinfo {title} {Connection between
  quantum-many-body scars and the affleck--kennedy--lieb--tasaki model from the
  viewpoint of embedded hamiltonians},\ }\href
  {https://iopscience.iop.org/article/10.1088/1742-5468/ab342e/meta} {\bibfield
   {journal} {\bibinfo  {journal} {Journal of Statistical Mechanics: Theory and
  Experiment}\ }\textbf {\bibinfo {volume} {2019}},\ \bibinfo {pages} {083103}
  (\bibinfo {year} {2019})}\BibitemShut {NoStop}%
\bibitem [{\citenamefont {Choi}\ \emph {et~al.}(2019)\citenamefont {Choi},
  \citenamefont {Turner}, \citenamefont {Pichler}, \citenamefont {Ho},
  \citenamefont {Michailidis}, \citenamefont {Papi\ifmmode~\acute{c}\else
  \'{c}\fi{}}, \citenamefont {Serbyn}, \citenamefont {Lukin},\ and\
  \citenamefont {Abanin}}]{ChoiTurner}%
  \BibitemOpen
  \bibfield  {author} {\bibinfo {author} {\bibfnamefont {S.}~\bibnamefont
  {Choi}}, \bibinfo {author} {\bibfnamefont {C.~J.}\ \bibnamefont {Turner}},
  \bibinfo {author} {\bibfnamefont {H.}~\bibnamefont {Pichler}}, \bibinfo
  {author} {\bibfnamefont {W.~W.}\ \bibnamefont {Ho}}, \bibinfo {author}
  {\bibfnamefont {A.~A.}\ \bibnamefont {Michailidis}}, \bibinfo {author}
  {\bibfnamefont {Z.}~\bibnamefont {Papi\ifmmode~\acute{c}\else \'{c}\fi{}}},
  \bibinfo {author} {\bibfnamefont {M.}~\bibnamefont {Serbyn}}, \bibinfo
  {author} {\bibfnamefont {M.~D.}\ \bibnamefont {Lukin}},\ and\ \bibinfo
  {author} {\bibfnamefont {D.~A.}\ \bibnamefont {Abanin}},\ }\bibfield  {title}
  {\bibinfo {title} {Emergent su(2) dynamics and perfect quantum many-body
  scars},\ }\href {https://doi.org/10.1103/PhysRevLett.122.220603} {\bibfield
  {journal} {\bibinfo  {journal} {Phys. Rev. Lett.}\ }\textbf {\bibinfo
  {volume} {122}},\ \bibinfo {pages} {220603} (\bibinfo {year}
  {2019})}\BibitemShut {NoStop}%
\bibitem [{\citenamefont {Ok}\ \emph {et~al.}(2019)\citenamefont {Ok},
  \citenamefont {Choo}, \citenamefont {Mudry}, \citenamefont {Castelnovo},
  \citenamefont {Chamon},\ and\ \citenamefont {Neupert}}]{OkSeulgi}%
  \BibitemOpen
  \bibfield  {author} {\bibinfo {author} {\bibfnamefont {S.}~\bibnamefont
  {Ok}}, \bibinfo {author} {\bibfnamefont {K.}~\bibnamefont {Choo}}, \bibinfo
  {author} {\bibfnamefont {C.}~\bibnamefont {Mudry}}, \bibinfo {author}
  {\bibfnamefont {C.}~\bibnamefont {Castelnovo}}, \bibinfo {author}
  {\bibfnamefont {C.}~\bibnamefont {Chamon}},\ and\ \bibinfo {author}
  {\bibfnamefont {T.}~\bibnamefont {Neupert}},\ }\bibfield  {title} {\bibinfo
  {title} {Topological many-body scar states in dimensions one, two, and
  three},\ }\href {https://doi.org/10.1103/PhysRevResearch.1.033144} {\bibfield
   {journal} {\bibinfo  {journal} {Phys. Rev. Research}\ }\textbf {\bibinfo
  {volume} {1}},\ \bibinfo {pages} {033144} (\bibinfo {year}
  {2019})}\BibitemShut {NoStop}%
\bibitem [{\citenamefont {Srivatsa}\ \emph {et~al.}(2020)\citenamefont
  {Srivatsa}, \citenamefont {Wildeboer}, \citenamefont {Seidel},\ and\
  \citenamefont {Nielsen}}]{wildeboer_scar_2020}%
  \BibitemOpen
  \bibfield  {author} {\bibinfo {author} {\bibfnamefont {N.~S.}\ \bibnamefont
  {Srivatsa}}, \bibinfo {author} {\bibfnamefont {J.}~\bibnamefont {Wildeboer}},
  \bibinfo {author} {\bibfnamefont {A.}~\bibnamefont {Seidel}},\ and\ \bibinfo
  {author} {\bibfnamefont {A.~E.~B.}\ \bibnamefont {Nielsen}},\ }\bibfield
  {title} {\bibinfo {title} {Quantum many-body scars with chiral topological
  order in two dimensions and critical properties in one dimension},\ }\href
  {https://doi.org/10.1103/PhysRevB.102.235106} {\bibfield  {journal} {\bibinfo
   {journal} {Phys. Rev. B}\ }\textbf {\bibinfo {volume} {102}},\ \bibinfo
  {pages} {235106} (\bibinfo {year} {2020})}\BibitemShut {NoStop}%
\bibitem [{\citenamefont {Wildeboer}\ \emph {et~al.}(2021)\citenamefont
  {Wildeboer}, \citenamefont {Seidel}, \citenamefont {Srivatsa}, \citenamefont
  {Nielsen},\ and\ \citenamefont {Erten}}]{wildeboer_scar_2021}%
  \BibitemOpen
  \bibfield  {author} {\bibinfo {author} {\bibfnamefont {J.}~\bibnamefont
  {Wildeboer}}, \bibinfo {author} {\bibfnamefont {A.}~\bibnamefont {Seidel}},
  \bibinfo {author} {\bibfnamefont {N.~S.}\ \bibnamefont {Srivatsa}}, \bibinfo
  {author} {\bibfnamefont {A.~E.~B.}\ \bibnamefont {Nielsen}},\ and\ \bibinfo
  {author} {\bibfnamefont {O.}~\bibnamefont {Erten}},\ }\bibfield  {title}
  {\bibinfo {title} {Topological quantum many-body scars in quantum dimer
  models on the kagome lattice},\ }\href
  {https://doi.org/10.1103/PhysRevB.104.L121103} {\bibfield  {journal}
  {\bibinfo  {journal} {Phys. Rev. B}\ }\textbf {\bibinfo {volume} {104}},\
  \bibinfo {pages} {L121103} (\bibinfo {year} {2021})}\BibitemShut {NoStop}%
\bibitem [{\citenamefont {Lee}\ \emph {et~al.}(2020)\citenamefont {Lee},
  \citenamefont {Melendrez}, \citenamefont {Pal},\ and\ \citenamefont
  {Changlani}}]{Lee}%
  \BibitemOpen
  \bibfield  {author} {\bibinfo {author} {\bibfnamefont {K.}~\bibnamefont
  {Lee}}, \bibinfo {author} {\bibfnamefont {R.}~\bibnamefont {Melendrez}},
  \bibinfo {author} {\bibfnamefont {A.}~\bibnamefont {Pal}},\ and\ \bibinfo
  {author} {\bibfnamefont {H.~J.}\ \bibnamefont {Changlani}},\ }\bibfield
  {title} {\bibinfo {title} {Exact three-colored quantum scars from geometric
  frustration},\ }\href {https://doi.org/10.1103/PhysRevB.101.241111}
  {\bibfield  {journal} {\bibinfo  {journal} {Phys. Rev. B}\ }\textbf {\bibinfo
  {volume} {101}},\ \bibinfo {pages} {241111} (\bibinfo {year}
  {2020})}\BibitemShut {NoStop}%
\bibitem [{\citenamefont {Banerjee}\ and\ \citenamefont
  {Sen}(2021)}]{Banerjee}%
  \BibitemOpen
  \bibfield  {author} {\bibinfo {author} {\bibfnamefont {D.}~\bibnamefont
  {Banerjee}}\ and\ \bibinfo {author} {\bibfnamefont {A.}~\bibnamefont {Sen}},\
  }\bibfield  {title} {\bibinfo {title} {Quantum scars from zero modes in an
  abelian lattice gauge theory on ladders},\ }\href
  {https://doi.org/10.1103/PhysRevLett.126.220601} {\bibfield  {journal}
  {\bibinfo  {journal} {Phys. Rev. Lett.}\ }\textbf {\bibinfo {volume} {126}},\
  \bibinfo {pages} {220601} (\bibinfo {year} {2021})}\BibitemShut {NoStop}%
\bibitem [{\citenamefont {Biswas}\ \emph {et~al.}(2022)\citenamefont {Biswas},
  \citenamefont {Banerjee},\ and\ \citenamefont {Sen}}]{Biswas22}%
  \BibitemOpen
  \bibfield  {author} {\bibinfo {author} {\bibfnamefont {S.}~\bibnamefont
  {Biswas}}, \bibinfo {author} {\bibfnamefont {D.}~\bibnamefont {Banerjee}},\
  and\ \bibinfo {author} {\bibfnamefont {A.}~\bibnamefont {Sen}},\ }\bibfield
  {title} {\bibinfo {title} {{Scars from protected zero modes and beyond in
  $U(1)$ quantum link and quantum dimer models}},\ }\href
  {https://doi.org/10.21468/SciPostPhys.12.5.148} {\bibfield  {journal}
  {\bibinfo  {journal} {SciPost Phys.}\ }\textbf {\bibinfo {volume} {12}},\
  \bibinfo {pages} {148} (\bibinfo {year} {2022})}\BibitemShut {NoStop}%
\bibitem [{\citenamefont {Bluvstein}\ \emph {et~al.}(2021)\citenamefont
  {Bluvstein}, \citenamefont {Omran}, \citenamefont {Levine}, \citenamefont
  {Keesling}, \citenamefont {Semeghini}, \citenamefont {Ebadi}, \citenamefont
  {Wang}, \citenamefont {Michailidis}, \citenamefont {Maskara}, \citenamefont
  {Ho}, \citenamefont {Choi}, \citenamefont {Serbyn}, \citenamefont {Greiner},
  \citenamefont {Vuletić},\ and\ \citenamefont {Lukin}}]{bluvstein2021}%
  \BibitemOpen
  \bibfield  {author} {\bibinfo {author} {\bibfnamefont {D.}~\bibnamefont
  {Bluvstein}}, \bibinfo {author} {\bibfnamefont {A.}~\bibnamefont {Omran}},
  \bibinfo {author} {\bibfnamefont {H.}~\bibnamefont {Levine}}, \bibinfo
  {author} {\bibfnamefont {A.}~\bibnamefont {Keesling}}, \bibinfo {author}
  {\bibfnamefont {G.}~\bibnamefont {Semeghini}}, \bibinfo {author}
  {\bibfnamefont {S.}~\bibnamefont {Ebadi}}, \bibinfo {author} {\bibfnamefont
  {T.~T.}\ \bibnamefont {Wang}}, \bibinfo {author} {\bibfnamefont {A.~A.}\
  \bibnamefont {Michailidis}}, \bibinfo {author} {\bibfnamefont
  {N.}~\bibnamefont {Maskara}}, \bibinfo {author} {\bibfnamefont {W.~W.}\
  \bibnamefont {Ho}}, \bibinfo {author} {\bibfnamefont {S.}~\bibnamefont
  {Choi}}, \bibinfo {author} {\bibfnamefont {M.}~\bibnamefont {Serbyn}},
  \bibinfo {author} {\bibfnamefont {M.}~\bibnamefont {Greiner}}, \bibinfo
  {author} {\bibfnamefont {V.}~\bibnamefont {Vuletić}},\ and\ \bibinfo
  {author} {\bibfnamefont {M.~D.}\ \bibnamefont {Lukin}},\ }\bibfield  {title}
  {\bibinfo {title} {Controlling quantum many-body dynamics in driven rydberg
  atom arrays},\ }\href {https://doi.org/10.1126/science.abg2530} {\bibfield
  {journal} {\bibinfo  {journal} {Science}\ }\textbf {\bibinfo {volume}
  {371}},\ \bibinfo {pages} {1355} (\bibinfo {year} {2021})},\ \Eprint
  {https://arxiv.org/abs/https://www.science.org/doi/pdf/10.1126/science.abg2530}
  {https://www.science.org/doi/pdf/10.1126/science.abg2530} \BibitemShut
  {NoStop}%
\bibitem [{\citenamefont {Langlett}\ \emph {et~al.}(2022)\citenamefont
  {Langlett}, \citenamefont {Yang}, \citenamefont {Wildeboer}, \citenamefont
  {Gorshkov}, \citenamefont {Iadecola},\ and\ \citenamefont {Xu}}]{ourrainbow}%
  \BibitemOpen
  \bibfield  {author} {\bibinfo {author} {\bibfnamefont {C.~M.}\ \bibnamefont
  {Langlett}}, \bibinfo {author} {\bibfnamefont {Z.-C.}\ \bibnamefont {Yang}},
  \bibinfo {author} {\bibfnamefont {J.}~\bibnamefont {Wildeboer}}, \bibinfo
  {author} {\bibfnamefont {A.~V.}\ \bibnamefont {Gorshkov}}, \bibinfo {author}
  {\bibfnamefont {T.}~\bibnamefont {Iadecola}},\ and\ \bibinfo {author}
  {\bibfnamefont {S.}~\bibnamefont {Xu}},\ }\bibfield  {title} {\bibinfo
  {title} {Rainbow scars: From area to volume law},\ }\href
  {https://doi.org/10.1103/PhysRevB.105.L060301} {\bibfield  {journal}
  {\bibinfo  {journal} {Phys. Rev. B}\ }\textbf {\bibinfo {volume} {105}},\
  \bibinfo {pages} {L060301} (\bibinfo {year} {2022})}\BibitemShut {NoStop}%
\bibitem [{\citenamefont {Schindler}\ \emph {et~al.}(2022)\citenamefont
  {Schindler}, \citenamefont {Regnault},\ and\ \citenamefont
  {Bernevig}}]{schindler2022exact}%
  \BibitemOpen
  \bibfield  {author} {\bibinfo {author} {\bibfnamefont {F.}~\bibnamefont
  {Schindler}}, \bibinfo {author} {\bibfnamefont {N.}~\bibnamefont
  {Regnault}},\ and\ \bibinfo {author} {\bibfnamefont {B.~A.}\ \bibnamefont
  {Bernevig}},\ }\bibfield  {title} {\bibinfo {title} {Exact quantum scars in
  the chiral nonlinear luttinger liquid},\ }\href@noop {} {\bibfield  {journal}
  {\bibinfo  {journal} {Physical Review B}\ }\textbf {\bibinfo {volume}
  {105}},\ \bibinfo {pages} {035146} (\bibinfo {year} {2022})}\BibitemShut
  {NoStop}%
\bibitem [{\citenamefont {Srivatsa}\ \emph {et~al.}(2022)\citenamefont
  {Srivatsa}, \citenamefont {Yarloo}, \citenamefont {Moessner},\ and\
  \citenamefont {Nielsen}}]{srivatsa2022mobility}%
  \BibitemOpen
  \bibfield  {author} {\bibinfo {author} {\bibfnamefont {N.}~\bibnamefont
  {Srivatsa}}, \bibinfo {author} {\bibfnamefont {H.}~\bibnamefont {Yarloo}},
  \bibinfo {author} {\bibfnamefont {R.}~\bibnamefont {Moessner}},\ and\
  \bibinfo {author} {\bibfnamefont {A.~E.}\ \bibnamefont {Nielsen}},\
  }\bibfield  {title} {\bibinfo {title} {Mobility edges through inverted
  quantum many-body scarring},\ }\href@noop {} {\bibfield  {journal} {\bibinfo
  {journal} {arXiv preprint arXiv:2208.01054}\ } (\bibinfo {year}
  {2022})}\BibitemShut {NoStop}%
\bibitem [{\citenamefont {Ram{\'{\i}}rez}\ \emph {et~al.}(2014)\citenamefont
  {Ram{\'{\i}}rez}, \citenamefont {Rodr{\'{\i}}guez-Laguna},\ and\
  \citenamefont {Sierra}}]{Ram_rez_2014}%
  \BibitemOpen
  \bibfield  {author} {\bibinfo {author} {\bibfnamefont {G.}~\bibnamefont
  {Ram{\'{\i}}rez}}, \bibinfo {author} {\bibfnamefont {J.}~\bibnamefont
  {Rodr{\'{\i}}guez-Laguna}},\ and\ \bibinfo {author} {\bibfnamefont
  {G.}~\bibnamefont {Sierra}},\ }\bibfield  {title} {\bibinfo {title} {From
  conformal to volume law for the entanglement entropy in exponentially
  deformed critical spin 1/2 chains},\ }\href
  {https://doi.org/10.1088/1742-5468/2014/10/p10004} {\bibfield  {journal}
  {\bibinfo  {journal} {Journal of Statistical Mechanics: Theory and
  Experiment}\ }\textbf {\bibinfo {volume} {2014}},\ \bibinfo {pages} {P10004}
  (\bibinfo {year} {2014})}\BibitemShut {NoStop}%
\bibitem [{\citenamefont {Vitagliano}\ \emph {et~al.}(2010)\citenamefont
  {Vitagliano}, \citenamefont {Riera},\ and\ \citenamefont
  {Latorre}}]{Vitagliano_2010}%
  \BibitemOpen
  \bibfield  {author} {\bibinfo {author} {\bibfnamefont {G.}~\bibnamefont
  {Vitagliano}}, \bibinfo {author} {\bibfnamefont {A.}~\bibnamefont {Riera}},\
  and\ \bibinfo {author} {\bibfnamefont {J.~I.}\ \bibnamefont {Latorre}},\
  }\bibfield  {title} {\bibinfo {title} {Volume-law scaling for the
  entanglement entropy in spin-1/2 chains},\ }\href
  {https://doi.org/10.1088/1367-2630/12/11/113049} {\bibfield  {journal}
  {\bibinfo  {journal} {New Journal of Physics}\ }\textbf {\bibinfo {volume}
  {12}},\ \bibinfo {pages} {113049} (\bibinfo {year} {2010})}\BibitemShut
  {NoStop}%
\bibitem [{\citenamefont {Yang}(1989)}]{YangPRL}%
  \BibitemOpen
  \bibfield  {author} {\bibinfo {author} {\bibfnamefont {C.~N.}\ \bibnamefont
  {Yang}},\ }\bibfield  {title} {\bibinfo {title} {\ensuremath{\eta} pairing
  and off-diagonal long-range order in a hubbard model},\ }\href
  {https://doi.org/10.1103/PhysRevLett.63.2144} {\bibfield  {journal} {\bibinfo
   {journal} {Phys. Rev. Lett.}\ }\textbf {\bibinfo {volume} {63}},\ \bibinfo
  {pages} {2144} (\bibinfo {year} {1989})}\BibitemShut {NoStop}%
\bibitem [{\citenamefont {Li}(2020)}]{Li}%
  \BibitemOpen
  \bibfield  {author} {\bibinfo {author} {\bibfnamefont {K.}~\bibnamefont
  {Li}},\ }\bibfield  {title} {\bibinfo {title} {$\ensuremath{\eta}$-pairing in
  correlated fermion models with spin-orbit coupling},\ }\href
  {https://doi.org/10.1103/PhysRevB.102.165150} {\bibfield  {journal} {\bibinfo
   {journal} {Phys. Rev. B}\ }\textbf {\bibinfo {volume} {102}},\ \bibinfo
  {pages} {165150} (\bibinfo {year} {2020})}\BibitemShut {NoStop}%
\bibitem [{\citenamefont {Cottrell}\ \emph {et~al.}(2019)\citenamefont
  {Cottrell}, \citenamefont {Freivogel}, \citenamefont {Hofman},\ and\
  \citenamefont {Lokhande}}]{blackhole2}%
  \BibitemOpen
  \bibfield  {author} {\bibinfo {author} {\bibfnamefont {W.}~\bibnamefont
  {Cottrell}}, \bibinfo {author} {\bibfnamefont {B.}~\bibnamefont {Freivogel}},
  \bibinfo {author} {\bibfnamefont {D.~M.}\ \bibnamefont {Hofman}},\ and\
  \bibinfo {author} {\bibfnamefont {S.~F.}\ \bibnamefont {Lokhande}},\
  }\bibfield  {title} {\bibinfo {title} {How to build the thermofield double
  state},\ }\href {https://doi.org/10.1007/JHEP02(2019)058} {\bibfield
  {journal} {\bibinfo  {journal} {Journal of High Energy Physics}\ }\textbf
  {\bibinfo {volume} {2019}},\ \bibinfo {pages} {1} (\bibinfo {year}
  {2019})}\BibitemShut {NoStop}%
\bibitem [{\citenamefont {Hartman}\ and\ \citenamefont
  {Maldacena}(2013)}]{hartman2013time}%
  \BibitemOpen
  \bibfield  {author} {\bibinfo {author} {\bibfnamefont {T.}~\bibnamefont
  {Hartman}}\ and\ \bibinfo {author} {\bibfnamefont {J.}~\bibnamefont
  {Maldacena}},\ }\bibfield  {title} {\bibinfo {title} {Time evolution of
  entanglement entropy from black hole interiors},\ }\href
  {https://doi.org/10.1007/JHEP05(2013)014} {\bibfield  {journal} {\bibinfo
  {journal} {Journal of High Energy Physics}\ }\textbf {\bibinfo {volume}
  {2013}},\ \bibinfo {pages} {1} (\bibinfo {year} {2013})}\BibitemShut
  {NoStop}%
\bibitem [{\citenamefont {Papadodimas}\ and\ \citenamefont
  {Raju}(2015)}]{papadodimas2015local}%
  \BibitemOpen
  \bibfield  {author} {\bibinfo {author} {\bibfnamefont {K.}~\bibnamefont
  {Papadodimas}}\ and\ \bibinfo {author} {\bibfnamefont {S.}~\bibnamefont
  {Raju}},\ }\bibfield  {title} {\bibinfo {title} {Local operators in the
  eternal black hole},\ }\href {https://doi.org/10.1103/PhysRevLett.115.211601}
  {\bibfield  {journal} {\bibinfo  {journal} {Phys. Rev. Lett.}\ }\textbf
  {\bibinfo {volume} {115}},\ \bibinfo {pages} {211601} (\bibinfo {year}
  {2015})}\BibitemShut {NoStop}%
\bibitem [{\citenamefont {Schuster}\ \emph {et~al.}(2021)\citenamefont
  {Schuster}, \citenamefont {Kobrin}, \citenamefont {Gao}, \citenamefont
  {Cong}, \citenamefont {Khabiboulline}, \citenamefont {Linke}, \citenamefont
  {Lukin}, \citenamefont {Monroe}, \citenamefont {Yoshida},\ and\ \citenamefont
  {Yao}}]{schuster2021many}%
  \BibitemOpen
  \bibfield  {author} {\bibinfo {author} {\bibfnamefont {T.}~\bibnamefont
  {Schuster}}, \bibinfo {author} {\bibfnamefont {B.}~\bibnamefont {Kobrin}},
  \bibinfo {author} {\bibfnamefont {P.}~\bibnamefont {Gao}}, \bibinfo {author}
  {\bibfnamefont {I.}~\bibnamefont {Cong}}, \bibinfo {author} {\bibfnamefont
  {E.~T.}\ \bibnamefont {Khabiboulline}}, \bibinfo {author} {\bibfnamefont
  {N.~M.}\ \bibnamefont {Linke}}, \bibinfo {author} {\bibfnamefont {M.~D.}\
  \bibnamefont {Lukin}}, \bibinfo {author} {\bibfnamefont {C.}~\bibnamefont
  {Monroe}}, \bibinfo {author} {\bibfnamefont {B.}~\bibnamefont {Yoshida}},\
  and\ \bibinfo {author} {\bibfnamefont {N.~Y.}\ \bibnamefont {Yao}},\
  }\bibfield  {title} {\bibinfo {title} {Many-body quantum teleportation via
  operator spreading in the traversable wormhole protocol},\ }\href
  {https://arxiv.org/abs/2102.00010} {\bibfield  {journal} {\bibinfo  {journal}
  {arXiv preprint arXiv:2102.00010}\ } (\bibinfo {year} {2021})}\BibitemShut
  {NoStop}%
\bibitem [{\citenamefont {Nezami}\ \emph {et~al.}(2021)\citenamefont {Nezami},
  \citenamefont {Lin}, \citenamefont {Brown}, \citenamefont {Gharibyan},
  \citenamefont {Leichenauer}, \citenamefont {Salton}, \citenamefont
  {Susskind}, \citenamefont {Swingle},\ and\ \citenamefont
  {Walter}}]{nezami2021quantum}%
  \BibitemOpen
  \bibfield  {author} {\bibinfo {author} {\bibfnamefont {S.}~\bibnamefont
  {Nezami}}, \bibinfo {author} {\bibfnamefont {H.~W.}\ \bibnamefont {Lin}},
  \bibinfo {author} {\bibfnamefont {A.~R.}\ \bibnamefont {Brown}}, \bibinfo
  {author} {\bibfnamefont {H.}~\bibnamefont {Gharibyan}}, \bibinfo {author}
  {\bibfnamefont {S.}~\bibnamefont {Leichenauer}}, \bibinfo {author}
  {\bibfnamefont {G.}~\bibnamefont {Salton}}, \bibinfo {author} {\bibfnamefont
  {L.}~\bibnamefont {Susskind}}, \bibinfo {author} {\bibfnamefont
  {B.}~\bibnamefont {Swingle}},\ and\ \bibinfo {author} {\bibfnamefont
  {M.}~\bibnamefont {Walter}},\ }\bibfield  {title} {\bibinfo {title} {Quantum
  gravity in the lab: teleportation by size and traversable wormholes, part
  ii},\ }\href {https://arxiv.org/abs/2102.01064} {\bibfield  {journal}
  {\bibinfo  {journal} {arXiv preprint arXiv:2102.01064}\ } (\bibinfo {year}
  {2021})}\BibitemShut {NoStop}%
\bibitem [{\citenamefont {Lesanovsky}(2011)}]{Lesanovsky2011many}%
  \BibitemOpen
  \bibfield  {author} {\bibinfo {author} {\bibfnamefont {I.}~\bibnamefont
  {Lesanovsky}},\ }\bibfield  {title} {\bibinfo {title} {Many-body spin
  interactions and the ground state of a dense rydberg lattice gas},\ }\href
  {https://doi.org/10.1103/PhysRevLett.106.025301} {\bibfield  {journal}
  {\bibinfo  {journal} {Phys. Rev. Lett.}\ }\textbf {\bibinfo {volume} {106}},\
  \bibinfo {pages} {025301} (\bibinfo {year} {2011})}\BibitemShut {NoStop}%
\bibitem [{\citenamefont {Khemani}\ \emph {et~al.}(2019)\citenamefont
  {Khemani}, \citenamefont {Laumann},\ and\ \citenamefont
  {Chandran}}]{scartheo8}%
  \BibitemOpen
  \bibfield  {author} {\bibinfo {author} {\bibfnamefont {V.}~\bibnamefont
  {Khemani}}, \bibinfo {author} {\bibfnamefont {C.~R.}\ \bibnamefont
  {Laumann}},\ and\ \bibinfo {author} {\bibfnamefont {A.}~\bibnamefont
  {Chandran}},\ }\bibfield  {title} {\bibinfo {title} {Signatures of
  integrability in the dynamics of rydberg-blockaded chains},\ }\href
  {https://doi.org/10.1103/PhysRevB.99.161101} {\bibfield  {journal} {\bibinfo
  {journal} {Phys. Rev. B}\ }\textbf {\bibinfo {volume} {99}},\ \bibinfo
  {pages} {161101} (\bibinfo {year} {2019})}\BibitemShut {NoStop}%
\bibitem [{\citenamefont {Lin}\ and\ \citenamefont
  {Motrunich}(2019)}]{scartheo9}%
  \BibitemOpen
  \bibfield  {author} {\bibinfo {author} {\bibfnamefont {C.-J.}\ \bibnamefont
  {Lin}}\ and\ \bibinfo {author} {\bibfnamefont {O.~I.}\ \bibnamefont
  {Motrunich}},\ }\bibfield  {title} {\bibinfo {title} {Exact quantum many-body
  scar states in the rydberg-blockaded atom chain},\ }\href
  {https://doi.org/10.1103/PhysRevLett.122.173401} {\bibfield  {journal}
  {\bibinfo  {journal} {Phys. Rev. Lett.}\ }\textbf {\bibinfo {volume} {122}},\
  \bibinfo {pages} {173401} (\bibinfo {year} {2019})}\BibitemShut {NoStop}%
\bibitem [{\citenamefont {Hirsch}(1989)}]{HIRSCH1989326}%
  \BibitemOpen
  \bibfield  {author} {\bibinfo {author} {\bibfnamefont {J.}~\bibnamefont
  {Hirsch}},\ }\bibfield  {title} {\bibinfo {title} {Bond-charge repulsion and
  hole superconductivity},\ }\href
  {https://doi.org/https://doi.org/10.1016/0921-4534(89)90225-6} {\bibfield
  {journal} {\bibinfo  {journal} {Physica C: Superconductivity and its
  Applications}\ }\textbf {\bibinfo {volume} {158}},\ \bibinfo {pages} {326}
  (\bibinfo {year} {1989})}\BibitemShut {NoStop}%
\bibitem [{\citenamefont {Hirsch}\ and\ \citenamefont
  {Marsiglio}(1989)}]{Marsiglio}%
  \BibitemOpen
  \bibfield  {author} {\bibinfo {author} {\bibfnamefont {J.~E.}\ \bibnamefont
  {Hirsch}}\ and\ \bibinfo {author} {\bibfnamefont {F.}~\bibnamefont
  {Marsiglio}},\ }\bibfield  {title} {\bibinfo {title} {Superconducting state
  in an oxygen hole metal},\ }\href {https://doi.org/10.1103/PhysRevB.39.11515}
  {\bibfield  {journal} {\bibinfo  {journal} {Phys. Rev. B}\ }\textbf {\bibinfo
  {volume} {39}},\ \bibinfo {pages} {11515} (\bibinfo {year}
  {1989})}\BibitemShut {NoStop}%
\bibitem [{\citenamefont {Duan}(2007)}]{duan2007general}%
  \BibitemOpen
  \bibfield  {author} {\bibinfo {author} {\bibfnamefont {L.-M.}\ \bibnamefont
  {Duan}},\ }\bibfield  {title} {\bibinfo {title} {General hubbard model for
  strongly interacting fermions in an optical lattice and its phase
  detection},\ }\href {https://doi.org/10.1209/0295-5075/81/20001} {\bibfield
  {journal} {\bibinfo  {journal} {EPL (Europhysics Letters)}\ }\textbf
  {\bibinfo {volume} {81}},\ \bibinfo {pages} {20001} (\bibinfo {year}
  {2007})}\BibitemShut {NoStop}%
\bibitem [{\citenamefont {Bhattacharyya}\ and\ \citenamefont
  {Sil}(1999)}]{bhattacharyya1999hubbard}%
  \BibitemOpen
  \bibfield  {author} {\bibinfo {author} {\bibfnamefont {B.}~\bibnamefont
  {Bhattacharyya}}\ and\ \bibinfo {author} {\bibfnamefont {S.}~\bibnamefont
  {Sil}},\ }\bibfield  {title} {\bibinfo {title} {The hubbard model with
  bond-charge interaction on a triangular lattice: a renormalization group
  study},\ }\href {https://doi.org/10.1088/0953-8984/11/17/309} {\bibfield
  {journal} {\bibinfo  {journal} {Journal of Physics: Condensed Matter}\
  }\textbf {\bibinfo {volume} {11}},\ \bibinfo {pages} {3513} (\bibinfo {year}
  {1999})}\BibitemShut {NoStop}%
\bibitem [{\citenamefont {Trefzger}\ \emph {et~al.}(2009)\citenamefont
  {Trefzger}, \citenamefont {Menotti},\ and\ \citenamefont
  {Lewenstein}}]{trefzger2009pair}%
  \BibitemOpen
  \bibfield  {author} {\bibinfo {author} {\bibfnamefont {C.}~\bibnamefont
  {Trefzger}}, \bibinfo {author} {\bibfnamefont {C.}~\bibnamefont {Menotti}},\
  and\ \bibinfo {author} {\bibfnamefont {M.}~\bibnamefont {Lewenstein}},\
  }\bibfield  {title} {\bibinfo {title} {Pair-supersolid phase in a bilayer
  system of dipolar lattice bosons},\ }\href
  {https://doi.org/10.1103/PhysRevLett.103.035304} {\bibfield  {journal}
  {\bibinfo  {journal} {Phys. Rev. Lett.}\ }\textbf {\bibinfo {volume} {103}},\
  \bibinfo {pages} {035304} (\bibinfo {year} {2009})}\BibitemShut {NoStop}%
\bibitem [{\citenamefont {Sun}\ \emph {et~al.}(2015)\citenamefont {Sun},
  \citenamefont {Wen}, \citenamefont {Liu}, \citenamefont
  {Juzeli\ifmmode~\bar{u}\else \={u}\fi{}nas},\ and\ \citenamefont
  {Ji}}]{sun2015tunneling}%
  \BibitemOpen
  \bibfield  {author} {\bibinfo {author} {\bibfnamefont {Q.}~\bibnamefont
  {Sun}}, \bibinfo {author} {\bibfnamefont {L.}~\bibnamefont {Wen}}, \bibinfo
  {author} {\bibfnamefont {W.-M.}\ \bibnamefont {Liu}}, \bibinfo {author}
  {\bibfnamefont {G.}~\bibnamefont {Juzeli\ifmmode~\bar{u}\else
  \={u}\fi{}nas}},\ and\ \bibinfo {author} {\bibfnamefont {A.-C.}\ \bibnamefont
  {Ji}},\ }\bibfield  {title} {\bibinfo {title} {Tunneling-assisted spin-orbit
  coupling in bilayer bose-einstein condensates},\ }\href
  {https://doi.org/10.1103/PhysRevA.91.033619} {\bibfield  {journal} {\bibinfo
  {journal} {Phys. Rev. A}\ }\textbf {\bibinfo {volume} {91}},\ \bibinfo
  {pages} {033619} (\bibinfo {year} {2015})}\BibitemShut {NoStop}%
\bibitem [{\citenamefont {Sowi\ifmmode~\acute{n}\else \'{n}\fi{}ski}\ \emph
  {et~al.}(2012)\citenamefont {Sowi\ifmmode~\acute{n}\else \'{n}\fi{}ski},
  \citenamefont {Dutta}, \citenamefont {Hauke}, \citenamefont {Tagliacozzo},\
  and\ \citenamefont {Lewenstein}}]{sowi2012dipolar}%
  \BibitemOpen
  \bibfield  {author} {\bibinfo {author} {\bibfnamefont {T.}~\bibnamefont
  {Sowi\ifmmode~\acute{n}\else \'{n}\fi{}ski}}, \bibinfo {author}
  {\bibfnamefont {O.}~\bibnamefont {Dutta}}, \bibinfo {author} {\bibfnamefont
  {P.}~\bibnamefont {Hauke}}, \bibinfo {author} {\bibfnamefont
  {L.}~\bibnamefont {Tagliacozzo}},\ and\ \bibinfo {author} {\bibfnamefont
  {M.}~\bibnamefont {Lewenstein}},\ }\bibfield  {title} {\bibinfo {title}
  {Dipolar molecules in optical lattices},\ }\href
  {https://doi.org/10.1103/PhysRevLett.108.115301} {\bibfield  {journal}
  {\bibinfo  {journal} {Phys. Rev. Lett.}\ }\textbf {\bibinfo {volume} {108}},\
  \bibinfo {pages} {115301} (\bibinfo {year} {2012})}\BibitemShut {NoStop}%
\bibitem [{\citenamefont {Wang}(2007)}]{wang2007quantum}%
  \BibitemOpen
  \bibfield  {author} {\bibinfo {author} {\bibfnamefont {D.-W.}\ \bibnamefont
  {Wang}},\ }\bibfield  {title} {\bibinfo {title} {Quantum phase transitions of
  polar molecules in bilayer systems},\ }\href
  {https://doi.org/10.1103/PhysRevLett.98.060403} {\bibfield  {journal}
  {\bibinfo  {journal} {Phys. Rev. Lett.}\ }\textbf {\bibinfo {volume} {98}},\
  \bibinfo {pages} {060403} (\bibinfo {year} {2007})}\BibitemShut {NoStop}%
\bibitem [{\citenamefont {Young}\ \emph {et~al.}(2022)\citenamefont {Young},
  \citenamefont {Eckner}, \citenamefont {Schine}, \citenamefont {Childs},\ and\
  \citenamefont {Kaufman}}]{young22}%
  \BibitemOpen
  \bibfield  {author} {\bibinfo {author} {\bibfnamefont {A.~W.}\ \bibnamefont
  {Young}}, \bibinfo {author} {\bibfnamefont {W.~J.}\ \bibnamefont {Eckner}},
  \bibinfo {author} {\bibfnamefont {N.}~\bibnamefont {Schine}}, \bibinfo
  {author} {\bibfnamefont {A.~M.}\ \bibnamefont {Childs}},\ and\ \bibinfo
  {author} {\bibfnamefont {A.~M.}\ \bibnamefont {Kaufman}},\ }\bibfield
  {title} {\bibinfo {title} {Tweezer-programmable 2d quantum walks in a
  hubbard-regime lattice},\ }\href@noop {} {\bibfield  {journal} {\bibinfo
  {journal} {arXiv:2202.01204}\ } (\bibinfo {year} {2022})}\BibitemShut
  {NoStop}%
\bibitem [{\citenamefont {Porras}\ and\ \citenamefont
  {Cirac}(2004)}]{porras04}%
  \BibitemOpen
  \bibfield  {author} {\bibinfo {author} {\bibfnamefont {D.}~\bibnamefont
  {Porras}}\ and\ \bibinfo {author} {\bibfnamefont {J.~I.}\ \bibnamefont
  {Cirac}},\ }\bibfield  {title} {\bibinfo {title} {Effective quantum spin
  systems with trapped ions},\ }\href@noop {} {\bibfield  {journal} {\bibinfo
  {journal} {Phys. Rev. Lett.}\ }\textbf {\bibinfo {volume} {92}},\ \bibinfo
  {pages} {207901} (\bibinfo {year} {2004})}\BibitemShut {NoStop}%
\bibitem [{\citenamefont {Deng}\ \emph {et~al.}(2008)\citenamefont {Deng},
  \citenamefont {Porras},\ and\ \citenamefont {Cirac}}]{deng08}%
  \BibitemOpen
  \bibfield  {author} {\bibinfo {author} {\bibfnamefont {X.~L.}\ \bibnamefont
  {Deng}}, \bibinfo {author} {\bibfnamefont {D.}~\bibnamefont {Porras}},\ and\
  \bibinfo {author} {\bibfnamefont {J.~I.}\ \bibnamefont {Cirac}},\ }\bibfield
  {title} {\bibinfo {title} {Quantum phases of interacting phonons in ion
  traps},\ }\href@noop {} {\bibfield  {journal} {\bibinfo  {journal} {Phys.
  Rev. A}\ }\textbf {\bibinfo {volume} {77}},\ \bibinfo {pages} {033403}
  (\bibinfo {year} {2008})}\BibitemShut {NoStop}%
\bibitem [{\citenamefont {Serafini}\ \emph {et~al.}(2009)\citenamefont
  {Serafini}, \citenamefont {Retzker},\ and\ \citenamefont
  {Plenio}}]{serafini09}%
  \BibitemOpen
  \bibfield  {author} {\bibinfo {author} {\bibfnamefont {A.}~\bibnamefont
  {Serafini}}, \bibinfo {author} {\bibfnamefont {A.}~\bibnamefont {Retzker}},\
  and\ \bibinfo {author} {\bibfnamefont {M.~B.}\ \bibnamefont {Plenio}},\
  }\bibfield  {title} {\bibinfo {title} {Manipulating the quantum information
  of the radial modes of trapped ions: linear phononics, entanglement
  generation, quantum state transmission and non-locality tests},\ }\href@noop
  {} {\bibfield  {journal} {\bibinfo  {journal} {New J. Phys.}\ }\textbf
  {\bibinfo {volume} {11}},\ \bibinfo {pages} {023007} (\bibinfo {year}
  {2009})}\BibitemShut {NoStop}%
\bibitem [{\citenamefont {Debnath}\ \emph {et~al.}(2018)\citenamefont
  {Debnath}, \citenamefont {Linke}, \citenamefont {Wang}, \citenamefont
  {Figgatt}, \citenamefont {Landsman}, \citenamefont {Duan},\ and\
  \citenamefont {Monroe}}]{debnath18}%
  \BibitemOpen
  \bibfield  {author} {\bibinfo {author} {\bibfnamefont {S.}~\bibnamefont
  {Debnath}}, \bibinfo {author} {\bibfnamefont {N.~M.}\ \bibnamefont {Linke}},
  \bibinfo {author} {\bibfnamefont {S.-T.}\ \bibnamefont {Wang}}, \bibinfo
  {author} {\bibfnamefont {C.}~\bibnamefont {Figgatt}}, \bibinfo {author}
  {\bibfnamefont {K.~A.}\ \bibnamefont {Landsman}}, \bibinfo {author}
  {\bibfnamefont {L.-M.}\ \bibnamefont {Duan}},\ and\ \bibinfo {author}
  {\bibfnamefont {C.}~\bibnamefont {Monroe}},\ }\bibfield  {title} {\bibinfo
  {title} {Observation of hopping and blockade of bosons in a trapped ion spin
  chain},\ }\href@noop {} {\bibfield  {journal} {\bibinfo  {journal} {Phys.
  Rev. Lett.}\ }\textbf {\bibinfo {volume} {120}},\ \bibinfo {pages} {073001}
  (\bibinfo {year} {2018})}\BibitemShut {NoStop}%
\bibitem [{\citenamefont {Katz}\ and\ \citenamefont {Monroe}(2022)}]{katz22c}%
  \BibitemOpen
  \bibfield  {author} {\bibinfo {author} {\bibfnamefont {O.}~\bibnamefont
  {Katz}}\ and\ \bibinfo {author} {\bibfnamefont {C.}~\bibnamefont {Monroe}},\
  }\bibfield  {title} {\bibinfo {title} {Programmable quantum simulations of
  bosonic systems with trapped ions},\ }\href@noop {} {\bibfield  {journal}
  {\bibinfo  {journal} {arXiv:2207.13653}\ } (\bibinfo {year}
  {2022})}\BibitemShut {NoStop}%
\bibitem [{\citenamefont {Bhattacharya}\ and\ \citenamefont
  {Meystre}(2008)}]{bhattacharya08}%
  \BibitemOpen
  \bibfield  {author} {\bibinfo {author} {\bibfnamefont {M.}~\bibnamefont
  {Bhattacharya}}\ and\ \bibinfo {author} {\bibfnamefont {P.}~\bibnamefont
  {Meystre}},\ }\bibfield  {title} {\bibinfo {title} {Multiple membrane cavity
  optomechanics},\ }\href@noop {} {\bibfield  {journal} {\bibinfo  {journal}
  {Phys. Rev. A}\ }\textbf {\bibinfo {volume} {78}},\ \bibinfo {pages} {041801}
  (\bibinfo {year} {2008})}\BibitemShut {NoStop}%
\bibitem [{\citenamefont {Chang}\ \emph {et~al.}(2011)\citenamefont {Chang},
  \citenamefont {Safavi-Naeini}, \citenamefont {Hafezi},\ and\ \citenamefont
  {Painter}}]{chang11b}%
  \BibitemOpen
  \bibfield  {author} {\bibinfo {author} {\bibfnamefont {D.~E.}\ \bibnamefont
  {Chang}}, \bibinfo {author} {\bibfnamefont {A.~H.}\ \bibnamefont
  {Safavi-Naeini}}, \bibinfo {author} {\bibfnamefont {M.}~\bibnamefont
  {Hafezi}},\ and\ \bibinfo {author} {\bibfnamefont {O.}~\bibnamefont
  {Painter}},\ }\bibfield  {title} {\bibinfo {title} {Slowing and stopping
  light using an optomechanical crystal array},\ }\href@noop {} {\bibfield
  {journal} {\bibinfo  {journal} {New J. Phys.}\ }\textbf {\bibinfo {volume}
  {13}},\ \bibinfo {pages} {023003} (\bibinfo {year} {2011})}\BibitemShut
  {NoStop}%
\bibitem [{\citenamefont {Raeisi}\ and\ \citenamefont
  {Marquardt}(2020)}]{raeisi20}%
  \BibitemOpen
  \bibfield  {author} {\bibinfo {author} {\bibfnamefont {S.}~\bibnamefont
  {Raeisi}}\ and\ \bibinfo {author} {\bibfnamefont {F.}~\bibnamefont
  {Marquardt}},\ }\bibfield  {title} {\bibinfo {title} {Quench dynamics in
  one-dimensional optomechanical arrays},\ }\href@noop {} {\bibfield  {journal}
  {\bibinfo  {journal} {Phys. Rev. A}\ }\textbf {\bibinfo {volume} {101}},\
  \bibinfo {pages} {023814} (\bibinfo {year} {2020})}\BibitemShut {NoStop}%
\bibitem [{\citenamefont {Mehta}\ \emph {et~al.}(2022)\citenamefont {Mehta},
  \citenamefont {Kuzmin}, \citenamefont {Ciuti},\ and\ \citenamefont
  {Manucharyan}}]{mehta22b}%
  \BibitemOpen
  \bibfield  {author} {\bibinfo {author} {\bibfnamefont {N.}~\bibnamefont
  {Mehta}}, \bibinfo {author} {\bibfnamefont {R.}~\bibnamefont {Kuzmin}},
  \bibinfo {author} {\bibfnamefont {C.}~\bibnamefont {Ciuti}},\ and\ \bibinfo
  {author} {\bibfnamefont {V.~E.}\ \bibnamefont {Manucharyan}},\ }\bibfield
  {title} {\bibinfo {title} {Down-conversion of a single photon as a probe of
  many-body localization},\ }\href@noop {} {\bibfield  {journal} {\bibinfo
  {journal} {arXiv:2203.17186}\ } (\bibinfo {year} {2022})}\BibitemShut
  {NoStop}%
\bibitem [{\citenamefont {Hartmann}\ \emph {et~al.}(2008)\citenamefont
  {Hartmann}, \citenamefont {Brandao},\ and\ \citenamefont
  {Plenio}}]{hartmann09}%
  \BibitemOpen
  \bibfield  {author} {\bibinfo {author} {\bibfnamefont {M.~J.}\ \bibnamefont
  {Hartmann}}, \bibinfo {author} {\bibfnamefont {F.~G. S.~L.}\ \bibnamefont
  {Brandao}},\ and\ \bibinfo {author} {\bibfnamefont {M.~B.}\ \bibnamefont
  {Plenio}},\ }\bibfield  {title} {\bibinfo {title} {Quantum many-body
  phenomena in coupled cavity arrays},\ }\href@noop {} {\bibfield  {journal}
  {\bibinfo  {journal} {Laser \& Photon. Rev.}\ }\textbf {\bibinfo {volume}
  {2}},\ \bibinfo {pages} {527} (\bibinfo {year} {2008})}\BibitemShut {NoStop}%
\bibitem [{\citenamefont {Tomadin}\ and\ \citenamefont
  {Fazio}(2010)}]{tomadin10}%
  \BibitemOpen
  \bibfield  {author} {\bibinfo {author} {\bibfnamefont {A.}~\bibnamefont
  {Tomadin}}\ and\ \bibinfo {author} {\bibfnamefont {R.}~\bibnamefont
  {Fazio}},\ }\bibfield  {title} {\bibinfo {title} {Many-body phenomena in
  qed-cavity arrays},\ }\href@noop {} {\bibfield  {journal} {\bibinfo
  {journal} {J. Opt. Soc. Am. B}\ }\textbf {\bibinfo {volume} {27}},\ \bibinfo
  {pages} {A130} (\bibinfo {year} {2010})}\BibitemShut {NoStop}%
\bibitem [{\citenamefont {Peropadre}\ \emph {et~al.}(2016)\citenamefont
  {Peropadre}, \citenamefont {Guerreschi}, \citenamefont {Huh},\ and\
  \citenamefont {Aspuru-Guzik}}]{peropadre16}%
  \BibitemOpen
  \bibfield  {author} {\bibinfo {author} {\bibfnamefont {B.}~\bibnamefont
  {Peropadre}}, \bibinfo {author} {\bibfnamefont {G.~G.}\ \bibnamefont
  {Guerreschi}}, \bibinfo {author} {\bibfnamefont {J.}~\bibnamefont {Huh}},\
  and\ \bibinfo {author} {\bibfnamefont {A.}~\bibnamefont {Aspuru-Guzik}},\
  }\bibfield  {title} {\bibinfo {title} {Proposal for microwave boson
  sampling},\ }\href@noop {} {\bibfield  {journal} {\bibinfo  {journal} {Phys.
  Rev. Lett.}\ }\textbf {\bibinfo {volume} {117}},\ \bibinfo {pages} {140505}
  (\bibinfo {year} {2016})}\BibitemShut {NoStop}%
\bibitem [{\citenamefont {Preiss}\ \emph {et~al.}(2015)\citenamefont {Preiss},
  \citenamefont {Ma}, \citenamefont {Tai}, \citenamefont {Simon},\ and\
  \citenamefont {Greiner}}]{Preiss15}%
  \BibitemOpen
  \bibfield  {author} {\bibinfo {author} {\bibfnamefont {P.~M.}\ \bibnamefont
  {Preiss}}, \bibinfo {author} {\bibfnamefont {R.}~\bibnamefont {Ma}}, \bibinfo
  {author} {\bibfnamefont {M.~E.}\ \bibnamefont {Tai}}, \bibinfo {author}
  {\bibfnamefont {J.}~\bibnamefont {Simon}},\ and\ \bibinfo {author}
  {\bibfnamefont {M.}~\bibnamefont {Greiner}},\ }\bibfield  {title} {\bibinfo
  {title} {Quantum gas microscopy with spin, atom-number, and multilayer
  readout},\ }\href {https://doi.org/10.1103/PhysRevA.91.041602} {\bibfield
  {journal} {\bibinfo  {journal} {Phys. Rev. A}\ }\textbf {\bibinfo {volume}
  {91}},\ \bibinfo {pages} {041602} (\bibinfo {year} {2015})}\BibitemShut
  {NoStop}%
\bibitem [{\citenamefont {Leghtas}\ \emph {et~al.}(2015)\citenamefont
  {Leghtas}, \citenamefont {Touzard}, \citenamefont {Pop}, \citenamefont {Kou},
  \citenamefont {Vlastakis}, \citenamefont {Petrenko}, \citenamefont {Sliwa},
  \citenamefont {Narla}, \citenamefont {Shankar}, \citenamefont {Hatridge},
  \citenamefont {Reagor}, \citenamefont {Frunzio}, \citenamefont {Schoelkopf},
  \citenamefont {Mirrahimi},\ and\ \citenamefont {Devoret}}]{leghtas15}%
  \BibitemOpen
  \bibfield  {author} {\bibinfo {author} {\bibfnamefont {Z.}~\bibnamefont
  {Leghtas}}, \bibinfo {author} {\bibfnamefont {S.}~\bibnamefont {Touzard}},
  \bibinfo {author} {\bibfnamefont {I.~M.}\ \bibnamefont {Pop}}, \bibinfo
  {author} {\bibfnamefont {A.}~\bibnamefont {Kou}}, \bibinfo {author}
  {\bibfnamefont {B.}~\bibnamefont {Vlastakis}}, \bibinfo {author}
  {\bibfnamefont {A.}~\bibnamefont {Petrenko}}, \bibinfo {author}
  {\bibfnamefont {K.~M.}\ \bibnamefont {Sliwa}}, \bibinfo {author}
  {\bibfnamefont {A.}~\bibnamefont {Narla}}, \bibinfo {author} {\bibfnamefont
  {S.}~\bibnamefont {Shankar}}, \bibinfo {author} {\bibfnamefont {M.~J.}\
  \bibnamefont {Hatridge}}, \bibinfo {author} {\bibfnamefont {M.}~\bibnamefont
  {Reagor}}, \bibinfo {author} {\bibfnamefont {L.}~\bibnamefont {Frunzio}},
  \bibinfo {author} {\bibfnamefont {R.~J.}\ \bibnamefont {Schoelkopf}},
  \bibinfo {author} {\bibfnamefont {M.}~\bibnamefont {Mirrahimi}},\ and\
  \bibinfo {author} {\bibfnamefont {M.~H.}\ \bibnamefont {Devoret}},\
  }\bibfield  {title} {\bibinfo {title} {Confining the state of light to a
  quantum manifold by engineered two-photon loss},\ }\href@noop {} {\bibfield
  {journal} {\bibinfo  {journal} {Science}\ }\textbf {\bibinfo {volume}
  {347}},\ \bibinfo {pages} {853} (\bibinfo {year} {2015})}\BibitemShut
  {NoStop}%
\bibitem [{\citenamefont {Zhang}\ \emph {et~al.}(2021)\citenamefont {Zhang},
  \citenamefont {Chen}, \citenamefont {Yao},\ and\ \citenamefont
  {Chin}}]{zhang21f}%
  \BibitemOpen
  \bibfield  {author} {\bibinfo {author} {\bibfnamefont {Z.}~\bibnamefont
  {Zhang}}, \bibinfo {author} {\bibfnamefont {L.}~\bibnamefont {Chen}},
  \bibinfo {author} {\bibfnamefont {K.-X.}\ \bibnamefont {Yao}},\ and\ \bibinfo
  {author} {\bibfnamefont {C.}~\bibnamefont {Chin}},\ }\bibfield  {title}
  {\bibinfo {title} {Transition from an atomic to a molecular bose--einstein
  condensate},\ }\href@noop {} {\bibfield  {journal} {\bibinfo  {journal}
  {Nature}\ }\textbf {\bibinfo {volume} {592}},\ \bibinfo {pages} {708}
  (\bibinfo {year} {2021})}\BibitemShut {NoStop}%
\bibitem [{\citenamefont {Verresen}\ \emph {et~al.}(2021)\citenamefont
  {Verresen}, \citenamefont {Lukin},\ and\ \citenamefont
  {Vishwanath}}]{Verresen21}%
  \BibitemOpen
  \bibfield  {author} {\bibinfo {author} {\bibfnamefont {R.}~\bibnamefont
  {Verresen}}, \bibinfo {author} {\bibfnamefont {M.~D.}\ \bibnamefont
  {Lukin}},\ and\ \bibinfo {author} {\bibfnamefont {A.}~\bibnamefont
  {Vishwanath}},\ }\bibfield  {title} {\bibinfo {title} {Prediction of toric
  code topological order from rydberg blockade},\ }\href
  {https://doi.org/10.1103/PhysRevX.11.031005} {\bibfield  {journal} {\bibinfo
  {journal} {Phys. Rev. X}\ }\textbf {\bibinfo {volume} {11}},\ \bibinfo
  {pages} {031005} (\bibinfo {year} {2021})}\BibitemShut {NoStop}%
\bibitem [{\citenamefont {Semeghini}\ \emph {et~al.}(2021)\citenamefont
  {Semeghini}, \citenamefont {Levine}, \citenamefont {Keesling}, \citenamefont
  {Ebadi}, \citenamefont {Wang}, \citenamefont {Bluvstein}, \citenamefont
  {Verresen}, \citenamefont {Pichler}, \citenamefont {Kalinowski},
  \citenamefont {Samajdar}, \citenamefont {Omran}, \citenamefont {Sachdev},
  \citenamefont {Vishwanath}, \citenamefont {Greiner}, \citenamefont
  {Vuleti{\'{c}}},\ and\ \citenamefont {Lukin}}]{Semeghini21}%
  \BibitemOpen
  \bibfield  {author} {\bibinfo {author} {\bibfnamefont {G.}~\bibnamefont
  {Semeghini}}, \bibinfo {author} {\bibfnamefont {H.}~\bibnamefont {Levine}},
  \bibinfo {author} {\bibfnamefont {A.}~\bibnamefont {Keesling}}, \bibinfo
  {author} {\bibfnamefont {S.}~\bibnamefont {Ebadi}}, \bibinfo {author}
  {\bibfnamefont {T.~T.}\ \bibnamefont {Wang}}, \bibinfo {author}
  {\bibfnamefont {D.}~\bibnamefont {Bluvstein}}, \bibinfo {author}
  {\bibfnamefont {R.}~\bibnamefont {Verresen}}, \bibinfo {author}
  {\bibfnamefont {H.}~\bibnamefont {Pichler}}, \bibinfo {author} {\bibfnamefont
  {M.}~\bibnamefont {Kalinowski}}, \bibinfo {author} {\bibfnamefont
  {R.}~\bibnamefont {Samajdar}}, \bibinfo {author} {\bibfnamefont
  {A.}~\bibnamefont {Omran}}, \bibinfo {author} {\bibfnamefont
  {S.}~\bibnamefont {Sachdev}}, \bibinfo {author} {\bibfnamefont
  {A.}~\bibnamefont {Vishwanath}}, \bibinfo {author} {\bibfnamefont
  {M.}~\bibnamefont {Greiner}}, \bibinfo {author} {\bibfnamefont
  {V.}~\bibnamefont {Vuleti{\'{c}}}},\ and\ \bibinfo {author} {\bibfnamefont
  {M.~D.}\ \bibnamefont {Lukin}},\ }\bibfield  {title} {\bibinfo {title}
  {Probing topological spin liquids on a programmable quantum simulator},\
  }\href {https://doi.org/10.1126/science.abi8794} {\bibfield  {journal}
  {\bibinfo  {journal} {Science}\ }\textbf {\bibinfo {volume} {374}},\ \bibinfo
  {pages} {1242} (\bibinfo {year} {2021})}\BibitemShut {NoStop}%
\bibitem [{\citenamefont {Bluvstein}\ \emph {et~al.}(2022)\citenamefont
  {Bluvstein}, \citenamefont {Levine}, \citenamefont {Semeghini}, \citenamefont
  {Wang}, \citenamefont {Ebadi}, \citenamefont {Kalinowski}, \citenamefont
  {Keesling}, \citenamefont {Maskara}, \citenamefont {Pichler}, \citenamefont
  {Greiner} \emph {et~al.}}]{bluvstein2022epr}%
  \BibitemOpen
  \bibfield  {author} {\bibinfo {author} {\bibfnamefont {D.}~\bibnamefont
  {Bluvstein}}, \bibinfo {author} {\bibfnamefont {H.}~\bibnamefont {Levine}},
  \bibinfo {author} {\bibfnamefont {G.}~\bibnamefont {Semeghini}}, \bibinfo
  {author} {\bibfnamefont {T.~T.}\ \bibnamefont {Wang}}, \bibinfo {author}
  {\bibfnamefont {S.}~\bibnamefont {Ebadi}}, \bibinfo {author} {\bibfnamefont
  {M.}~\bibnamefont {Kalinowski}}, \bibinfo {author} {\bibfnamefont
  {A.}~\bibnamefont {Keesling}}, \bibinfo {author} {\bibfnamefont
  {N.}~\bibnamefont {Maskara}}, \bibinfo {author} {\bibfnamefont
  {H.}~\bibnamefont {Pichler}}, \bibinfo {author} {\bibfnamefont
  {M.}~\bibnamefont {Greiner}}, \emph {et~al.},\ }\bibfield  {title} {\bibinfo
  {title} {A quantum processor based on coherent transport of entangled atom
  arrays},\ }\href {https://doi.org/10.1038/s41586-022-04592-6} {\bibfield
  {journal} {\bibinfo  {journal} {Nature}\ }\textbf {\bibinfo {volume} {604}},\
  \bibinfo {pages} {451} (\bibinfo {year} {2022})}\BibitemShut {NoStop}%
\bibitem [{\citenamefont {Agarwal}\ \emph {et~al.}(2022)\citenamefont
  {Agarwal}, \citenamefont {Langlett},\ and\ \citenamefont {Xu}}]{Agarwal22}%
  \BibitemOpen
  \bibfield  {author} {\bibinfo {author} {\bibfnamefont {L.}~\bibnamefont
  {Agarwal}}, \bibinfo {author} {\bibfnamefont {C.~M.}\ \bibnamefont
  {Langlett}},\ and\ \bibinfo {author} {\bibfnamefont {S.}~\bibnamefont {Xu}},\
  }\href@noop {} {\bibinfo {title} {Long-range bell states from local
  measurements and many-body teleportation without time-reversal}} (\bibinfo
  {year} {2022}),\ \Eprint {https://arxiv.org/abs/arXiv:2205.02782}
  {arXiv:2205.02782} \BibitemShut {NoStop}%
\bibitem [{\citenamefont {Oganesyan}\ and\ \citenamefont
  {Huse}(2007)}]{Poisson}%
  \BibitemOpen
  \bibfield  {author} {\bibinfo {author} {\bibfnamefont {V.}~\bibnamefont
  {Oganesyan}}\ and\ \bibinfo {author} {\bibfnamefont {D.~A.}\ \bibnamefont
  {Huse}},\ }\bibfield  {title} {\bibinfo {title} {Localization of interacting
  fermions at high temperature},\ }\href
  {https://doi.org/10.1103/PhysRevB.75.155111} {\bibfield  {journal} {\bibinfo
  {journal} {Phys. Rev. B}\ }\textbf {\bibinfo {volume} {75}},\ \bibinfo
  {pages} {155111} (\bibinfo {year} {2007})}\BibitemShut {NoStop}%
\bibitem [{Note1()}]{Note1}%
  \BibitemOpen
  \bibinfo {note} {The Bose-Hubbard bilayer model is the only model that
  utilizes periodic boundary conditions in the respective numerics. For all
  other bilayer models we use open boundary conditions.}\BibitemShut {Stop}%
\end{thebibliography}%
\end{document}